\documentclass[11pt]{article}
\newcommand{\Comment}[1]{{}}
\usepackage{amsfonts,amsthm,amsmath,amssymb}
\usepackage[textwidth = 480 pt, textheight = 650 pt]{geometry}

\usepackage{color}
\definecolor{MyDarkBlue}{rgb}{0.15,0.15,0.45}
\usepackage[linktocpage=true]{hyperref}
\hypersetup{
colorlinks=true,
citecolor=MyDarkBlue,
linkcolor=MyDarkBlue,
urlcolor=MyDarkBlue,
pdfauthor={Horatiu Nastase, Constantinos Papageorgakis and Sanjaye Ramgoolam},
pdftitle={The fuzzy S^2 structure of M2-M5 systems in ABJM membrane theories},
pdfsubject={hep-th}
}

\usepackage[numbers,sort&compress]{natbib}
\usepackage{hypernat}

\newcommand\ignore[1]{}
\def\one{{\,\hbox{1\kern-.8mm l}}}

\def\Tr{{\rm Tr\, }}

\newcommand{\SO}{\mathrm{SO}} 
\newcommand{\SU}{\mathrm{SU}} \newcommand{\U}{\mathrm{U}}
\newcommand{\ie}{\emph{i.e.}\:}
\newcommand{\eg}{\emph{e.g.}\;}

\def\a{\alpha}\def\b{\beta}
\def\g{\gamma}

\def\s{\sigma}

\def\d{\partial}

\def\R{\mathbb{R}}

\def\hJ{ \hat{J} }\def\hJb{ \hat{ \bar { J } } }
\def\bV{ {\bf{V}} }\def\bVp{ {\bf{V}}^+ }
\def\bVm{ {\bf{V}}^- }\def\Rd{ R^{\dagger} } 
\def\Td{ T^{\dagger} } 

\newcommand{\hF}{\hat F}
\newcommand{\cA}{\mathcal A}

\newcommand{\cF}{\mathcal F}

\newcommand{\cL}{\mathcal L}
\newcommand{\cN}{\mathcal N}\newcommand{\cO}{\mathcal O}
\newcommand{\cP}{\mathcal P}

\newcommand{\be}{\begin{equation}}
\newcommand{\bea}{\begin{eqnarray}}

\newcommand{\ee}{\end{equation}}
\newcommand{\eea}{\end{eqnarray}}

\newcommand{\nn}{\nonumber}

\def\hE{\hat{E}}

\def\Gd{ G^{\dagger} }

\def\ts{ \tilde \sigma }
\def\Gd{ G^{\dagger} }

\def\Cd{ C^{\dagger} }

\def\Xone{ X^{(1)} } 
\def\Xtwo{ X^{(2)} } 

\begin{document}

\makeatletter
\@addtoreset{equation}{section}
\makeatother
\renewcommand{\theequation}{\thesection.\arabic{equation}}

\rightline{TIT/HEP-593}
\rightline{TIFR/TH/09-07}
\rightline{QMUL-PH-09-04}
   \vspace{1.8truecm}

\vspace{15pt}

%%%%%%%%%%%%%%%%%

\centerline{\LARGE \bf The fuzzy $S^2$ structure of   M2-M5 systems } 
\centerline{ \LARGE\bf  in ABJM membrane theories} \vspace{1truecm}
\thispagestyle{empty} \centerline{
    {\large \bf Horatiu Nastase${}^{a,}$}\footnote{E-mail address: \href{mailto:nastase.h.aa@m.titech.ac.jp}{\tt nastase.h.aa@m.titech.ac.jp}},
    {\large \bf Constantinos Papageorgakis${}^{b,}$}\footnote{E-mail address:
                                 \href{mailto:costis@theory.tifr.res.in}{\tt costis@theory.tifr.res.in}}
    {\bf and}
    {\large \bf Sanjaye Ramgoolam${}^{c,}$}\footnote{E-mail address: \href{mailto:s.ramgoolam@qmul.ac.uk}{\tt s.ramgoolam@qmul.ac.uk}}
                                                       }

\vspace{.4cm}
\centerline{{\it ${}^a$ Global Edge Institute, Tokyo institute of Technology,}}
\centerline{{\it Tokyo 152-8550, Japan}}

\vspace{.4cm}
\centerline{{\it ${}^b$ Department of Theoretical Physics, Tata
    Institute of Fundamental Research},} \centerline{{\it Homi Bhabha
    Road, Mumbai 400 005, India}}

\vspace{.4cm}
\centerline{{\it ${}^c$ Centre for Research in String Theory, Department of Physics},}
\centerline{{ \it Queen Mary, University of London},} \centerline{{\it
    Mile End Road, London E1 4NS, UK}} 

\vspace{1.4truecm}

%%%%%%%%%%%%%%%%%
\thispagestyle{empty}

\centerline{\bf ABSTRACT}

\vspace{.4truecm}

\noindent
We analyse the fluctuations of the ground-state/funnel solutions proposed to describe M2-M5 systems in the level-$k$ mass-deformed/pure Chern-Simons-matter ABJM theory of multiple membranes. We show that in the large $N$ limit the fluctuations approach the space of functions on the 2-sphere rather than the naively expected 3-sphere.  This is a novel realisation of the fuzzy 2-sphere in the context of Matrix Theories, which uses bifundamental instead of adjoint scalars. Starting from the multiple M2-brane action, a $\U(1)$ Yang-Mills theory on $\R^{2,1} \times S^2 $ is recovered at large $N$, which is consistent with a single D4-brane interpretation in Type IIA string theory. This is as expected at large $k$, where the semiclassical analysis is valid. Several aspects of the fluctuation analysis, the ground-state/funnel solutions and the mass-deformed/pure ABJM equations can be understood in terms of a discrete noncommutative realisation of the Hopf fibration.  We discuss the implications for the possibility of finding an M2-brane worldvolume derivation of the classical $S^3$ geometry of the M2-M5 system.  Using a rewriting of the equations of the $\SO(4)$-covariant fuzzy 3-sphere construction, we also directly compare this fuzzy 3-sphere against the ABJM ground-state/funnel solutions and show them to be different.

\vspace{.5cm}

\setcounter{page}{0}
\setcounter{tocdepth}{2}

\newpage

\tableofcontents

\setcounter{footnote}{0}

\linespread{1.1}
\parskip 4pt

{}~
{}~

\section{Introduction}\label{intro}
 
M2 worldvolume theories have been at the centre of intense recent activity after the discovery of manifestly $\cN = 8$ superconformal three dimensional Lagrangians by Bagger-Lambert and Gustavsson (BLG) \cite{Bagger:2006sk,Bagger:2007jr,Bagger:2007vi,Gustavsson:2007vu}, following earlier insights of \cite{Schwarz:2004yj,Basu:2004ed}.  Inspired by subsequent related developments \cite{Mukhi:2008ux,Bandres:2008vf,VanRaamsdonk:2008ft,Lambert:2008et,Distler:2008mk} and independent advances in supersymmetric Chern-Simons-matter theories \cite{Gaiotto:2008sd,Hosomichi:2008jd}, ABJM \cite{Aharony:2008ug} have proposed an $\cN = 6$ Chern-Simons-matter theory with $\U(N) \times \U(N) $ gauge group, $\SO(6)$ R-symmetry and equal but opposite Chern-Simons (CS) levels $(k,-k)$, to capture the dynamics of the low-energy limit of multiple M2-branes on an M-theory orbifold, $\mathbb C^4/\mathbb Z_k$.  At $k=1$ the theory is strongly coupled and is conjectured to describe membranes in flat space. The supersymmetry and R-symmetry were argued to be nonperturbatively enhanced to $\cN = 8$ and $\SO(8)$ respectively.

An important test for any candidate theory of multiple membranes in flat space is that it should reproduce the physics of M2$\perp$M5 intersections.  These are M2-spike solutions of the M5-brane worldvolume theory \cite{Howe:1997ue}, generalising the D1-spike solutions on D3 worldvolumes.  In the case of D1$\perp$D3 systems, there is a detailed understanding of the physics, both from the D1 and the D3-worldvolume perspectives \cite{Constable:1999ac}.\footnote{Similar studies for the   D1$\perp$D5 and D1$\perp$D7 systems have also been undertaken in \cite{Constable:2001ag,     Cook:2003rx}.} They can be captured equivalently by a $\U(N)$ theory of multiple D1-branes and a $\U(1)$ theory of a single D3-brane with $N$ units of magnetic flux.  From the point of view of the $\U(N)$ theory the D1$\perp$D3 system is described in terms of a solution involving fuzzy 2-spheres \cite{Hoppe:1982,Hoppe:1988gk,Madore:1991bw} through the Myers effect \cite{Myers:1999ps}. These are related to families of matrices obeying
 \bea\label{SU(2)condition} 
[ X^{i} , X^{j} ] = 2i \epsilon^{ijk } X^k\;.
\eea 
The $X^i$ enter the physics as ans\"atze for solving the equations of motion and their commutator action on the space of all $N \times N $ matrices organises these matrices into representations of $\SU(2) \simeq \SO(3)$.  An important aspect of the geometry of the fuzzy 2-sphere involves the construction of fuzzy (matrix) spherical harmonics in $\SU(2)$ representations, which approach the space of all classical $S^2$ spherical harmonics in the limit of large matrices \cite{Hoppe:1988gk}. This construction of fuzzy spherical harmonics allows the analysis of fluctuations in a nonabelian theory of Dp-branes to be expressed at large $N$ in terms of an abelian higher dimensional theory of a D(p+2) brane. At finite $N$ the higher dimensional worldvolume theory becomes a noncommutative $\U(1)$ with a UV cutoff \cite{Iso:2001mg,Dasgupta:2002hx,Dasgupta:2002ru,Papageorgakis:2005xr}.

The relation (\ref{SU(2)condition}) also appears as an F-flatness condition for a particular mass deformation of $\cN=4$ SYM, called $\cN=1^*$, where all three adjoint chiral multiplets have acquired equal mass. This theory has a discrete set of  vacua that satisfy the above condition, which from the string theory point of view can be interpreted as $N$ D3 branes blowing up into concentric D5 branes with flux that adds up to $N$ \cite{Polchinski:2000uf}. The matrix structure of these solutions is the same as for the D1$\perp$D3 spikes, with the relative transverse directions between the D3 and the D5 forming a fuzzy $S^2$ at finite $N$, which can again be seen by analysing small fluctuations around the  vacua \cite{Dorey:2002ad,Andrews:2005cv,Andrews:2006aw}.  The lift of this system to M-theory has been discussed in \cite{Bena:2000zb}.  The natural generalisation of the D3-D5 system is M2-M5, and a theory of multiple membranes should include such solutions.

With the advent of the ABJM action it is possible to investigate both these kinds of M2-M5 configurations in greater detail.  An interesting class of mass deformations of the ABJM theory were given in \cite{Hosomichi:2008jb, Gomis:2008vc}. An important difference between the latter theories and $\cN=1^*$ is that they preserve the same $\cN = 6$ amount of supersymmetry as the parent theory but still break R-symmetry, down to $\SU(2)\times \SU(2)\times \U(1)$, as well as conformal invariance.  The authors of \cite{Gomis:2008vc} also presented a set of  ground-state solutions.  Closely related solutions describe M2$\perp$M5 funnels in the undeformed ABJM theory \cite{Gomis:2008vc} and have subsequently been discussed in \cite{Terashima:2008sy,   Hanaki:2008cu, Fujimori:2008ga, Arai:2008kv,Low:2009kv}.  The explicit form of the matrices for these solutions was interpreted as exhibiting a fuzzy $S^3$ structure. Our aim here will be to revisit this claim by examining the algebra obeyed by the matrices. Hence, the study of these solutions and their associated fuzzy sphere geometry will be at the central focus of this paper.

Our main result will be that the above ground-state/funnel solutions actually describe fuzzy $S^2$'s, as opposed to fuzzy $S^3$'s, albeit realised in a new fashion involving the noncommutative base of the $S^1\hookrightarrow S^3\stackrel{\pi}{\rightarrow} S^2$ Hopf fibration. We verify this by analysing small fluctuations around the ground-states at large $N,k$ and showing that they can be organised in terms of a $\U(1)$ theory on $\mathbb R^{2,1}\times S^2$ consistent with an interpretation as a D4-brane in Type IIA.  The calculation involves a combination of techniques familiar from the Matrix Theory literature and the novel Higgs mechanism \cite{Mukhi:2008ux} that relates $\U(N)\times\U(N)$ CS-matter theories with bifundamental matter to $\U(N)$ YM theories with adjoint matter.

It is interesting that the equations of motion defining the ground-state solutions of \cite{Gomis:2008vc} are the same as the BPS equations for the fuzzy funnel of the ABJM model (once the dependence on the worldvolume coordinate along the funnel is factored out). These equations are in turn a generalisation of the BPS equations of \cite{Basu:2004ed} that inspired Bagger and Lambert to define their theory. Our result will imply that the special case of the Bagger-Lambert $ \cA_4$-theory is as of yet unique in describing a fuzzy $S^3$; in all other cases of BLG/ABJM-type three dimensional Lagrangians one is dealing with fuzzy $S^2$ solutions, despite the original motivation.

A natural definition of the fuzzy $S^3$ is that it should involve an
 $\SO(4)$-covariant construction using matrices or some other appropriate
 algebraic structure, generalising the fuzzy $S^2$ matrix construction.
 Accordingly, a fully $\SO(4)$-covariant matrix construction of fuzzy
 3-spheres should exhibit matrix spherical harmonics transforming under
 $\SO(4)$, which approach the classical spherical harmonics on $S^3$ in
 a large $N$ limit.  Such a construction was given by Guralnik and one of
 the present authors in
 \cite{Guralnik:2000pb,Ramgoolam:2001zx,Ramgoolam:2002wb}. The Guralnik-Ramgoolam (GR) construction
 has its mysteries, in that the matrix algebra contains the spherical harmonics only as a subset of
 all the matrices. A complete construction of a finite $N$ geometry of fuzzy 3-spheres has to take
 account of the projection which does away with the extra degrees of freedom. This leads to 
non-associativity \cite{Ramgoolam:2001zx}. 
Related developments  on fuzzy odd spheres appear in 
\cite{Azuma:2002zi,Janssen:2003ri,Dolan:2003kq,Nair:2003st,Nastase:2004te,
SheikhJabbari:2004ik,Janssen:2004cd,Lozano:2005kf,Papageorgakis:2006ed,Lozano:2006jr}.

In Section\,\ref{BLGABJM}, we begin by writing the equations obeyed by the generating matrices of the GR fuzzy $S^3$ in a new form, appropriate for comparison with the equations of motion of ABJM and BLG.  We find that the GR fuzzy $S^3$ is a solution for the $N=2$ case of the ABJM theory. This $\U(2) \times \U(2)$ theory is closely related to the $\cA_4$-theory of BLG, which is the same as the $\SU(2)\times \SU(2)$ ABJM theory. However, with our rewriting of the equations obeyed by the matrices of the GR fuzzy $S^3$, we will explicitly see that these become different from the solutions of ABJM that were found in \cite{Gomis:2008vc} for $ N > 2 $. Given that the mass-deformed theory of \cite{Gomis:2008vc} does not contain an $\SO(4)$ R-symmetry factor the absence of the `usual' fuzzy $S^3$ should perhaps not come as a surprise.\footnote{We note that for the extreme strongly-coupled region $k=1,2$ maximal supersymmetry and $\SO(8)$ R-symmetry are expected to be restored. In particular, the $k=1$ case corresponds to M2-branes in flat space. Since this region is well beyond our semiclassical, large-$k$ analysis, one cannot rule out the emergence of the  $S^3$ when the full quantum theory is taken into account. An example of  an emergent $S^5$ from quantum physics can be found in \cite{Maldacena:2002rb}.} The fact that the GR construction does work for $N=2$ is closely related to its role in the work of Basu and Harvey \cite{Basu:2004ed} and the BLG interpretation of the Basu-Harvey equation in terms of the of the $ \cA_4$ 3-algebra. Developments on fuzzy M-brane intersections following \cite{Basu:2004ed} are reviewed in \cite{Berman:2007bv,Copland:2007by}.

In Section\,\ref{fuzzystr}, we return to the ground-state solutions of ABJM theory at general $N$ given in \cite{Gomis:2008vc} in terms of matrices $G^{\alpha} , G^{\dagger}_{\alpha} $, and we make the fuzzy $S^2$ structure apparent. Since the ABJM theory has $ \U(N) \times \U(N) $ gauge symmetry, the Lie algebra is a direct sum which can be viewed as matrices acting on a direct sum of vector spaces $ \bVp \oplus \bVm $.  The ABJM action has a global $\SU(4)$ symmetry, which is broken by the mass deformation to $ \SU(2)\times \SU(2)\times \U(1) $ and the complex scalars decompose as $C^I = (R^\alpha,Q^{\dot\alpha})$. One of the $\SU(2)$'s acts on the coordinates $R^{\alpha } $ which are nonzero on the solutions.  The other $\SU(2)$ acts on the $Q^{\dot \alpha}$'s, which are zero on the solutions. The first $\SU(2)$ is relevant to the geometry described by the solution and we will call it the geometric $\SU(2)$.  The generators of this $ \SU(2)$ can be constructed from bilinears in $ G^\alpha , \Gd_\alpha $. We describe how the $N^2$-dimensional space of matrices acting on $\bVp$, called $End ( \bVp ) $, can be decomposed in terms of representations of this $\SU(2)$. Likewise we give the decomposition of $ End ( \bVm ) $.  Interestingly these turn out to be different.  We complete the $\SU(2)$ decompositions with those of $ Hom ( \bVp, \bVm )$ and $ Hom ( \bVm, \bVp )$, that is matrices mapping $ \bVp \rightarrow \bVm $ and $ \bVm \rightarrow \bVp $ respectively. These matrices are needed to describe the fluctuations of the matter fields which transform in the bifundamental of the gauge group. We obtain that the fuzzy spherical harmonics for all physical fluctuations are consistent with a fuzzy $S^2$, and reduce to the usual spherical harmonics, with no other factors, in the `classical' (large $N$) limit.

In Section\,\ref{matter}, we calculate the action for the matter fluctuations to quadratic order around the fuzzy solution, and take their classical limit. In Section\,\ref{CSHiggs}, we then analyse the gauge fields, for which we use the Higgs mechanism of \cite{Mukhi:2008ux} to turn the nonpropagating 3d CS-type action for $A^{(1)}$ and $A^{(2)}$ into the propagating 5d YM action. The full action for the fluctuations takes the form of an abelian YM theory on $ \R^{2,1} \times S^2 $ and we explain this result in terms of a D4 wrapping the $S^2$ in Type IIA string theory. In Section\,\ref{sechopf} we show how the IIA D4-brane interpretation is expected by considering the M2-M5 system in the large $k$ limit of the M-theory quotient of \cite{Aharony:2008ug}. The action of the quotient on the $ S^1$ Hopf fibre of the $S^3$ cross-section of the M2-M5 system leads to a IIA reduction.  The structure of the ground-state solutions and their fluctuations are retrospectively determined by considering possible finite matrix realisations of the fuzzy $S^2$ base of the $S^1\hookrightarrow S^3\stackrel{\pi}{\rightarrow} S^2$ Hopf fibration. In this discussion it becomes apparent that the $S^1$ fibre has (in a sense that we explain) been reduced to two points.  The appendices contain some formulae and details of the calculations in Section\,\ref{matter}.

Related work pertaining to M5-branes in the more general context of 3-algebra theories includes \cite{Berman:2008be,Ho:2008ei,Ho:2008ve,Ho:2008nn,Bandos:2008fr, Chu:2009iv,Bonelli:2008kh,Krishnan:2008zm,Krishnan:2008xh}.

\section{The $\SO(4)$-covariant fuzzy three-sphere in  BLG and ABJM theories}\label{BLGABJM}

In this section, we will rewrite the definition of the GR fuzzy $S^3$ in a way that can 
be embedded in BLG/ABJM theory and then see that it only matches the BPS fuzzy sphere 
solution of BLG and ABJM theory in the $r=1$ case of the $\Gamma$-matrix representation.

{\em Notation:} We will denote by $M,N,...=1,8$ the 8 real indices of coordinates
 transverse to an M2 brane $X^I$, by $m,n,...=1,4$ the indices of 4 real coordinates
 $X_m$ that will form a fuzzy $S_3$, with $X_m X^m=1$, by $a,b,...$ the indices of 
generators of algebras $T_a$, for both regular Lie algebras and 3-algebras, by
 $I,J,...=1,4$ the indices of 4 complex coordinates $C^I$ transverse to an M2, 
splitting into $\alpha,\beta,...=1,2$ and $\dot{\a},\dot{\b},...=1,2$, by
 $A,B,...=1,6$ the $\SO(6)$ indices of ${\cal N}=6$ supersymmetry 
$\epsilon^A$ and by $i,j,...=1,3$ the indices defining an $S^2$, 
with $X_iX^i=1$. Also, for $\SU(N)$ matrices that can be written 
as tensor products of $\Gamma$-matrix representations, we will denote 
by  $r$ the number of such representations.

\subsection{The $\SO(4)$-covariant (Guralnik-Ramgoolam) fuzzy $S^3$}

The $\SO(4)$-covariant matrix construction of the fuzzy $S^3$, in terms of tensor products
of $\Gamma$-matrices, was done by Guralnik and Ramgoolam (GR) in
\cite{Guralnik:2000pb, Ramgoolam:2001zx,Ramgoolam:2002wb}. One splits the vector
space of spinor representations of $\SO(4)$ under $\SO(4)\simeq\SU(2)\times
\SU(2)$ as $V=V_+\oplus V_-$. Take the subspace ${\cal R}_r$ of
$Sym(V^{\otimes r})$, acting as the basis vector space for the fuzzy
three-sphere representation. ${\cal R}_r$ is defined as ${\cal R}_r^+\oplus{\cal
  R}_r^-$, where ${\cal R}_r^+$ and ${\cal R}_r^-$ are spaces of
$\SU(2)\times \SU(2)$ with labels $(\frac{r+1}{2}, \frac{r-1}{2})$ and
$(\frac{r-1}{2}, \frac{r+1}{2})$ respectively, denoting the number of
$V_+$ and $V_-$ factors in each. Then the fuzzy $S^3$ coordinates
$X_m$ ($\SU(N)$ matrices) are given by 
\be\label{s3coords}
 X_m= {\cal P}_{{\cal
    R}_r}\hat{X}_m{\cal P}_{{\cal R}_r}= {\cal P}_{{\cal
    R}_r^+}\hat{X}_m{\cal P}_{{\cal R}_r^-}+ {\cal P}_{{\cal
    R}_r^-}\hat{X}_m{\cal P}_{{\cal R}_r^+}=X_m^++X_m^-\;,
 \ee
 where ${\cal P}_{{\cal R}_r}={\cal P}_{{\cal R}_r^+}+{\cal P}_{{\cal
    R}_r^-}$, with ${\cal P}_{{\cal R}_r^+}$, ${\cal P}_{{\cal
    R}_r^-}$ being projectors onto the subspaces ${\cal R}_r^+$ and ${\cal
    R}_r^-$ respectively and 
\be
 \hat{X}_m=\sum_l
\rho_l(\Gamma_m)\;,
 \ee
with $\rho_l$ denoting which factor of $Sym(V^{\otimes r})$ the $\Gamma$-matrix acts on.
Thus the coordinates of the fuzzy $S^3$ are represented by the matrices $X_m=X_m^++X_m^-$, where
\be
X_m^+= {\cal P}_{{\cal R}_r^-}\sum_l \rho_l(\Gamma_m P_+)
{\cal P}_{{\cal R}_r^+}\qquad\textrm{and}\qquad
X_m^-= {\cal P}_{{\cal R}_r^+}\sum_l \rho_l(\Gamma_m P_-)
{\cal P}_{{\cal R}_r^-}\;.
\ee
These turn out not to be sufficient 
to describe the fuzzy $S^3$; one  also needs to consider the matrices $Y_m=X_m^+-X_m^-$. Equivalently, one 
can  instead use the coordinates $X_m^+$ and $X_m^-$. Commutation relations for these objects were found in 
\cite{Ramgoolam:2002wb} and were simplified in \cite{Nastase:2004te}. 
These can be understood to act as equations of motion for the matrix fields, or 
alternatively as a defining algebra. The simplified set found in \cite{Nastase:2004te} is 
\be
[J_{mn},X_n]=6X_m\;,  \qquad [J_{mn},Y_n]=6Y_m\label{defining}\;,
\ee
where $X_m$ are the sphere coordinates, satisfying $X_mX_m=N\one$, while 
\be
J_{mn}=\alpha[X_m,X_n]+\beta\epsilon_{mnpq}\{ X_p,Y_q\}\label{defini}\;,
\ee
with
\be
\alpha=-\frac{2}{(r+1)(r+3)}\qquad\textrm{and}\qquad
\beta=\frac{r+2}{(r+1)(r+3)}
\ee
and also the constraint
\be
[X_m,X_n]=-[Y_m,Y_n]\label{constra}\;.
\ee

We would now like to write this definition of the fuzzy $S^3$ in a form that can be straightforwardly compared with the BLG/ABJM theories.

In \cite{Papageorgakis:2006ed}
 it was proved that the fuzzy $S^3$ also satisfies the Matrix model type equation 
\be
[[X_m,X_n],X_n]=\Big((r+1)(r+3) +4\Big)X_m\;,
\ee
which means that the $X_m$ equations in (\ref{defining}) become 
\be
\epsilon^{mnpq}[\{X_p, Y_q\},X_n]=8\Big(\frac{(r+1)(r+3)+1}{r+2}\Big) X_m\label{necessary}\;.
\ee
The second equation in (\ref{defining}) gives the same result with $X_m$ and $Y_m$ interchanged. However, since $X_m=X_m^++X_m^-$ and $Y_m=X_m^+-X_m^-$ while $X_m^+X_n^+=X_m^-X_n^-=0$, we have the 
more restrictive constraints (as compared to (\ref{constra}))
\bea
X_mX_n&=&-Y_mY_n\cr
X_mY_n&=&-Y_mX_n\label{constraints}\;.
\eea
This in turn means that (\ref{necessary}) becomes 
\be
\epsilon^{mnpq}X_n Y_p X_q=2\Big(\frac{(r+1)(r+3)+1}{r+2}\Big) X_m\label{fuzzydef}
\ee
and there also exists a corresponding relation with $X_m$ interchanged with $Y_m$. Then (\ref{fuzzydef}) and the constraints
(\ref{constraints}) can be taken as a definition of the fuzzy $S^3$, where 
\be
X_mX_m=X_m^+X_m^-+X_m^-X_m^+=\frac{(r+1)(r+3)}{2}\equiv N \label{xsquare}
\ee
in a similar way to the fuzzy $S^2$, for which 
the defining relation is $X_iX_i=R^2$ together with the $\SU(2)$ algebra
\be
[X_i,X_j]=2i\epsilon_{ijk}X_k\;.
\ee

Another way to define the fuzzy $S^3$ is by writing the equations in terms of variables going from 
the $+$ space to the $-$ space or vice-versa, \ie $X_m^+$ and $X_m^-$. We obtain 
\bea
\epsilon^{mnpq}X^+_nX^-_pX^+_q&=&2\Big(\frac{(r+1)(r+3)+1}{r+2}\Big)X^+_m\nonumber\\
\epsilon^{mnpq}X^-_nX^+_pX^-_q&=&2\Big(\frac{(r+1)(r+3)+1}{r+2}\Big)X^-_m\label{defrel}\;,
\eea
which are to be supplemented with the sphere condition (\ref{xsquare}) and the constraints
\be
X^+_mX^+_n=X^-_mX^-_n=0 \label{newconstra}\;,
\ee
since in that case the constraints (\ref{constraints}) are automatically satisfied.

The relations (\ref{defrel}) together with the sphere condition (\ref{xsquare}) and the constraints (\ref{newconstra}) give a new definition of the GR fuzzy $S^3$ that can be easily matched against BLG/ABJM.

\subsection{Bagger-Lambert-Gustavsson theory}

The fuzzy $S^3$ solution that was proposed in the model found by Bagger-Lambert \cite{Bagger:2006sk} and independently Gustavsson \cite{Gustavsson:2007vu} is similar to the one defined above and we will see that the two match for the case of the $\SO(4)$, or $\cA_4$-algebra. However, since $\cA_4$ is the only example of a finite dimensional 3-algebra theory with positive definite metric \cite{Papadopoulos:2008sk,Gauntlett:2008uf}, this is somewhat disappointing as we would like to find the $S^3$ as a solution to the theory for any number $N$ of M2-branes. In retrospect, one needs to look instead at the ABJM theory, but we will see that there one also encounters problems in identifying the GR fuzzy $S^3$.

In \cite{Basu:2004ed}  Basu and Harvey constructed a fuzzy
$S^3$ funnel for the $X^m$ coordinates 
of M5-branes, as obtained from the hypothetical worldvolume theory of multiple M2-branes, with BPS equation
\be
\frac{dX^m}{ds}+\frac{k}{4!}\epsilon^{mnpq}[G,X^n,X^p,X^q]=0\;.
\ee
This is solved by the ansatz
\be
 X^m(s)\propto \frac{1}{\sqrt{s}}G^m
\ee
and the corresponding fuzzy 3-sphere relation
\be
G^m \sim \epsilon^{mnpq}[G,G^n,G^p,G^q]\label{basuharvey}\;,
\ee
where $G$ is a fixed matrix and $s$ is the direction along which the membranes extend away from the fivebrane. Subsequently, Bagger and Lambert \cite{Bagger:2006sk,Bagger:2007vi}
proposed the fuzzy funnel (BPS) equation 
\be
\frac{dX^m}{ds}=ik\epsilon^{mnpq}[X^n,X^p,X^q]\;.
\ee
Here $m,n,p,q$ are directions transverse to the M2s and defining the
fuzzy 3-funnel parallel to the M5 
by $X_mX_m=1$. The full set of M2 transverse scalars is decomposed as
$X^M=X^{M}_aT^a$ ($M=1,..,8$ are all the coordinates transverse 
to M2, including the 4 coordinates $m,n,p,q$), where the $T^a$ are generators that define a 3-algebra \cite{Bagger:2007jr}
\be
[T^a,T^b,T^c]={f^{abc}}_dT^d\label{3algebra}
\ee
and the metric $h^{ab}=\Tr(T^aT^b)$, used to raise and lower indices, is assumed to be positive definite.\footnote{If the metric is Lorentzian \cite{Benvenuti:2008bt,Gomis:2008uv,Ho:2008ei} one has to deal with potential ghosts in the quantum theory that could lead to violation of unitarity. These have been shown to decouple by gauging a certain shift symmetry but the theory is then on-shell equivalent to usual $\cN = 8$ SYM in 3d  \cite{Bandres:2008kj, Gomis:2008be, Ezhuthachan:2008ch,Mukhi:2008ux}. There are various interpretations of the `un-gauged' version \cite{Verlinde:2008di,Cecotti:2008qs}, which has been shown to be obtained from ABJM through a particular scaling if one also adds a decoupled abelian `ghost' multiplet to the latter \cite{Antonyan:2008jf,Honma:2008jd,Honma:2008ef,Kluson:2009tz}.} 
Substituting the 3-algebra (\ref{3algebra}), 
we obtain the BPS equation for the commuting fields $X^m_a$
\be
\partial_s X^m_a=ik \epsilon^{mnpq} {f^{bcd}}_a X_b^nX_c^pX_d^q\label{scala}\;.
\ee
The corresponding fuzzy $S^3$ equation, which is an extension of the Basu-Harvey  equation (\ref{basuharvey}), is
\be
R^2X^m=ik\epsilon^{mnpq}[X^n,X^p,X^q]\;.
\ee
It would be natural to think of this equation, with $X_m$ 
satisfying $X_mX_m=R^2$, as the analogue of (\ref{fuzzydef}). However, note that the spacetime index $m$ 
of the commuting fields $X^m_a$ in (\ref{scala})
is contracted with $\epsilon^{mnpq}$, while the `noncommuting' (\ie matrix or algebra) 
index $a$ is contracted with ${f^{abc}}_d$, which makes it {\it a priori} different from (\ref{fuzzydef})
for general ${f^{abc}}_d$. 
Only in the case of the $\cA_4$ (or $\SO(4)$) algebra, when 
\be
f^{abcd}={f^{abc}}_eh^{de}=\tilde f\epsilon^{abcd}
\ee
do we have a similar equation, and in fact that is the only case for which the Bagger-Lambert equation is equivalent to the GR fuzzy 3-sphere. 

We can see this equivalence explicitly by using van Raamsdonk's reformulation \cite{VanRaamsdonk:2008ft}
of the $\cA_4$-theory in terms of 
a usual $\SO(4)\simeq\SU(2)\times \SU(2)$ CS gauge theory with bifundamental matter fields. The reformulation starts with
the observation that the objects $\tau^a=(i\sigma_j,{\one})$ obey
\be
\tau^{[a}\tau^{\dagger b}\tau^{c]}=-\epsilon^{abcd}\tau^d=\epsilon^{dabc}\tau^d\label{fundam}\;.
\ee
This nontrivial relation is obtained because  $\tau^a=(i\sigma_j,{\one})$ have 
different properties under conjugation: $(i\sigma_j)^\dagger=-i\sigma_j$, whereas ${\one}^\dagger={\one}$. Then
define
\be
X^M=\frac{1}{2}X^M_a\tau^a=\frac{1}{2}(X_4^M{\one}+iX_i^M\sigma^i)\;,
\ee
obeying the reality condition
\be
(X^M)^*=-\epsilon X^M\epsilon\;,
\ee
where $\epsilon=-i\sigma_2$ is the antisymmetric 2d symbol. These fields are now in a representation of 
$\SO(4)\simeq\SU(2)_1\times \SU(2)_2$, as bifundamental fields with one index in $\SU(2)_1$ and another one in $\SU(2)_2$, \ie ${(X^I)^\alpha}_{\dot{\alpha}}$, with $\alpha\in \SU(2)_1$ and $\dot{\alpha}\in \SU(2)_2$. 
Then (\ref{scala}) becomes 
\be
\partial_s X^m=-ik\tilde{f}\epsilon^{mnpq}X^nX^{\dagger p}X^q\label{vanR}\;.
\ee
The corresponding fuzzy $S^3$ equation
\be
R^2X^m=-ik\epsilon^{mnpq}X^nX^{\dagger p}X^q\label{fuzzvanR}
\ee
is thus the natural analogue of (\ref{defrel}) for $r=1$,
since just as  $X^m$ and $X^{\dagger m}$ 
\be
X_m^+=X^m\;\; {\rm and}\;\; X_m^-=X^{\dagger m}\label{identif}
\ee
are bifundamental and conjugate bifundamental fields respectively. The constraints (\ref{newconstra}), originating from the fact that $X_m^+$ and $X_m^-$ go from $V_+$ to $V_-$ and vice-versa, are 
also immediately satisfied due to the bifundamental nature of $X^m$.

Thus for $r=1$ the GR fuzzy three-sphere, as seen in the new description (\ref{defrel}), (\ref{newconstra}), matches with the fuzzy $S^3$ solution of the $\cA_4$-theory.  This is not too surprising since for $r=1$ the fuzzy 3-sphere coordinates (\ref{s3coords}) reduce to just $\SO(4)$ $\Gamma$-matrices. Yet, as can be easily seen from the above, the identification will not extend to $r>1$ (or equivalently, 3-algebras other than $\cA_4$ which are not necessarily $\cN =8$), since the $\SO(4)$ relation (\ref{fundam}) was crucial in turning (\ref{scala}) into (\ref{vanR}), which can in turn be related to the GR fuzzy three-sphere.

\subsection{ABJM theory}

To have any hope of seeing the $r>1$ realisation of the GR fuzzy $S^3$, one should depart from the $\cA_4$-theory and go to the ABJM model \cite{Aharony:2008ug}. These are superconformal CS-matter theories generalising \cite{VanRaamsdonk:2008ft} beyond $\SU(2)\times \SU(2)$ gauge groups, which in turn break $\cN=8$ supersymmetry and $\SO(8)$ R-symmetry. The choice leading to the theory dual to M2-branes on a $\mathbb C_4/\mathbb Z_k$ singularity is the one with $\U(N)\times \U(N)$, which is the case that we will be focusing on. One could still consider the $\SU(N)\times \SU(N)$ case, which reduces to the $\cA_4$-theory for $N=2$, but for which no M-theory dual interpretation has yet been found. Both these generically have explicit $\cN = 6$ supersymmetry and $\SU(4)$ R-symmetry. Different kinds of gauge groups like $\U(M)\times \U(N)$ and $\mathrm{O}(M)\times \mathrm{Sp}(N)$ have also been considered in the literature \cite{Hosomichi:2008jb,Aharony:2008gk}. A classification of all possibilities was made by \cite{Bagger:2008se,Schnabl:2008wj}. Since in this case one has bifundamental matter fields right from the beginning, they will automatically satisfy (\ref{newconstra}) and we might hope that we can obtain the GR fuzzy $S^3$.

The authors of \cite{Benna:2008zy} expressed the original ABJM Lagrangian in component form. It employs $\U(N)\times \U(N)$ bifundamental matrix
fields $X^M$, organised into the complex variables 
\be
Z^1=X^1+iX^2; \;\; Z^2=X^3+iX^4;\;\;\;
W_1={X^5}^\dagger+i{X^6}^\dagger;\;\;\;
W_2={X^7}^\dagger+i{X^8}^\dagger\label{complex}\;,
\ee
that is $Z^\a,Z^\dagger_\a,W_{\dot{\a}},W^{\dot{\a}\dagger}$, with $\a,\dot{\a}=1,2$. Then, in terms of 
$C^I=(Z^\a,W^{\dot{\a}\dagger})$ and $C^{\dagger}_I$, one gets 
the BPS  fuzzy funnel equation (setting $W_{\dot{\a}}=0$)
\be
\partial_s Z^\a=-4a(Z^\b Z_\b^\dagger Z^\a-Z^\a Z_\b^\dagger Z^\b)\label{BPS}\;,
\ee
where $a$ some normalisation constant. It was suggested \cite{Gomis:2008vc,Terashima:2008sy,Hanaki:2008cu}
that this gives a fuzzy $S^3$ funnel, so 
we will  now test whether one can  obtain the GR fuzzy $S^3$ from it.

Since we have seen that for the $\cA_4$-theory we employed the identification (\ref{identif}), and we know how to obtain it as a particular case of the $\SU(N)\times \SU(N)$ ABJM theory for $N=2$, we write the ansatz
\bea
&& Z^1=X_1^++iX_2^+\;,\;\;\; Z_1^\dagger=X_1^--iX_2^-\nonumber\\
&& Z^2=X_3^++iX_4^+\;,\;\;\; Z_2^\dagger=X_3^--iX_4^-\label{ansatz}\;,
\eea
which should certainly be correct for the $r=1$ fuzzy sphere, but hopefully more. 

Now, consider the fuzzy $S^3$ equivalent of (\ref{BPS}), studied in \cite{Gomis:2008vc,Terashima:2008sy,Hanaki:2008cu}, which can be obtained through the  ansatz
\be
Z^\a=\frac{1}{\sqrt{8a s}}G^\a
\ee
and leads to 
\be
G^\a= G^\b G_\b^\dagger G^\a-G^\a G_\b^\dagger G^\b\label{equatio}\;.
\ee
The real part for $\a=1$ then gives 
\be
{\rm Re}\{G^1\}= {\rm Re}\{G^\b G_\b^\dagger G^1-G^1G_\b^\dagger G^\b\}\label{realpart}\;.
\ee
But with the ansatz (\ref{ansatz}) for $G^\a$, the right hand side becomes 
\bea
{\rm Re}\{ G^\b G^\dagger_\b G^1-G^1G^\dagger _\b G^\b\}&=& \;X_2^+X_3^-X_4^+-X_2^+X_4^-X_3^+
+X_3^+X_4^-X_2^+-X_4^+X_3^-X_2^+  \nonumber\\
&&+X_3^+X_3^-X_1^++X_4^+X_4^-X_1^+-X_1^+X_3^-X_3^+-X_1^+X_4^-X_4^+\label{rhsi}\;.
\eea
For $r=1$, when the fuzzy three-sphere coordinates reduce to $\SO(4)$ $\Gamma$-matrices, one has that $X_m^+X_n^-=-X_n^+X_m^-$ and $X_m^-X_n^+=-X_n^-X_m^+$ for $m\neq n$, while zero for $m=n$. This in turn means that (\ref{rhsi}) reduces to just $4X_2^+X_3^-X_4^+$. Compared to the left hand side of the first equation in (\ref{defrel}) for $m=1$, the difference is only a relative numerical factor of 4.  Hence up to a constant rescaling there is a matching between (\ref{realpart}) and (\ref{defrel}), as expected.

Since we have just seen the matching for $r=1$, it will serve as a test for the formalism for $r>1$. However, we now have an $r$-dependent factor on the right hand side of (\ref{defrel}), which means (\ref{rhsi}) should end up being proportional to
\bea
\frac{r+2}{2[(r+1)(r+3)+1]}\epsilon^{1mnp}X_m^+X_n^-X_p^+
&=&\frac{r+2}{2[(r+1)(r+3)+1]}\Big(X_2^+X_3^-X_4^+-X_2^+X_4^-X_3^+\cr
&&+X_3^+X_4^-X_2^+-X_4^+X_3^-X_2^+ +X_4^+X_2^-X_3^+-X_3^+X_2^-X_4^+\Big)\;.\nn\\
\label{relatio}
\eea
The first line in (\ref{rhsi}) matches the first four terms of (\ref{relatio}), but that is where the agreement stops. The best case scenario would be 
 if the final result were proportional to 
(\ref{relatio}), yet still give  $4X_2^+X_3^-X_4^+$ for the $r=1$ case, \ie if the following were true 
\bea
&&X_3^+X_3^-X_1^++X_4^+X_4^-X_1^+-X_1^+X_3^-X_3^+-X_1^+X_4^-X_4^+\nonumber\\
&&\hspace{-1cm}\stackrel{?}{=}\frac{2}{3}[X_4^+X_2^-X_3^+-X_3^+X_2^-X_4^+]-\frac{1}{3}[
X_2^+X_3^-X_4^+-X_2^+X_4^-X_3^+
+X_3^+X_4^-X_2^+-X_4^+X_3^-X_2^+]\;,\nn\\
\eea
where the relative factor of 2 between the two brackets is such that the right hand side vanishes for $r=1$, and then the $\frac{2}{3}$ and $\frac{1}{3}$ factors are fixed by the requirement that the final result is proportional to (\ref{relatio}). But even if that were the case, one would end up with the same numerical factor on the right hand side of (\ref{rhsi}) for any value of $r>1$, instead of the $r$-dependent factor required by (\ref{relatio}).

Thus the GR fuzzy $S^3$ is not a solution to (\ref{equatio}) for any $r>1$, in agreement with the expectation that one would have due to the absence of $\SO(4)$ R-symmetry for the above solution, which instead only sports an $\SU(2)\times \U(1)$. In the following sections we will show that for $r>1$ the geometry of these solutions is instead a fuzzy $S^2$.

\section{Fuzzy two-sphere structure of M2-M5  ABJM solutions }\label{fuzzystr}

As we have discussed in the previous section, the BPS equation for the ABJM scalars (\ref{BPS}) was conjectured to give a fuzzy 3-funnel, with the proposed 3-sphere equation (\ref{equatio}). Equivalently, the authors of \cite{Gomis:2008vc} obtained a mass deformation of the ABJM-theory preserving maximal ($\cN = 6$) supersymmetry, but breaking R-symmetry to $\SU(2)\times\SU(2)\times\U(1)\subset\SU(4)$. The mass-deformed theory has two sets of ground states written in terms of two sets of scalars $R^\a$ and $Q^{\dot{\a}}$, the equivalent of $Z^\a$ and $W_{\dot{\a}}$ for the undeformed ABJM theory.\footnote{We will give the ABJM and mass-deformed ABJM actions in due course. At the moment we wish to focus entirely on the matrix structure of the funnel/ground-state solutions.} One set of ground-states corresponds to $Q^{\dot \alpha}=0$ and $R^\alpha$ satisfying
\be\label{grvvvac}
-R^\a = \frac{2\pi}{\mu k}\Big( R^\b R_\b^\dagger R^\a-R^\a R^\dagger_\b R^\b\Big)\;.
\ee
The other set has $R^\alpha=0$ with $Q^\dagger_{\dot\alpha}$ satisfying the equation (\ref{equatio}). Both sets were given the interpretation of a 3-sphere. Note that (\ref{grvvvac}) is the same equation as (\ref{equatio}). Hence, the above are directly related to the funnel solutions of the undeformed ABJM theory and there is no difference in looking at the proposed `fuzzy 3-funnel' solution of ABJM or the proposed `fuzzy 3-sphere' ground-states of the mass-deformed theory. As a result, the latter solutions cannot admit the GR 3-sphere interpretation either. In the following we will switch to the study of the ground-state solutions of \cite{Gomis:2008vc} for concreteness and examine their symmetries to find that they are actually the ones of a fuzzy 2-sphere. We will later extrapolate the results to the funnel.  

{\em Notation:} In this section, we will denote by $k,l,m,n$ the matrix indices/indices of states in a vector space, while keeping $i,j=1,...,3$ as vector indices on the fuzzy $S^2$. We will also use $j$ for the $\SU(2)$ spin and $Y_{lm}$ for $S^2$ spherical harmonics, following the standard notation. The distinction should be clear by the context.

\subsection{Ground-state matrices and symmetries}

The ground-state solutions to (\ref{equatio}) were found in \cite{Gomis:2008vc} and are given by the following set of matrices 
\bea\label{BPSmatrices}
&& ( G^1)_{m,n }    = \sqrt { m- 1 } ~\delta_{m,n} \cr
&& ( G^2)_{m,n} = \sqrt { ( N-m ) } ~\delta_{ m+1 , n } \cr
&& (G_1^{\dagger} )_{m,n} = \sqrt { m-1} ~\delta_{m,n} \cr
&& ( G_2^{\dagger} )_{m,n} = \sqrt { (N-n ) } ~\delta_{ n+1 , m }\;.
\eea
Using the decomposition (\ref{complex}) of the fuzzy complex
coordinates $G^\a$ into real  
coordinates $X_p$, these satisfy 
\be
\sum_{p=1}^4 X_pX^p\equiv G^\a G_\a^\dagger=N-1\;,
\ee
which at first glance would seem to indicate a fuzzy $S^3$ structure.

However, note that in the above $G^1=G_1^\dagger$ for the ground-state solution. With the help of (\ref{complex}) this seems to suggest that $X_2=0$ which is instead indicative of a fuzzy $S^2$ structure. Nevertheless, given the fact that the $r=1$ solution is in fact a fuzzy $S^3$, one might still be open to the possibility that the symmetries and fluctuations of this solution could tell a different story.

In the following, and as is usual in the case of fuzzy sphere constructions, the matrices $G^\a$ will be used to construct both the symmetry operators (as bilinears in $G, G^\dagger$ and acting on $G^\a$ themselves) and fuzzy coordinates (used to expand in terms of spherical harmonics on the fuzzy sphere).

\subsubsection{$ G \Gd $ relations }
We start by calculating the $G \Gd$ bilinears
\bea\label{GGDrels}  
&& ( G^1 G_1^{\dagger} )_{m,n} = (m-1) ~\delta_{mn} \cr
&& ( G^2 \Gd_2 )_{mn} = ( N-m ) ~\delta_{mn} \cr
&& ( G^{1} \Gd_2 )_{ mn} = \sqrt { ( m-1) ( N-m+1) } ~\delta_{ m , n+1 } \cr
&& ( G^2 \Gd_1 )_{mn} = \sqrt { ( N-m ) m } ~\delta_{m+1 , n }  \cr
&&  ( G^\alpha G_\alpha^{\dagger} )_{mn} = ( N-1 ) ~ \delta_{mn}
\eea
 and examining their commutation relations.
Defining $ J^{\a}_{ \b } = G^{\a } \Gd_{ \b }  $ we get
\bea
[ J^{\a}_{\b}   , J^{ \mu }_{ \nu }  ] = \delta^{\mu }_{\b } J^{\a }_{\nu}
     - \delta^{\a}_{\nu} J^{\mu }_{\b }\;.
\eea
These are commutation relations of the generators of $\U(2) $.  The
traceless combinations $ \hat J^{\a}_{\b} = J^{\a}_{\b} - { 1 \over 2
} J \delta^{\a}_{\b} $ are generators of $\SU(2) $ and can equivalently
be given in terms of the usual angular momentum-type variables $ J_{i} = ( \tilde \s_i )^{\a}_{\b} J_{\a}^{\b} $,\footnote{This observation was independently made by D. Rodr\'iguez-G\'omez.} where $\tilde\sigma_i=\sigma_i^T$
are the transpose of the Pauli matrices. Note that more correctly, we should have written 
${J^\a}_\b=G^\a G_\b^\dagger$ and
\be
J_i={(\tilde\s_i)^\a}_\b{J^\b}_\a={(\s_i)_\b}^\a {J^\b}_\a\;,
\ee
but in the following we will stick to the notation $J^\a_\b$, as what kind of matrix multiplication  we have will be made clear from the context. The commutation relations of $J_i$ can be calculated as 
\bea\label{normsu2} 
 [ J_i , J_j ] = 2i \epsilon_{ijk} J_k \;.
\eea
Note that the trace $J\equiv J^\a_\a=N-1$ is a trivial $\U(1)\simeq\U(2)/\SU(2)$ generator, commuting with everything.

\subsubsection{$ \Gd G$  relations }

Next, we calculate the $ \Gd G$  matrices
\bea\label{GDGrels} 
&&  (  \Gd_1  G^1  )_{mn}   = (m-1) \delta_{ mn} \cr
&&  (  \Gd_2  G^2  )_{mn}  = (N-m+1) \delta_{mn} - N \delta_{m1} \delta_{n1}
\cr
&& ( \Gd_1  G^2  )_{mn} = \sqrt { ( m-1 ) ( N-m ) } \delta_{m+1 , n } \cr
&& ( \Gd_2 G^1 )_{mn} = \sqrt { ( m-2)( N-m+1 ) } \delta_{ m,n+1 }  \cr
&& ( \Gd_\alpha G^\alpha )_{mn}  = N \delta_{mn } - N \delta_{ m1 } \delta_{m1 }
\eea
and define $ \bar J_{\a}^{\b} = \Gd_{\a} G^{\b}$. The commutation relations then are
\bea
 [ \bar J^{\a}_{\b} , \bar J^{\mu }_{\nu } ] = \delta_{\b}^{\mu } \bar
 J^{\a}_{\nu }  - \delta_{\nu  }^{\a} \bar J^{\mu }_{\b }\;.
\eea
Similarly, we define 
traceless combinations $ \hat{\bar{ J}}^{\a}_{\b} = \bar{J}^{\a}_{\b} - { 1 \over 2
} \bar{J} \delta^{\a}_{\b} $ which are generators of $\SU(2)$ and 
their angular momentum-type variable counterparts 
$ \bar{J}_{i} = ( \tilde \s_i )^{\a}_{\b} \bar{J}_{\a}^{\b} $ satisfy
\bea
 [ \bar{J}_i , \bar{J}_j ] = 2i \epsilon_{ijk} \bar{J}_k\;. 
\eea
Again, note that we should have written ${\bar{J}_\a\,}^\b=\Gd_\a G^\b$ which emphasises that for $\bar{J}$, the 
lower index is the first matrix index, and 
\be
\bar{J}_i={(\tilde{\s}_i)^\a}_\b {\bar{J}_\a\,}^\b={(\sigma_i)_\b}^\a{\bar{J}_\a\,}^\b\;,
\ee
which emphasises that as matrices, $\bar{J}_i$ is defined with the Pauli matrices, whereas $J_i$ was defined with
their transpose. However, we will keep the notation $\bar{J}^\b_\a$, as which 
matrix contraction one has will once again be made clear from the context .
The trace 
\be
(\bar{J})_{mn}=(\bar{J}^\a_\a)_{mn}=N\delta_{mn}-N\delta_{m1}\delta_{n1}\label{barJ}\;,
\ee
which is a $\U(1)\simeq\U(2)/\SU(2)$ generator, commutes with the $\SU(2)$ generators $\bar{J}_i$, though as a matrix does not commute with the generators $J^1_2$ and $J^2_1$ of the first set of $\SU(2)$ generators.

At this point, is seems that we have two $\SU(2)$'s, \ie $\SO(4)\simeq\SU(2)\times \SU(2)$ as expected for a 3-sphere, but we will see that these are not independent.

\subsection{Fuzzy $S^2$ harmonics  from $\U(N)\times\U(\bar N)$ with bifundamentals}

The first important observation is that $G^\a$ and $G^\dagger_\a$ are bifundamental matrices of $\U(N)\times\U(\bar N)$, mapping between two different vector spaces. However, one can construct combinations of the above that map each vector space back to itself, \eg the form that we have just seen, $G\Gd$, being examples of fundamental matrices for the vector space corresponding to the group $\U(N)$ and $G^\dagger G$ for the vector space corresponding to  $\U(\bar{N})$. We next analyse this structure.

\subsubsection{The adjoint of $\U(N)$}

The matrices $ G \Gd $ are acting on an $N$ dimensional vector space that we will call $ {\bf V}^+ $. The space of linear maps from $ {\bf V}^+ $ back to itself, $ End ( { \bf V }_+ ) $, is the adjoint of the $\U(N)$ factor in the $ \U(N) \times \U( \bar N ) $ gauge group and $G \Gd$ are examples of matrices belonging to it.  The space $ {\bf V}^+ $ forms an irreducible representation of $ \SU(2) $ of spin $ j ={ N-1 \over 2 }$, denoted by $V_N$
\bea\label{bVpdec}  
 \bVp = V_N \;.
\eea 
The set of all operators of the form $  G \Gd ,  G \Gd G \Gd  , \ldots  $ belong in $ End ( { \bf V }_+ ) $ and can be expressed in terms of a basis of `fuzzy spherical harmonics'
defined using the $\SU(2)$ structure. Through the $\SU(2) $ generators $J_i$ we can form the fuzzy spherical harmonics as
\bea\label{fuzzysphharm} 
&& Y^{0}  = 1 \cr
&& Y^1_{ i } = J_i \cr
&& Y^2_{ (i_1 i_2 )} = J_{ {(} i_1 } J_{ i_2 {)} } \cr
&& Y^l_{ (i_1 \cdots i_l ) } = J_{ ( i_1 } \cdots J_{ i_l  ) }\;.
\eea
The brackets $ ( i_1 \cdots i_l ) $ denote symmetrisation.
It is known that symmetrised elements with $l > 2j = N-1  $ can be reexpressed
in terms of lower $l$ elements. The complete space of
$N\times N $  matrices can be expressed in terms of the fuzzy
spherical harmonics with $ 0 \le l \le 2j = N-1  $.  One indeed checks that
\bea
N^2 = \sum_{l=0}^{2j} ( 2l+1 )^2\;.
\eea
Then, a general matrix in the adjoint of $\U(N)$ is expanded as 
\be
A=\sum_{l=0}^{N-1}a^{lm} Y_{lm}(J_i)\;,
\ee
where 
\be
 Y_{lm}(J_i) = \sum_i f_{lm}^{(i_1\ldots i_l)} J_{i_1}\ldots J_{i_l}\;.
\ee
The $ Y_{lm}(J_i)$ become the usual spherical harmonics in the `classical' limit, when $N\to\infty$ and the cut-off in the angular momentum is removed.

In conclusion, all the matrices of $\U(N)$ can be organised into irreps of $\SU(2)$ constructed out of $J_i$, which form the fuzzy spherical harmonics $ Y_{lm}(J_i)$.

\subsubsection{The adjoint of $  \U ( \bar  N ) $ }

Similarly to the $\U(N)$ case, the matrices $ \Gd G , \Gd G \Gd G , \ldots $, are linear endomorphisms of $ {\bf V}^-$.  These matrices are in the adjoint of the $ \U ( \bar N ) $ factor of the $ \U(N) \times \U ( \bar N ) $ gauge group, and will be organised into irreps of the $\SU(2)$ constructed out of $\bar{J}_i$.

However,  we now have an extra ingredient: We have already noticed in (\ref{barJ}) that the $\U(1)$ generator $\bar{J}$ is nontrivial. We can express it as
\be
\bar J = \Gd_{\a} G^{\a} = N - N \bar E_{11}\;.
\ee
This means that $ End ( {\bf V}_- )$ contains in addition to the identity matrix, the matrix $ \bar E_{11} $ which is invariant under $ \SU(2) $. If we label the basis states in $ { \bf V }^- $ as $ | e^-_{k} \rangle  $
 with $ k =1,..., N $, then $ \bar E_{11} = |e^{-}_1\rangle \langle e^{-}_1 | $. This in turn means that ${\bf V}^- $ is a reducible representation
\bea\label{bVmdec}  
 {\bf V}^- = V^-_{N-1 } \oplus  V^-_{1 }\;.
\eea 
The first direct summand is the irrep of $\SU(2) $ with dimension $N-1$ while the second is the one-dimensional irrep.  Indeed, one checks that the $ \bar J_i $'s annihilate the state $ | e^-_1 \rangle $, which is necessary for the identification with the one-dimensional irrep to make sense.  This follows because $ G^1 | e^-_1 \rangle = G^2 | e^-_1 \rangle = 0 $.  Equivalently one has
\bea \label{Eidentity}
G^{\a} \bar E_{11} = 0 =  \bar E_{11} \Gd_{\beta } \;.
\eea 

As a result, the space $ End ( {\bf V}^- )$ decomposes as follows
\bea End ( {\bf V}^- ) = End ( V^-_{N-1 } ) \oplus End ( V^-_1 )
\oplus Hom ( V^-_{N-1 } , V^-_1 ) \oplus Hom ( V^-_{1} , V^-_{N-1} )\;.
\eea
 The first summand has a decomposition
 in terms of another set of fuzzy spherical harmonics
\be
 Y_{lm}(\bar J_i) =  \sum_i  f_{lm}^{(i_1\ldots i_l)} \bar J_{i_1}\ldots \bar J_{i_l}\;,
\ee
 for $ l $ going from $0$ to $N-2$, since
\bea 
(N-1)^2 = \sum_{l=0}^{N-2}  ( 2l + 1 ) \;.
\eea 
However, in this case one is only getting the 
matrices in the $(N-1)$ block, \ie the  $End ( V^-_{N-1 } )$. 
The second summand
is just one matrix transforming in the trivial irrep.
The remaining two
$N-1$ dimensional spaces of matrices cannot be expressed
as products of $ \bar J_i$. They are spanned by
\bea
\bar E_{1k} &=& | e^-_1 \rangle \langle e^-_k | \;\equiv \; g^{--}_{1k } \cr
\bar E_{k1} &=& | e^-_k \rangle \langle e^-_1 | \; \equiv  \; g^{--}_{k 1}\;, 
\eea
which are the equivalent of zero mode 
spherical harmonics for $Hom ( V^-_{N-1 } , V^-_1 ) \oplus Hom ( V^-_{1} , V^-_{N-1} )$.

Indeed, usual spherical harmonics $Y_{lm}$ are eigenfunctions of the
symmetry operators $\vec{L}^2$ and $L_3$, whereas 
$\bar{E}_{1k}=g^{--}_{1k}$ is a zero mode eigenfunction of the $\U(1)$
symmetry operator $\bar{J}$, \ie  
\be
 \bar{J} \bar{E}_{1k}=0\cdot \bar{E}_{1k}
\ee
just as on a circle, we have\footnote{$\bar{J}$ is the $\U(1)$ Lie
algebra element, just like $R\d_x$.} 
\be
R\d_xe^{\frac{2\pi inx}{R}}=0\cdot e^{\frac{2\pi inx}{R}}\Rightarrow n=0\;.
\ee
Hence we have the `zero mode spherical harmonics' $g^{--}_{1k}$ and $g^{--}_{k1}$
\be
\bar{J}g^{--}_{1k}=0\cdot g^{--}_{1k};\;\;\;\;
g^{--}_{k1}\bar{J}=0\cdot g^{--}_{k1}\;.
\ee
These also transform in the $N-1$ dimensional irrep
of $\SU(2) $ under the adjoint action of $ \bar J_i $.

Therefore, one can expand a general matrix in the adjoint of $\U ( \bar N ) $ as
\bea 
\bar{A} = \bar a_{0} \bar E_{11} + \sum_{ l=0}^{N-2} \bar a_{l  m } 
Y_{lm}(\bar J_i) + \sum_{k=2 }^N b_{k } g^{--}_{1k } +
\sum_{k=2}^N \bar b_k g^{--}_{k 1} \label{ubarnexp}
\eea
and note that we could have replaced $\bar{E}_{11}$ with the $\U(1)$ generator
$\bar{J}$ by redefining $\bar{a}_0$ and $\bar{a}_{00}$.

In the large $N $ limit the $ Y_{lm}(\bar J_i)$ become ordinary spherical harmonics of a scalar field on $S^2$, just like $Y_{lm}(J_i)$. There are order $N^2$ of these modes, which is appropriate if one roughly think of the fuzzy $S^2$ as a space with linear dimensions discretised in $N$ units.  The mode $\bar a_0$ can be neglected at large $N$. The modes $b_k$ and $\bar b_k $ have order $N$ degrees of freedom, and we will see that they will also become irrelevant at large $N$, as expected.

\subsubsection{Physical fluctuations: bifundamental $(N,\bar{N})$ matrices}

All the $(N,\bar{N})$ bifundamental  scalar matrices are of the type
$ G , G G^{\dagger} G , G G^{\dagger} G G^{\dagger} G , \ldots  $ and form  maps from ${\bf V}^-$ to ${\bf V}^+$, \ie the space $ Hom ( \bV^-  , \bV^+ ) $. Since we intend to perform a fluctuation analysis of the mass-deformed ABJM theory, in which the physical degrees of freedom are  bifundamental, we will also need to find out how the symmetries act on these matrices.

It is easy to check that the matrices $G^a$ satisfy
\bea\label{basrels}
G^1 G^{\dagger}_2 G^2 - G^2 G^{\dagger}_2 G^1 &=& G^1 \cr
 G^2 G^{\dagger}_1 G^1 - G^1 G^{\dagger}_1 G^2 &=& G^2\;.
\eea
Using the definitions $J_3 = J^1_1 - J^2_2 $, $J_+ = J_1 + i J_2= J^1_2$, $J_- =J_1-iJ_2= J^2_1$  and likewise 
 $ \bar J_{3}  = \bar J^1_1 - \bar J^2_2$, $ \bar J_+ = \bar  J^1_2$, 
$ \bar J_-  =  \bar  J^2_1 $, we find 
\bea\label{symm}
 J_+ G^2 - G^{2} \bar J_+ &=& G^1 \cr
 J_- G^1 - G^1 \bar J_- &=& G^2\;.
\eea
Notice that $J_i$ by itself does not give a
nice transformation of $G^{\a} $. The $G^{\a}$
form a nice representation of an action constructed from
both $J_i $ and $\bar J_i $. This means that the geometry
we will be constructing from bifundamental fluctuation modes has a single $\SU(2)$ symmetry, as opposed to two.

In fact, it is easy to see how we would have needed the above to be modified in order to obtain the full $\SU(2)\times \SU(2)\simeq\SO(4)$ symmetry: The $J_i$ and $\bar{J}_i$ ought to act independently. For instance, in \cite{Ishii:2008ib}, the fuzzy $S^3$ was obtained from the BMN Plane Wave Matrix Model \cite{Berenstein:2002jq} through the intermediate step of a fuzzy $S^2$. There, the matrices $X_i$ were divided into  $N_s\times N_t$ subsets $X_i^{(s,t)}$. Then the $\SU(2)$ generators in the  $j_s$ representation, $L_i^{(j_s)}$,  acted separately on the two sides, giving an action expressed in terms of
\be
L_i\circ X_i^{(s,t)}=L_i^{(j_s)}X_i^{(s,t)}-X_i^{(s,t)}L_i^{(j_t)}\;,
\ee
with an infinite set of $s,t$ values for each side (in the large $N$ limit), that can be used to create an extra coordinate.

Under this single $\SU(2)$ symmetry, Eq.\,(\ref{symm}), $G^1$ is the up-state of spin $\frac{1}{2}$, while $G^2$ is the down-state. This implies that we should also have
\bea
 J_+G^1 - G^{1} \bar J_+ &=& 0\cr
 J_-G^2 - G^2 \bar J_- &=& 0 \;,
\eea
which follow trivially by writing out the $\SU(2)$
generators. One also checks that
\bea
 J_3 G^1 - G^1 \bar J_3 &=& G^1 \cr
 J_3 G^2 - G^2 \bar J_3 &=& -  G^2\;.
\eea
These in fact follow from (\ref{basrels}).
We can summarise all of the  above relations by writing
\bea
J_i G^{\a } - G^{\a}  \bar J_i =  ( \tilde \sigma_i )^{\a}_{\b} G^{\b} \label{offdiagtrans}\;.
\eea
The $G^1 , G^2 $
transform like the $ (1,0) $ and $ (0,1)$ column
vectors of the spin-$\frac{1}{2}$ rep. Note that
the $J$'s and $ \bar J $'s are matrices in the
$ \U(N) \times\U( \bar N)$ Lie algebra. 

By taking Hermitian conjugates in (\ref{offdiagtrans}), we find that the
$G^\dagger_\a$ transform as
\bea
G^{\dagger}_{\a}  J_i - \bar J_i G^{\dagger}_{\a}  = 
 G^{\dagger}_{\b}   ( \tilde \sigma_i )^{\b}_{\a }\;.
\eea
To make the relation between the above  and the $\U(N) \times \U(\bar N) $ gauge group clear, it is useful to construct matrices acting on ${\bf V}^+\oplus {\bf V}^-$, 
\ie $2N\times 2N$ matrices. The $G^\a$ and $G^\dagger_\alpha$ matrices are bifundamental, so we have
\bea
{ {\bf G}^{ \a}  } =  \begin{pmatrix}
0 & G^{ \a }  \\
G^{\dagger}_{ \a}  & 0 \end{pmatrix}\;,
\eea
while the $J_i,\bar{J}_i$ matrices are adjoint and
\bea
{ \bf J}_i  = \begin{pmatrix}  J_i & 0 \\
                               0 & \bar J_i \end{pmatrix}\;.
\eea
Hence, the $G^\a$ and $G^\dagger_a$ transformation laws can be summarised as a single law acting in the ${\bf V}^+\oplus 
{\bf V}^-$ space, by
\bea
[ {\bf J}_i , {\bf G}^{ \a}  ]  &=&   \begin{pmatrix} 0 & J_i G^{ \a}  - G^\a \bar{J}_i \\
                      \bar J_i  G^{\dagger}_\a - G^{\dagger}_\a J_i  & 0
                          \end{pmatrix} \cr
 &=&  \begin{pmatrix} 0 &  ( \tilde \sigma_i )^\a_{\b} G^\b  \\
                          - G^{\dagger}_\b  ( \tilde \sigma_i )^\b_{\a} & 0
                                 \end{pmatrix}\;.
\eea
The matrices $ {\bf J}_i $ are elements of the $ \U(N) \times \U(\bar N)$ gauge Lie algebra. Their action on $ G^{\alpha } $ is precisely the gauge symmetry action corresponding to the fact that the matrices $ R^{\alpha } $ transform as bifundamentals.  This embedding of a global symmetry into the gauge symmetry is an example of the mixing of symmetries in soliton physics and has interesting physical consequences \cite{Jackiw:1976xx}.  In this case, the global $ \SU(2)$ rotations $ \int d^3 x R^{\alpha } D_0 R^{\dagger}_{ \beta } $ do not leave the solution invariant. But combining the global symmetries with rotations generated by $ J_{i}$ gives symmetries which do leave the solution invariant.  Hence the geometric $\SU(2)$ does survive as an invariance of the solution and acts on the space of fluctuations.

\subsubsection{Action of the full $\SU(2) \times \U(1) $  } 

Writing the action of the full $\SU(2) \times \U(1) $ on the $G^\a$, including the $\U(1)$ trace $\bar{J}$, we obtain 
\bea 
 J^{\a}_{\b} G^{\g } - G^{\g} \bar J^{\a}_{\b} 
 = \delta^{\g}_{\b} G^{ \a} - \delta^{\a}_{\b} G^{\g}\label{u2}\;.
\eea 
This implies the relations\footnote{To obtain the first line one uses $ \bar E_{11} \bar J^{\a}_{\b} = 0 $, since $  \bar E_{11} \Gd_{\b} = 0 $.}
\bea 
 \Gd_{\g} J^{\a}_{\b} G^{\g} &=& ( N+1 ) \bar J^{\a}_{\b} -
 N \delta^{\a}_{\b} + N \delta^{\a}_{\b} \bar E_{11} \cr 
 G^{\g} \bar J^{\a}_{\b} \Gd_{\g } &=& ( N-2) J^{\a}_{\b} +
 ( N-1) \delta^{\a}_{\b}  \;.
\eea 
In terms of the  $J_i$'s these equations are simpler 
\bea 
 \Gd_{\g} J_i  G^{\g}&=&  ( N+1 ) \bar J_i \cr 
 G^{\g} \bar J_i  \Gd_{\g } &=& ( N-2) J_i \;,
\eea 
while taking Hermitian conjugates of (\ref{u2}) we obtain the $\U(2)$ transformation of $\Gd_\a$,
\bea 
\bar J^{\a}_{\b} \Gd_{\g} - \Gd_{\g}  J^{\a}_{\b} 
=  - \delta^{\a}_{\g} \Gd_{ \b} + \delta^{\a}_{\b} \Gd_{\g}\;.
\eea 
The consequence of the above equations is that $G^{ \a } $ has charge 
$1$ under the $\U(1)$ generator $ \bar J $. This means that 
a global $\U(1)$ symmetry  action on $R^{\a}$ does not leave 
the solution invariant, but when combined with the action of 
$ \bar J$ from the gauge group does lead to an invariance.

\subsection{$\SU(2)$ harmonic decomposition of bifundamental matrices}\label{harmdec} 

As in the case of the $\U(\bar{N})$ matrices, the bifundamental matrices of the form $ G , G \Gd G , \ldots $ giving physical fluctuating fields, are not enough to completely fill $ Hom ( \bV^- , \bV^+ ) $.  Given the decomposition $ \bVm = V_{N-1}^- \oplus V_{1 }^- $, we decompose $ Hom ( \bV^- , \bV^+ ) $ as
\bea
Hom ( \bV^-  , \bV^+ ) = Hom ( V_{N-1}^- , V^+_N  ) \oplus Hom ( V_1^- ,V^+_N  )\;.
\eea
The first summand has dimension $ N ( N-1) $.
The  second has dimension $ N $.
The $N$ matrices in $ Hom ( V_1^- ,V^+_N  )$ form
an irreducible representation of $\SU(2)$ of dimension $N$.

Since the $ V_{N-1}^- $ and  $V^+_N$ are irreps. of $\SU(2)$
we can label the states with the eigenvalue of $\bar J_3 , J_3  $
respectively. Given the normalisation of the $\SU(2)$ generators 
in (\ref{normsu2}), the conventionally
 defined spin is $ { J_3^{max} \over 2 } $. 
The matrices in $ Hom ( V_{N-1}^- , V^+_N )$ are of the form $ | e^+_m \rangle\langle e^-_n | $, where $ m = { -N+1\over 2 } ,{ -N +3\over 2 } ,...,
{  N-1\over  2 }  $,  $ n = { -N +2\over 2 } , { -N+4\over 2 } ,..., { N-2 \over 2 }  $  denote the eigenvalues of $ {J_3 \over 2 }$.  
  These are spanned by matrices of the form $ G ( \bar J_{i_1} ) ( \bar J_{i_2} ) \cdots ( \bar J_{i_l} )$, \ie the matrix $G$ times matrices in $End(V_{N-1}^-)$. To understand the cutoff on the maximum value of $l$ in the above `spherical harmonics', note that acting on the lowest weight state
 $ |e^{-}_{{ -N+2\over 2 }  } \rangle $ we can have $ G^1 (\bar J_+)^{N-2} $
to obtain the highest weight state $ |e^+_{{ N-1\over 2 }  } \rangle $.
 Operators of the form $ G^2 ( \bar J_+ )^{N-1}$ can be rewritten
 to have $G^1 $ first by using the relations (\ref{symm}).

The upshot is that operators in $ Hom ( V_{N-1}^- , V^+_N )$ transform 
in representations of spin  $ l+\frac{1}{2}$ for $ l = 0,..., N-2 $. 
The dimensions of these representations add up to
\bea
\sum_{ l=0 }^{ N-2} ( 2l + 2  )  = N ( N - 1 )\;.
\eea
This then gives the $\SU(2) $ decomposition of  $ Hom ( V_{N-1}^- , V^+_N  )$
as follows 
\bea
Hom ( V_{N-1}^- , V^+_N  ) = \bigoplus_{l=0}^{N-2}  V_{l+ 1/2 }\;.
\eea

On the other hand, the matrices in $ Hom ( V_1^- , V^+_N ) $ cannot be written in terms of the $G$'s and $G^\dagger$'s alone. Indeed, matrices $ |e^+_{k}\rangle \langle e^-_1 | \equiv \hE_{k 1 }\in Hom ( V_1^- , V^+_N ) $ cannot be so expressed, because $G^{\a} $ acting on $| e^{-}_1\rangle $ gives zero. The index $k$ runs over the $N$ states in $ \bV^+$.  The operators  $\hE_{k1}$ are eigenfunctions of the operator $\bar { E}_{11} $
\be
\hE_{k1}\bar {E}_{11}=\hE_{k1}\;,
\ee
with unit charge. They span the  $ Hom ( V_1^- , V^+_N ) $ matrices.

Combining all of the above, the fluctuations $ r^{ \alpha } $ of the bifundamental 
fields $ R^{\alpha } $ can be expanded in spherical harmonics as follows 
\be
r^\a=  r^{ \a }_{\b} G^\b  +\sum_{k=1}^Nt^\a_k\hE_{k1}\;,
\ee
with
\be
  r^{ \a }_{\b} = \sum_{l=0}^{N-2}(r^{lm})^\a_\b  Y_{lm}(J_i)\;.
\ee
We then decompose $ r^\a_\b $ into a trace and a traceless part
 and define the fluctuating fields
\bea
s^\a_\b &=&  r^{ \a }_{\b} - { 1 \over 2 } \delta^{\a}_{\b } r^{\g}_{\g}  \cr  
r&=&   r^{\g}_{\g}   \cr 
T^{ \a } &=& t^{ \alpha}_{ k } \hE_{k 1}\;.
\eea
Thus the complete expansion of the fluctuating field $r^\a$ is given simply in terms of 
\be
r^{ \a } = r G^{ \a } + s^{\a}_{ \b  } G^{ \b }+ T^{ \alpha} \label{fluctuati}\;.
\ee
We could also have used  equivalently 
\be
r^\a=\sum_{l=0}^{N-2}(r^{lm})^\a_\b G^\b  Y_{lm}(\bar J_i)+\sum_{k=1}^Nt^\a_k\hE_{k1}
\ee
using the spherical harmonics in $ \bar J$ in   (\ref{ubarnexp}). This a signature for a $\mathbb Z_2$ symmetry of the fluctuation action that will become manifest as we progress in our analysis.
We will choose, without loss of generality,  to work with (\ref{fluctuati}). 

Until now we have focused on matrices in $ Hom ( \bV^- , \bV^+ ) $ but the case of $ Hom ( \bV^+ , \bV^- ) $ is similar. The matrices $ \Gd , \Gd G \Gd , \ldots  $ will also form a representation of  $\SU(2)$ given by $\bar{J}\sim \Gd G $, times a $\Gd$ matrix. And again matrices $\hF_{ 1 k } \equiv |e^-_{1}\rangle \langle e^+_k | \in Hom(V_N^+,V_1^-)$ cannot be so expressed, so one needs to add an extra $ T^{\dagger}_{ \a } = ( t^{ \alpha}_{ k } )^* \hF_{1 k} $ fluctuation. In fact, the result for the complete fluctuating field can be obtained by taking a Hermitian conjugate of (\ref{fluctuati}), yielding 
\be
r^{ \dagger }_{ \alpha } =  \Gd_{ \a } r + \Gd_{ \b } s^{\b}_{\a} + T^{\dagger}_{\alpha} \label{fluctuatidag}\;.
\ee

\section{Action for matter fluctuations on the two-sphere}\label{matter}

So far we have analysed the symmetries of the $G^\alpha$ matrices of (\ref{BPSmatrices}) and found a single $\SU(2)$. In this section we will begin computing the action for fluctuations around these  ground-state solutions of the mass-deformed ABJM theory \cite{Gomis:2008vc}. Our aim in the following sections will be to write the resultant action for the fluctuations in terms of a higher-dimensional worldvolume theory, which will make the existence of a fuzzy $S^2$ manifest. 

The pure ABJM action is given by
\bea
S_{\mathrm{ABJM}}&=&\int d^3 x\left[\frac{k}{4\pi}\epsilon^{\mu\nu\lambda}{\rm Tr}\left(A_\mu^{(1)}\d_\nu A_\lambda^{(1)}+\frac{2i}{3}
A_\mu^{(1)} A_\nu^{(1)} A_\lambda^{(1)}-A_\mu^{(2)}\d_\nu A_\lambda^{(2)}-\frac{2i}{3}
A^{(2)}_\mu A^{(2)}_\nu A^{(2)}_\lambda\right)\right.\cr
&&-{\rm Tr}\Big( D_\mu C^\dagger_ID^\mu C^I\Big) -i {\rm Tr}\Big(\psi^{I\dagger}\gamma^\mu D_\mu \psi_I\Big) \nonumber\\
&&\left.+\frac{4\pi^2}{3k^2}{\rm Tr}\left(C^IC_I^\dagger C^JC_J^\dagger
C^KC_K^\dagger+C^\dagger_IC^IC^\dagger_JC^JC^\dagger_KC^K\right.\right.\nonumber\\ 
&&\left.\left.+4 C^IC^\dagger_JC^KC^\dagger_IC^JC^\dagger_K-6 C^IC^\dagger_JC^JC^\dagger_IC^KC^\dagger_K\right)\right.\nonumber\\
&&\left.+\frac{2\pi i}{k}{\rm Tr}\left(C^\dagger_IC^I\psi^{J\dagger}\psi_J-\psi^{\dagger J}C^IC^\dagger_I
\psi_J-2 C^\dagger_IC^J\psi^{\dagger I}\psi_J+2\psi^{\dagger J}C^IC^\dagger_J\psi_I\right.\right.\nonumber\\
&&\left.\left. +\epsilon^{IJKL}\psi_IC^\dagger_J\psi_KC^\dagger _L-\epsilon_{IJKL}\psi^{\dagger I}C^J
\psi^{\dagger K}C^L\right)\right]\;,
\label{abjmaction}
\eea
where on the first two lines we have the CS gauge fields and kinetic terms, on the 3rd and 4th lines $\int d^3 x (-V_6)$, where $V_6$ is the scalar bosonic potential (sextic), and on the last two lines the fermionic interactions. We will treat the CS and scalar kinetic terms in the first two lines separately in the next section, as these will participate in a version of the Higgs mechanism. Focusing on the purely bosonic sector will also prove enough for our purposes, hence we will not study the fluctuations for the fermion fields in the rest of this paper, although doing so should be straightforward. Therefore, in this section we will exclusively discuss the scalar potential terms. 

By splitting $C^I=(R^\a,Q^{\alpha})$, the mass deformation changes the potential to 
\be
V=|M^{ \alpha}|^2+|N^\a|^2\label{potential}\;,
\ee
where 
\bea
M^{ \alpha}&=& \mu Q^{ \alpha} +\frac{2\pi}{k}(2Q^{[ \alpha }Q^\dagger_{\b} Q^{\b ]}+R^{\b} R^\dagger_{\b} Q^{\a}-Q^{\a} R^\dagger_\b R^\b
+2Q^{\b} R^\dagger_\b R^\a)\nonumber\\
N^\a&=&-\mu R^\a +\frac{2\pi}{k}(2R^{[\a}R^\dagger_\b R^{\b ]}+Q^\b Q^\dagger_{\b} R^\a-R^\a Q^\dagger_{\b} Q^{\b}
+2R^\b Q^\dagger_{\b} Q^{\a})\label{potterms}\;.
\eea

In addition, the potential also involves a mass term $\mu$ for the fermions. At this point notice that the massive deformation couples $R^\a$ with $Q^\a$; that is the reason we have used the same index $\alpha$ for both, even though they had different kinds of indices in our  treatment of funnels in the pure ABJM theory. This coupling thus breaks the $\SO(6)$ invariance. Nevertheless, when writing down the full scalar potential the terms that couple $R^\alpha$ and $Q^\alpha$ vanish \cite{Gomis:2008vc}. As a result, we will keep the different notation in their respective indices that we used in the previous section with $R^\a$ and $Q^{\dot{\a}}$.

\subsection{Matrix fluctuation expansion}\label{matrixflucts}
We now proceed to compute the action for quadratic fluctuations around the solution $R^\a=f G^\a$. Satisfying the classical equations of motion sets $f^2 = \frac{\mu k}{2\pi} $. A general fluctuation of the fields is given by\footnote{For notational economy we use the same symbol for the gauge field and fermion fluctuations as for the classical fields. We hope that this will not cause confusion.}
\bea
 R^{ \a } = f G^{\a} + r^{\a}\;, && 
 R^{\dagger}_{\a}  = f \Gd_{\a}  + r^{\dagger}_{\a}  \cr
 Q^{ \dot \alpha}  = q^{\dot \alpha }\;,\qquad \quad&& 
 Q^{\dagger}_{\dot \alpha}  = q_{\dot \alpha}^{\dagger}  \cr
 A_{\mu} = A_{\mu}\;,\qquad \quad &&  \psi^{ \dagger I } = \psi^{\dagger I }\;.
\eea
This solution preserves an $\SU(2)\times\SU(2)\times\U(1)$ subgroup of the R-symmetry group and  $\cN = 6$ supersymmetry \cite{Hosomichi:2008jb,Gomis:2008vc}.   We will call these two $\SU(2)$ factors  `geometric' and `transverse' respectively, as the former will account for the geometric symmetry of the emergent fuzzy $S^2$ by rotating the scalars $R^\alpha$ entering the solution, while the transverse acts on the $Q^{\dot \a } $, which are zero on the solution.  The vector index $I$ of $\SU(4)$ decomposes into the $ (\frac{1}{2} , 0 ) \oplus ( 0 , \frac{1}{2} ) $ of $ \SU(2) \times \SU(2) $. This is clear since we have $C^I = ( R^{\a}, Q^{\dot \a }  ) $, as seen in Section\,\ref{BLGABJM}.

In writing the fluctuation action, the symmetry operators that we will use along with the fields are
\bea\label{identities}
 J^{ \alpha}_{ \beta } = G^{ \a } G^{\dagger}_{\b}\;, && {\hat
   J}^{\a}_{\b} = J^{\a}_{\b} - { 1 \over 2 } \delta^{\a}_{\b} J\;,
 \qquad  J = J^{\g}_{\g} =  N-1\;, \cr
\bar J^{\alpha}_{\b} = \Gd_{\b} G^{\a}\;, &&
 {\hat {\bar J }}^{\a}_{\b} =  \bar J^{\a}_{\b} - { 1 \over 2 }
\delta^{\a}_{\b}\bar J\;, \qquad \bar J = \bar J^\gamma_\gamma = N - N
\bar E_{11}
\eea
and
\be\label{geometricaction}
\cL_{\b}^{\a}  ( q^{\dot \g} ) = \hJ^{\a}_{\b}  q^{\dot\g} - q^{\dot
  \g} \hJb^{\a}_{\b}\;.
\ee
The $\cL_{\b}^{\a}$ give the action of the geometric $\SU(2)$ 
 on the  field $q^{\dot \alpha}$, which will become a derivative (translation operator)
in the classical limit. On the $r^\a_\beta$ and $A_\mu$ fluctuations, the same  will be given by the adjoint action of
\be
J_i =  ( \tilde\sigma_i )^{\a}_{\b
 } J^{\b}_{\a}\;,\qquad
J^{\a}_{\b}  = { ( N-1) \over 2 } \delta^{\a}_{\b} 
+   { 1\over 2 } J_i ( \tilde \s_{i} )^{\a}_{\b}\;.
\ee 
Then $[J_i,\cdot\;]$ acts as a derivative (translation operator) on the 2-sphere in the classical limit, while at finite $N$ it can be thought of as a `fuzzy derivative' operator.

As we have already seen in (\ref{fluctuati}) and (\ref{fluctuatidag}), the scalar  fluctuations $r^\a$ can be decomposed in terms of fields $r,s^\a_\b$ and $T^\a$. We will not consider $T^\a$ for the best part  of this section, as we will see towards the end that it will decouple when we keep fluctuations up to quadratic order. Thus, we write
\bea 
    r^{\a} &=&  r G^{\alpha }  + s^{\a}_{\b} G^{\b  } \cr 
   r_{\a }^{\dagger } &=&  \Gd_{ \alpha }   r + G^{\dagger}_{\b}  s^{\b}_{\a}\;,
\eea 
where $r$ and $s^{\a}_{\b}$ have an expansion in the fuzzy spherical
harmonics of $ End ( \bV^+ ) $, given in terms of $ 1 , G \Gd ,
\ldots $.  We have assumed that $ r^{\dagger } = r $ and
$s^{\dagger} = s$. Instead of $s^\a_\b$ we will also find it more convenient to use the vector field $s_i$ defined as 
\be
 s_i =  s_{\a}^{\b} (\tilde \s_i)^{\a}_{\b}\;, \qquad s^{\a}_{\b} =   { 1\over 2 }
 s_i (  \tilde\s_{i} )^{\a}_{\b} \;.   
\ee 
This way we have traded the two complex fluctuations $r^{\alpha } $
with four real fields on $S^2$. These will be like the fluctuations
$A_i $ of $ X_i + A_i $ in usual matrix realisations of the fuzzy
2-sphere \cite{Iso:2001mg,Papageorgakis:2005xr}. From the $s_i$ we will get a radial scalar $ J_i s_i + s_i
\bar J_i $ and a gauge field $A_a$ on the sphere.

\subsubsection{Fluctuations for the transverse scalars: sextic terms}

The ABJM potential for the scalars is purely sextic. However, the mass-deformed theory of \cite{Gomis:2008vc} or equivalently (as we will see shortly through the kinetic terms) the funnel solution, also have a quartic and a quadratic piece. We will evaluate these contributions separately, starting with the ABJM sextic potential $V_6 = \frac{4\pi^2}{3k^2}\hat V$, which is composed of 4 terms. We denote these by
\bea
\nn \hat V_1 &=&  - \Tr( C^I \Cd_I C^J \Cd_J C^K \Cd_K )\\
\nn \hat V_2 &=&  - \Tr ( \Cd_I C^I \Cd_J C^J \Cd_K C^K ) \\
 \nn \hat V_3 &=&  - 4 \Tr ( C^{I} \Cd_J C^K \Cd_I C^J \Cd_K ) \\
  \hat V_4 &=& ~ 6 \Tr ( C^I \Cd_J C^J \Cd_I C^K \Cd_K ) \;.
\eea 

By plugging  the transverse fluctuations ansatz into the above we find that the quadratic terms in $q^{\dot \a}$ give
\bea
\nn \hat V_1 &\to& - 3(N-1)^2\Tr(q^{\dot
  \beta}q_{\dot\beta}^\dagger)\\
\nn \hat V_2 &\to&
- 3N^2\Tr(q^\dagger_{\dot\alpha} q^{\dot\alpha})+  3N^2
\Tr(q^\dagger_{\dot\alpha }q^{\dot\alpha}\bar E_{11})\\
\nn \hat V_3 &\to& 6 \Tr (q_{\dot\beta}^\dagger 
 J ^\gamma_\alpha- \bar J^\gamma_\alpha
 q^\dagger_{\dot\beta})(J^\alpha_\gamma q^{\dot\beta} - q^{\dot\beta} \bar
 J_\gamma^\alpha)- 12N(N-1)\Tr (q^\dagger_{\dot\beta} q^{\dot\beta})\\
 \nn &&+6N(N-1)\Tr
 (q^\dagger_{\dot\beta} q^{\dot\beta} \bar E_{11})\\
 \hat V_4 &\to&  18 N(N-1)
\Tr(q^{\dot\alpha}q^\dagger_{\dot \alpha})
-12N(N-1)\Tr(q^{\dagger}_{\dot\beta}q^{\dot\beta} \bar E_{11})\;,
\eea
where we have made use of the definitions and identities in
(\ref{identities}) as well as (\ref{Eidentity}). Now, using the relation
\be
6 \Tr (q_{\dot\beta}^\dagger 
 J ^\gamma_\alpha- \bar J^\gamma_\alpha
 q^\dagger_{\dot\beta})(J^\alpha_\gamma q^{\dot\beta} - q^{\dot\beta} \bar
 J_\gamma^\alpha) =
 6\Tr\left(\cL^\gamma_\alpha(q^{\dot\beta})^\dagger\cL_\gamma^\alpha(q^{
     \dot\beta})  
 \right) 
 + 3 \Tr(q^{\dot\beta}q_{\dot\beta}^\dagger)
 +3N(N-2)\Tr(q_{\dot\beta}^\dagger q^{\dot\beta}\bar E_{11})
\ee 
one can easily see that the  mass terms cancel and that the final
answer for the transverse scalar fluctuations is
\be\label{transcal}
V^{\perp}_6 =  \frac{4\pi^2 f^4}{3k^2} 6 
\Tr\left(\cL^\gamma_\alpha(q^{\dot\beta})^\dagger\cL_\gamma^\alpha(q^{\dot
    \beta})
\right)\;.
\ee
Thus we are getting the gradient (kinetic) terms of the $q^{\dot\alpha}$
fields on the fuzzy $S^2$.  These 2 complex scalars are  massless at this level, but we have an explicit mass term for the full $C^I$ fields in the mass-deformed theory.

\subsubsection{Fluctuations for the parallel scalars: sextic terms}

Next we analyse the fluctuations $r^{\alpha}$. The calculations are rather involved, so here we present only the final results. Some identities and intermediate steps are presented in Appendix \ref{scalarids}.

It is useful to organise the parallel fluctuations into ones separately involving only
$r^2$, $s^2$ and $r$-$s$. Restricting to $r^2$ terms coming from $r^\alpha\sim r G^\alpha$ one has
\bea
\nn \hat V_1 &\to&  - 9 (N-2)\Tr\left(
   [J_i ,  r][J_i, r]\right) -3 (N-1) (17N^2-18N-15)\Tr(r^2)\\
\nn \hat V_2 &\to& 12 (N-1) \Tr\left(
   [J_i ,  r][J_i, r]\right) + N (N-1) (69N -72)\Tr(r^2)\\
\nn  \hat V_3 &\to& - 3 (N-2) \Tr\left(
   [J_i ,  r][J_i, r]\right) - 12 (N-1)(N^2-3N)\Tr(r^2)\\
  \hat V_4 &\to&  - 3 N \Tr\left(
   [J_i ,  r][J_i, r]\right) + 6 N (N-1) (N-3)\Tr(r^2)\;, 
\eea
where we have implemented the following relation
\be
\Tr\left( J^\alpha_\beta r J^\beta_\alpha r \right)= \frac{1}{2}\Tr\left(
   [J_i ,  r][J_i, r]\right) + N(N-1) \Tr (r^2)\;.
\ee
Therefore, the contribution  from these terms is
\be
 V^{r^2}_6 = -  \frac{4\pi^2}{3}\frac{f^4}{k^2}
\left[3(N-4)\Tr\left(
   [J_i ,  r][J_i,r]\right) - 45 (N-1) \Tr (r^2)\right]\;.
\ee

For the $s^2$ terms we obtain
\bea
\hat V_1 &\to& - \frac{3}{4} (N-1)\Tr ((s_i J_i s_j J_j+ J_i s_i J_j s_j)
-\frac{3}{2}(N-1)\Tr(J_i s_i s_j J_j)\cr
&& +\frac{3}{4}i \epsilon_{ijk} (N-1)^2 \Tr
(s_i J_j s_k )-\frac{3}{4}(N-1)^3 \Tr (s_i s_i)\cr
\hat V_2 &\to & - { 3  \over 4 } N ( N-1 ) (N-4) \Tr( s_i s_i)  + 
       { 3N \over 4 } ( 5 N -4 ) i \epsilon_{ijk}\Tr ( s_i J_j s_k )\cr
&&- { 3 \over 2 }N  \Tr (J_i s_i J_j s_j + s_i J_i s_j J_j ) - { 3  \over 4 }N  
\Tr (s_i \Box s_i  ) \cr
 \hat V_3 &\to & - \frac{3}{2}(N-1)(2N^2 -3N +5) \Tr( s_i s_i ) -6N \Tr
 (s_i J_i J_j s_j)\cr
&& +\frac{3}{2}(3N^2 -10N+5) i \epsilon_{ijk }\Tr(s_i
 J_j s_k )+\frac{3}{2}\Tr (J_i s_i J_j s_j)\cr
&&+\frac{9}{4}\Tr(s_i\Box s_i ) - \frac{3}{2}
 (2N-3)\Tr (J_i s_i J_j s_j  + s_i J_i s_j J_j)\cr 
&& -\frac{9}{4}i \epsilon_{ijk }\Tr((J_i  s_j J_k)(s_l J_l + J_l s_l))
+\frac{3}{8} i\epsilon_{ijk}\Tr(J_i s_j\Box s_k)\cr
\hat V_4 &\to &
 {  3\over 2 }  ( N-1)^2 ( 3N-4)   \Tr (s_i s_i ) 
   + { 3 \over 2 } ( N-1) \Tr (s_i  \Box s_i)\cr
&& +  { 3  \over 2 } (4N-3)  \Tr (J_i s_i J_j s_j + s_i J_i s_j J_j )
-  {  3 \over 2 }(N-1) ( 7N-4 )i \epsilon_{ijk}\Tr (s_i  J_j s_k)\cr
&& 
- { 3 \over 2  }i\epsilon_{ijk} \Tr ((J_i s_j J_k) ( J_l s_l + s_l J_l )) +
 3(2N-1) \Tr(s_i J_i J_j s_j)\;,
\eea    
where  we have defined the `fuzzy Laplacian'
\be
 \Box s_k \equiv  [ J_p , [ J_p , s_k ]]\;.\label{fuzzylap}
\ee
Adding up the above contributions and using identities in
(\ref{s2identities})  one arrives at
\be\label{sextics2}
\begin{split}
   V^{s^2}_6=  &-  \frac{4\pi^2}{3}\frac{f^4}{k^2}
\Big[ \frac{3}{4}(N-1)\Tr(s_i s_i) - \frac{3}{4} (N+1) \Tr
  ([J_i,s_i]^2) + \frac{3}{4}(4N-1)i \epsilon_{ijk}\Tr (s_i J_k  s_k)\\
  & -\frac{3}{4}N\Tr (s_i\Box s_i)  +\frac{3}{8} i \epsilon_{ijk}
  \Tr(J_i s_j \Box s_k + s_i J_j \Box s_k)\Big] \;.
\end{split}
\ee

For the $r$-$s$ fluctuations we get
\bea 
 \hat V_1 &\to&- \frac{15}{2} (N-1)^2  \Tr (r ( s.J + J.s )) \cr
\hat V_2 &\to & - \frac{3}{2} N (5N-4) \Tr (r (s_i J_i  + J_i s_i )) + 6 N i
\epsilon_{ijk} \Tr (r (J_i  s_j  J_k))\cr
\hat V_3 &\to&  3 \Tr (r \Box s_i J_i +r J_i \Box s_i) - 6 (5N^2-7N+1) \Tr
(r(J_i s_i + s_i J_i ))\cr
&&  + 12(N-1)i \epsilon_{ijk}\Tr (r (J_i  s_j  J_k))\cr
\hat V_4 &\to &  3 \Tr ((J_i r J_i)(s_j J_j  + J_j s_j )) - 18 (N-1)
i \epsilon_{ijk} \Tr (r(J_i  s_j  J_k)) \cr
&& + 21 (N-1)(2N-1)\Tr (r (s_i J_i + J_i s_i))\;.
 \eea    
Collecting these terms and using identities in (\ref{rsidentities})
\be 
 V^{r\textrm{-}s}_6 = - \frac{4\pi^2}{3}\frac{f^4}{k^2}\Big[ -{9\over 4 } \Tr( r (
\Box s_i J_i  + J_i  \Box s_i) ) + { 3 \over 4 } \Tr (r \Box ( J_i s_i
+ s_i J_i ))-  { 9 \over 2 }  \Tr( r ( s_i J_i  + J_i s_i ))\Big]\;.
\ee 

\subsubsection{Quartic bosonic potential terms}

At $Q^{\dot\alpha}=0$, the bosonic quartic term in the potential, linear in $\mu$, for the massive
deformation of \cite{Gomis:2008vc} is
\be
 V_4 =  \frac{ 8\pi \mu}{k} \Tr(R^{[\a}R^\dagger_\b R^{\b]}R^\dagger_\a)\;.
\ee
Using that $J^{[\a}_\b J^{\b]}_\a=\frac{(N-1)}{2}$, we get for the $r^2$
terms
\be
V^{r^2}_4=  \frac{ 8\pi \mu f^2}{k}
\Big[\frac{1}{2}\Tr([J_i,r][J_i,r])+3(N-1)\Tr(r^2)\Big] \;.
\ee
Using the second identity (\ref{quartictraced}) we get for the $s^2$ terms
\be\label{quads2}
 V_4^{s^2} =  \frac{ 8\pi \mu f^2}{k}
\Big[\frac{1}{8}\Tr([J_i,s_i]^2)+\frac{1}{8}\Tr(s_i\Box s_i)-\frac{1}{4}(N-1)
\Tr[s_i s_i] -\frac{1}{4}(2N-1)i\epsilon_{ijk} 
\Tr(s_i J_j s_k)\Big]\;.
\ee
 The $r$-$s$ terms are with the help of (\ref{cubic})
\be\label{quadrs}
 V_4^{r\textrm{-}s}=   \frac{ 8\pi \mu f^2}{k}
\Big[-i\epsilon_{ijk}\Tr(r J_i s_j J_k)+\frac{1}{2}\Tr(r(s_i J_i +J_i
s_i ))\Big]\;.
\ee
There is {\it a priori} also a quartic term $V_4^{q^2}$ involving the $q$'s from (\ref{potential})-(\ref{potterms}), but after a short calculation one can check that it in fact vanishes.

\subsubsection{Quadratic bosonic potential terms}

From (\ref{potterms}) the bosonic mass term is
\be
 V_2 = - \mu^2 \Tr [R^\alpha R^\dagger_\alpha+Q^{\dot{\a}}Q^\dagger_{\dot{\a}}] 
\ee
and can be easily evaluated on the solution.
From it we obtain
\bea
V_2^{r\textrm{-}s\textrm{-}q} &=& - \mu^2\Big[ \frac{1}{4}(N-1) \Tr(s_is_i)+(N-1)\Tr
(r^2)+\frac{1}{2}\Tr(r(s_iJ_i+J_is_i))\cr 
&&-\frac{1}{4}i\epsilon_{ijk}\Tr
(s_iJ_js_k )+q^{\dot{\a}}q^\dagger_{\dot{\a}}\Big]\;.
\eea

\subsubsection{The $T^\a$ fluctuations} 

Until now we have neglected the $T^\a$ terms. However, we have seen that the full expansion of the $R^\a$ scalars actually is 
\bea 
  R^{ \a } &=& f G^{ \alpha } + r G^{ \a } + s^{\a}_{ \b  } G^{ \b } +   
T^{ \alpha}   \cr 
  R^{ \dagger }_{ \alpha } &=& f \Gd_{ \a } + \Gd_{ \a } r + 
\Gd_{ \b } s^{\b}_{\a} + T^{\dagger}_{\alpha} \;.
\eea 
Since these fluctuations are of order $N$, while the ones we have already consider of order $N^2$, we assumed that they will not play a role in our large-$N$ calculation. However, in order to show that they have been rightfully neglected  let us look at some terms involving the $T^\alpha$'s. A mass term $ \Tr \Rd_{\a} R^{\a} $ gives the new contribution 
\bea 
 \mu^2 \Tr   (\Rd_{\a} R^{\a}) &\to& \mu^2 \Tr( T^{\dagger}_{\alpha}
T^{ \alpha}) \cr
 &=& \mu^2 \sum_k t_k^{\a}  ( t^{\a}_k )^*  \Tr (\hE_{k1} \hF_{1k }) \cr 
 &=&  \mu^2\sum_{k }  t_k^{\a}   ( t^{\a}_k )^*\;.
\eea 
Note that we do not get cross terms between 
 $T$ and the $r,s$ at quadratic order  in the fluctuations  because 
\bea\label{Gann1}   
&& G^{\alpha}  \hF_{1K } = 0 \cr 
&& \hE_{K1} \Gd_{\a} =0 \;,
\eea
which follow from the previously established relations
\bea\label{Gann} 
&& G^{\a} | e^-_1 \rangle = 0 \cr 
&& \langle e_1^- | \Gd_{\a} = 0 \;.
\eea   
Following the above through for all possible contributions it is clear that the combinations that appear are
\bea 
&& T^{\a} \Td_{\b} = \sum_{ k,l =1}^N t_k^{\a} { ( t_l^{\b})}^* 
    | e_k^+\rangle\langle e_l^+| \equiv K^\alpha_\beta\in End(\bV^+)
 \cr
&& \Td_{\a} T^{\b} = \sum_{ k } t^{\a }_{k}  ( t^{\b}_k )^*  | e_1^-
\rangle\langle e_1^-| \sim \bar E_{11} \in  End(V^-_1)\;.
\eea
However, because of equations (\ref{Gann1}) and (\ref{Gann}) one finds that in any expression that involves $G$'s the $\Td_{\a} T^{\b}$ yields zero, much in the same fashion as similar terms also did in the $r,s$ calculation. On the other hand $ T^{\a} \Td_{\b}\equiv K^\alpha_\beta $ does not give zero and should admit an expansion in terms of fuzzy spherical harmonics. By evaluating all contributions to the scalar potential
\be
V^T \to - \mu^2 [(N^2 -8N + 6 )\Tr (K^\alpha_\alpha) + (N-8) \Tr(K^\alpha_\beta J^\beta_\alpha) ]
\ee
and one explicitly finds no mixing between $T$ and $r,s$ at quadratic order in the fluctuating fields. Hence, the $T^\a$ fluctuation truly decouples from the rest of our discussion, as expected.

\subsection{Fluctuations and bosonic field theory on the classical $S^2$ }\label{fluctsbos}

To summarise, the action of fluctuations on the fuzzy $S^2$ coming from the bosonic potential is
\bea
V&=&V_6^{r^2}+V_6^{s^2}+V_6^{\perp}+V_6^{\textrm{r}\textrm{-}s}+V_2^{r\textrm{-}s\textrm{-}q}+V_4^{r\textrm{-}s}\cr
&=&-  \frac{4\pi^2}{3}\frac{f^4}{k^2}
\left[3(N-4)\Tr\left(
   [J_i ,  r][J_i,r]\right) - 45 (N-1) \Tr (r^2)\right]\cr
&& -  \frac{4\pi^2}{3}\frac{f^4}{k^2}
\Big[ \frac{3}{4}(N-1)\Tr(s_i s_i) - \frac{3}{4} (N+1) \Tr
  ([J_i,s_i]^2) + \frac{3}{4}(4N-1)i \epsilon_{ijk}\Tr (s_i J_k  s_k)\cr
  && -\frac{3}{4}N\Tr (s_i\Box s_i)  +\frac{3}{8} i \epsilon_{ijk}
  \Tr(J_i s_j \Box s_k + s_i J_j \Box s_k)\Big]+\frac{4\pi^2 f^4}{3k^2} 6 
\Tr\left(\cL^\gamma_\alpha(q^{\dot\beta})^\dagger\cL_\gamma^\alpha(q^{\dot    \beta})\right) \cr
&&- \frac{4\pi^2}{3}\frac{f^4}{k^2}\Big[ -{9\over 4 } \Tr( r (
\Box s_i J_i  + J_i  \Box s_i) ) + { 3 \over 4 } \Tr (r \Box ( J_i s_i
+ s_i J_i ))-  { 9 \over 2 }  \Tr( r ( s_i J_i  + J_i s_i ))\Big]\cr
&&- \mu^2\Big[ \frac{1}{4}(N-1) \Tr(s_is_i)+(N-1)\Tr
(r^2)+\frac{1}{2}\Tr(r(s_iJ_i+J_is_i))-\frac{1}{4}i\epsilon_{ijk}\Tr
(s_iJ_js_k ) \nn\\ 
&&+q^{\dot{\a}}q^\dagger_{\dot{\a}}\Big] +
 \frac{ 8\pi \mu f^2}{k}
\Big[-i\epsilon_{ijk}\Tr(r J_i s_j J_k)+\frac{1}{2}\Tr(r(s_i J_i +J_i
s_i ))\Big]\;.
\eea
We now turn to the understanding of the classical (large $N$) limit, when the fuzzy $S^2$ becomes a commutative $S^2$, in which the above expression simplifies  considerably. This makes use of a correspondence between the fuzzy sphere matrix algebra and the algebra of functions on $S^2$ that is familiar from Matrix Theory and has seen a number of similar applications \cite{Iso:2001mg,Dasgupta:2002hx,Dasgupta:2002ru,Papageorgakis:2005xr,Dorey:2002ad, Andrews:2005cv,Andrews:2006aw}.

\subsubsection{Geometric decomposition of $s_i$}

The fluctuations are fields on a fuzzy $S^2$ and
at large $N$ they become fields on $S^2$.
This $S^2$ is described as being embedded in
$\mathbb R^3$. We have already seen that the coordinates $J_i$ have the property
\bea
 J_i^2 &=& (N^2-1) \cr
 [ J_i , J_j ] &=& 2 i \epsilon_{ijk} J_k\;.
\eea
When we define $ x_i = { J_{i} \over \sqrt{ N^2-1} }  $ we get
\bea
 x_i^2 &=& 1 \cr
 [x_i , x_j ] &=&  {2 i \over \sqrt{ N^2-1} } \epsilon_{ijk}  x_k\;.
\eea
So these are the coordinates on the sphere that become commuting in
the large $N$ limit.  Moreover, the appropriately normalised Tr becomes an integral over the unit $S^2$
\be
\frac{1}{N}\Tr \to \int d^2\sigma \sqrt{\hat h}\;, 
\ee
where $\sigma^\alpha = (\theta,\phi)$. The adjoint action of $J_i$ also has a well defined large $N$ limit,
which we can identify with the action of Killing vectors \cite{Iso:2001mg,Papageorgakis:2005xr}
\be
[ J_i , \cdot\; ] = - 2i \epsilon_{ijk} x_j \partial_k =-2i K_i  = - 2i
K_i^a \partial_{a} \;.
\ee
The vector fields $ K_i = \epsilon_{ijk} x_j \partial_k$ obey $x_i K_i = 0 $. The index $a$ above runs over $ (\theta , \phi) $, and so we have an expansion in terms of partial derivatives of the angular coordinates on the sphere.  In the following $ \epsilon^{\theta \phi } = 1 $, $ \hat h_{ab} $ is the metric on the unit sphere, which we use to raise and lower indices, and $ \omega^{ab}= \epsilon^{ab}/\sqrt{\hat h}$ is the inverse of the symplectic form. For explicit expressions and identities we refer to Appendix \ref{decompid}.

One can easily see that the `fuzzy Laplacian' $\Box$ defined in
(\ref{fuzzylap}) can be related to the 
geometric Laplacian $\hat{\Box}$ on the sphere
\bea\label{laplace}
\Box = [J_i,[J_i,\cdot]] &=&-4 K_j^a \d_a K_j^b \d_b \cdot\cr &=& -4 \d_a \d^a -4
K_p^a   ( \d_a K_p^b ) \d_b
\cr
&=& -4 \frac{1}{\sqrt{\hat h}} \d_a (\sqrt{\hat h}\; \d^a )\equiv -4 \hat \Box\;,
\eea
with 
\be
\Box x_i = -4 \hat\Box x_i = 8 x_i\;.
\ee

We will then decompose the $s_i$ as \cite{Iso:2001mg,Papageorgakis:2005xr}
\be\label{decomposition}
s_i = K_i^a A_a + x_i \phi\;.
\ee
The $s_i$ transform as vectors of the $\SO(3)$ rotational symmetry of the embedding $\mathbb R^3$. The $A_a$ transform as vectors of the $\SO(2)$ tangent space group of the embedded sphere. We are decomposing $ 3 = 2 +1 $ to get a vector and scalar of $\SO(2)$: The vector will give gauge fields on the $S^2$ and the scalar will correspond to radial motions.

\subsubsection{ Transverse scalars  }\label{transscal} 

Recall from (\ref{transcal}) that we have
\be
V^{\perp}_6 =\frac{4\pi^2 f^4}{3k^2} 6
\Tr\left(\cL^\gamma_\alpha(q^{\dot\beta})^\dagger\cL_\gamma^\alpha(q^{\dot
    \beta}) 
\right)\;,
\ee
which can be rewritten as
\be
V_6^\bot=\mu^2\Tr[(q^\dagger_{\dot{\b}}J_i-\bar{J}_iq^\dagger_{\dot{\b}})(J_i q^{\dot{\b}}-q^{\dot{\b}}\bar{J}_i)]\;.
\label{formula}
\ee
The $q^{\dot\a}$ are still bifundamentals but are essentially different from the bifundamental fields that we studied in Section\,\ref{harmdec}, since they have a transverse spinor index $\dot\a$ and no indices (scalars) on the 
$S^2$. 
They can be decomposed in terms of spherical harmonics as
\be 
q^{\dot \a } =  \sum_{l=0}^{N-1} (Q^{ \dot \a }_\a)_{l m } Y_{ lm } ( J_i ) G^{\a} =   Q^{ \dot \a }_{\a } G^{\a}
\;,\label{sphhardecom}
\ee
where in the last equality we have transformed to a representation in terms of commuting spinors on $S^2$, $Q^{\dot \a}
_\a$, decomposed in usual spherical harmonics, and a $G^\a.$\footnote{We will 
see in  Section \ref{sechopf} that  $ G^\a$ is the matrix-version of   
 a `twistor-like' coordinate for the classical  Hopf fibration of $S^3$ over $S^2$. }

In terms of the scalar representation $q^{\dot\a}$, we can obtain an expression that takes the form of the usual massless scalar action.  To do this, we must first extend the definition of the usual angular derivatives $\d_a$ which act on $x_i$, to act on a larger algebra including $ G^{ \a }$. This will allow the definition of $\d_a q^{\dot\a} $.  We start by defining the adjoint action of $K_i$ on $G^\a$ as
\be
-2iK_i^a\d_a(G^\a)=-2i K_i(G^\a)\equiv N(x_iG^\a-G^\a\bar x_i)=[(\ts_i)^\a_\b-x_i\delta^\a_\b]G^\b\;.
\ee
The algebra generated by $ x_i , G^{\a} , \Gd_{\a} $ is constrained by 
$ x_i = \frac{1}{N}(\ts_j)^\b_\a G^\a G^\dagger_\b $ and one can check that the consistency condition 
\be
K_i(x_j)=K_i\left(\frac{1}{N}(\ts_j)^\b_\a G^\a G^\dagger_\b\right)=\epsilon_{ijk} x_k
\ee
is satisfied, as expected. In particular, using the projector 
\be
K_i^ah_{ab}K^b_j=\delta_{ij}-x_ix_j
\ee
we find 
\be
\d_a (G^\a)=\frac{1}{-2i}h_{ab}K_i^b (\ts_i)^\a_\b G^\b.
\ee
With these definitions,  the operator whose square appears in (\ref{formula}) becomes  in the large $N$ limit
\be\label{JqJ}
J_i q^{\dot \a}-q^{\dot \a}\bar J_i =  -2i K_i(q^{\dot\a})+x_iq^{\dot\a}\;.
\ee
The sextic term then gives
\be
V_6^\bot
\rightarrow 4\mu^2N \int d^2\sigma \sqrt{\hat h}
\left[ h^{ab}\d_a q^\dagger_{\dot \a} \d_b q^{\dot \a}+\frac{1}{4} q^\dagger_{\dot\a}q^{\dot\a}\right]\label{qsextic}\;.
\ee
The quadratic part of the potential cancels the above mass term, since
\be
V_2^\bot\rightarrow -\mu^2N \int d^2\sigma \sqrt{\hat h}\;q^\dagger_{\dot\a}q^{\dot\a}\;,
\ee
so that the full quadratic action for the scalars is just the usual massless term 
\be\label{fullquadratic}
V^\bot\rightarrow 4\mu^2N \int d^2\sigma \sqrt{\hat h}\; h^{ab}\;\d_a q^\dagger_{\dot \a} \d_b q^{\dot \a}\;.
\ee
Although the fields $ q^{\dot \a} $  contain the elements $ G^{\a}$ of an extended 
algebra, the Lagrangian above contains bilinears which can be expressed in terms of $x_i$. 
Hence the action obtained as an integral over angles on the 2-sphere is well-defined.    

In order to have fields depending only on the usual Cartesian sphere coordinates $x_i$, one needs to go to the commuting spinor representation using (\ref{sphhardecom}). At large $N$
\bea
J_i Q^{\dot\alpha}_\alpha G^\a- Q^{\dot\alpha}_\alpha G^{\a} \bar J_i
= [ J_i ,  Q^{\dot\alpha}_\alpha]  G^{ \a } + ( \ts_{i} )^{\a}_{\g} Q^{\dot\alpha}_\alpha G^{\g }\;,
\eea
which is equivalent to (\ref{JqJ}) and implies
\bea 
J_i Q^{\dot \a}_{ \a} G^\a -  Q^{\dot \a}_{ \a} G^\a \bar J_i 
&=& -2i K_i^a  ( \d_a  Q^{\dot \a}_{\a} )G^\alpha 
    +        ( \ts_{i} )^{\a}_{\g} Q_{\a}^{\dot \a}  G^{\g }\nonumber\\
& =&  -2 i K_i^a \bigl (  ( \nabla_a )^{\a}_{\g}   Q_{\a}^{\dot \a}  \bigr ) 
 G^{ \g}+ x_i Q^{\dot \alpha}_\alpha G^\alpha  \;,
\eea
with the definition of $\nabla$ given by
\bea 
  ( \nabla_a )^{\a}_{\g}   =
 \delta^{\a}_{\g}  \d_a  + \frac{1}{-2i} h_{ab} K_j^b ( \ts_{j} )^{\a}_{\g} \;.
\eea 
In these variables the sextic action for the transverse scalars then becomes
\bea
V_6^\bot
&\rightarrow& 4\mu^2N\int d^2 \s \sqrt { \hat h } \left[
\hat h^{ab } \left( (\nabla_{a})_{ \gamma}^{\alpha }    Q^{\dot \a }_{\alpha }   \right) J^\gamma_\beta
\left( ( \nabla_{ b })^{  \beta }_{\mu}Q_{\dot \a }^{ \mu  }\right)+ \frac{1}{4} Q^{\dot \alpha}_\alpha J^\alpha_\beta Q^\beta_{\dot\alpha}\right]\cr
&=& 2\mu^2N^2\int d^2 \s \sqrt { \hat h } \left[
\hat h^{ab } \left( (\nabla_{a})_{ \gamma}^{\alpha }    Q^{\dot \a }_{\alpha }   \right) (\delta^\gamma_\beta + x_i (\tilde \sigma_i)^\gamma_\beta)
\left( ( \nabla_{ b })^{  \beta }_{\mu}Q_{\dot \a }^{ \mu  }\right)+ \frac{1}{4} Q^{\dot \alpha}_\alpha (\delta^\alpha_\beta + x_i (\tilde \sigma_i)^\alpha_\beta) Q^\beta_{\dot\alpha}\right]\;.\nn \\
\eea
The spinor transformation property of $Q^{\dot \alpha}_\alpha$ under the geometric $\SU(2)$ is interesting and we will comment on it in Section\,\ref{discussion}.  The fact that the large $ N $ action for $ Q^{ \dot \a }_{ \a} $ can be written compactly, in this large $k$ semiclassical analysis, in terms of $ q^{ \dot \a} $ (which contains the $ G^{\a} $) will be interpreted in Section\,\ref{sechopf} in terms of a reduction of the Hopf fibre $S^1$ of $S^3$ to a $\mathbb Z_2$.

\subsubsection{Parallel scalars: The sextic $s^2$ terms}

From (\ref{sextics2}) we have for the sextic $s^2$ terms
\be
\begin{split}
V^{s^2}_6=  &-   \frac{4\pi^2}{3}\frac{f^4}{k^2}
\Big[ \frac{3}{4}(N-1)\Tr(s_i s_i) - \frac{3}{4} (N+1) \Tr
  ([J_i,s_i]^2) + \frac{3}{4}(4N-1)i \epsilon_{ijk}\Tr (s_i J_k  s_k)\\
  & -\frac{3}{4}N\Tr (s_i\Box s_i)  +\frac{3}{8} i \epsilon_{ijk}
  \Tr(J_i s_j \Box s_k + s_i J_j \Box s_k)\Big] \;.
\end{split}
\ee
The term $ \Tr (s_i s_i) $ contains mass terms for the gauge field and the radial scalar. However, there are also other hidden mass terms that could cancel the former.  In fact it would be surprising if the gauge field had a mass.  Note that $\Tr(s_i s_i) $ gives a mass to the {\it diagonal} of the $ \U(N) \times \U(\bar N)$ gauge field, which needs to cancel if the story is to be similar to \cite{Mukhi:2008ux}.  It was observed in \cite{Papageorgakis:2005xr} that there can be highly nontrivial cancellations of mass terms for gauge fields, some of which only become apparent after using the equations of motion.

Upon converting the terms using (\ref{decomposition}) we find the
following in the large $N$ limit: For the $ijk$-antisymmetric term
\be\label{CStotal}
  i \epsilon_{ijk}\Tr (J_i  s_j  s_k) 
 \to N \int d^2\sigma  \sqrt { \hat h} \bigl (- \hat h^{ab} A_a A_b +
 \omega^{ab}    \phi    F_{ab}  -2 \phi^2 \bigr )\;.
\ee
For $s_i s_i$ one easily gets
\be
 \Tr (s_i s_i) \to N \int d^2\sigma\sqrt{\hat h} (\hat h^{ab}A_a A_b + \phi^2)\;.
\ee
We also have an extra term of the type  $i \epsilon_{ijk}(J_i s_j \Box
s_k + s_i J_j \Box s_k)$ at leading order in $N$. However this will
yield a zero result since
\be
i \epsilon_{ijk}(J_i s_j \Box s_k + s_i J_j \Box s_k) \sim  N i
\epsilon_{ijk} (x_i K^a_j A_a + K_i^a A_a x_j) \hat\Box s_k = 0\;,
\ee
by relabelling and since the $K$'s and $x$'s commute. We are left with
the $s_i \Box s_i$ and $[J_i , s_i]^2 $ terms. One could evaluate
these directly. However, it is simpler to use the following
alternative method:
 
 First compute $ [ J_i , s_j ] - [ J_j , s_i ]$ 
 \bea\label{cFF}
& & \cF_{ij}  \equiv  [ J_i , s_j ] - [J_j , s_i ] \cr 
        && = 2 i \epsilon_{ijk} K_k^a A_a - 2 i K_i^a K_j^b F_{ab} 
         + 4i \epsilon_{ijk} x_k \phi + 2 i ( x_i K_j ( \phi  ) - x_j K_i( \phi ) )\;. 
\eea 
Now observe that
\bea 
 \Tr (\cF_{ij} \cF_{ij}) &=& 2 \Tr [ J_i , s_j ] [ J_i , s_j ]  - 
2 \Tr [ J_i , s_j ] [ J_j , s_i ]\cr
  &=&  - 2 \Tr   s_j \Box   s_j   - 2  \Tr [J_i,s_i]^2 
       + 2 i \epsilon_{ijk} \Tr s_j [ J_k , s_i ]\;, 
\eea
where in the above we have also made use of the Jacobi identity.  So we see that the
combination $ \Tr s_j \Box s_j + \Tr [J_i,s_i]^2$ that we would like
to evaluate can be directly related to $ \Tr \cF_{ij}^2 $.
Now using (\ref{CStotal}) and (\ref{cFF}) we can easily calculate $
\Tr (\cF_{ij}  \cF_{ij})$
 \be  
 \Tr (\cF_{ij} \cF_{ij }) \to N \int d^2\sigma \sqrt{\hat h}\left(
 -8 A_a A^a - 4 F_{ab}F^{ab } - 32 \phi^2 - 8 \d_a \phi \d^a \phi
      + 24 \omega^{ab} F_{ab } \phi \right)\;.
 \ee
 Combining all of the above and using (\ref{laplace}), we find that the
 $s^2$-type terms give, in the large $N$ limit
\be
V^{s^2}_6\to\frac{4\pi^2}{3}\frac{f^4}{k^2}N \int d^2\sigma\sqrt{\hat h}
\Big( 3N \phi \hat{\Box}\phi-\frac{3}{2}N
F^{ab}F_{ab}+9N\omega^{ab}\phi F_{ab}-\frac{45}{4}N\phi^2
-\frac{9}{4}N A^aA_a +{\cal O}(1) \Big)\;.
\ee

\subsubsection{Parallel scalars: The sextic $r^2$ and $r$-$s$ terms}

We follow similar steps for these terms. Note that
\be
\Tr [J_i,r][J_i,r]\to-N \int d^2\sigma \sqrt{\hat h} ( \d^a \tilde r \d_a
\tilde r)\;,
\ee
where $\tilde r = 2 r $,
\be
J_i s_i +s_i J_i  \to 2 \sqrt{N^2-1}~\phi+{\cal O}(1)=2 N \phi +{\cal O}(1)
\ee
and
\bea
i\epsilon_{ijk}\Tr[r(J_is_jJ_k)]&=&\frac{1}{4}r[\Box(s_iJ_i)-(\Box s_i)J_i]=\frac{1}{4}r[\Box(J_is_i)-J_i(\Box s_i)]
\nonumber\\
&\rightarrow & N^2\int d^2\sigma \sqrt{\hat{h}}[-\tilde{r}\phi+\frac{1}{2}\omega^{ab}F_{ab}\tilde{r}]\;.
\eea
Thus to leading order we obtain
\be
 V_6^{r^2,r\textrm{-}s}\to - \frac{4\pi^2}{3}\frac{f^4}{k^2}  12N^2 \int d^2\sigma
\sqrt{\hat h} \left(  \frac{1}{2}\tilde r\hat \Box \phi -\frac{3}{4}\tilde{r} \omega^{ab}F_{ab}-\frac{15}{8} \tilde
  r\phi -\frac{1}{4}\d^a
\tilde r \d_a \tilde r -\frac{15}{16}\tilde r^2 \right)+{\cal O}(N) \;.
\ee
Combining everything the sextic potential gives the following
quadratic fluctuations at leading order
\be
 V^{r,s(\textrm{quad})}_6
\to -6N^2\frac{4\pi^2}{3}\frac{f^4}{k^2}\int d^2\sigma
\sqrt{\hat h}  
\Big(\frac{F_{ab}^2}{4}+\frac{[\d_a(\tilde{r}+\phi)]^2}{2}-\frac{3}{2}
(\tilde{r}+\phi)\omega^{ab}
F_{ab}+\frac{3}{8}A^a A_a+\frac{15}{8}(\tilde r +\phi)^2\Big)\;.
\ee

\subsubsection{The quartic and quadratic terms}

For the quartic terms we find
\be
V_4 \to \frac{N^2}{4}\frac{ 8\pi\mu f^2}{k}\int d^2\sigma\sqrt{\hat h}
\left( A^a A_a  +3 ( \tilde r+ \phi)^2 - 2
  \omega^{ab}F_{ab} (\tilde{r}+\phi)\right)\;, 
\ee
where everything else is of order one. 

The quadratic terms are
\bea
 V_2 &\to &- \mu^2  \frac{N^2}{4} \int d^2\sigma \sqrt{\hat h}
(A^aA_a+(\tilde r +  \phi)^2 + \frac{4}{N} q^{\dot \a}  q^\dagger_{\dot \a})\nonumber\\
&=&  - \mu^2  \frac{N^2}{4} \int d^2\sigma \sqrt{\hat h}
(A^aA_a+(\tilde r +  \phi)^2 
+ 2 Q^{\dot \beta}_\alpha (\delta^\alpha_\beta + x_i(\tilde \sigma_i)^\alpha_\beta)  Q^\beta_{\dot \beta})
\;.
\eea

\subsubsection{Collecting all potential  terms} 

We remind that the equations of motion require that we have $f^2 = { k
  \mu \over 2 \pi } $. By taking this into account, the  mass terms for the gauge fields cancel and we obtain the following simple final result
\be
 V^{||}(r^\alpha_\beta)
\to  -2N^2 \mu^2\int d^2\sigma
\sqrt{\hat h}  
\Big(\frac{1}{4}F_{ab}^2 +\frac{1}{2}(\d_a(\tilde{r}+\phi))^2
+\frac{1}{2} (\tilde
r +\phi)^2 - \frac{1}{2} \omega^{ab} F_{a b}(\tilde{r}+\phi) \Big)
\ee
and
\bea
V^{\perp}(q^{\dot\alpha})&\to& 4 \mu^2 N \int d^2\sigma\sqrt{\hat h} \;
\hat h^{ab}\;\d_a q^\dagger_{\dot \a} \d_b q^{\dot \a}\nonumber\\
&=&2\mu^2N^2\int d^2 \s \sqrt { \hat h } \left[
\hat h^{ab } \left( (\nabla_{a})_{ \gamma}^{\alpha }    Q^{\dot \a }_{\alpha }   \right) (\delta^\gamma_\beta + x_i (\tilde \sigma_i)^\gamma_\beta)
\left( ( \nabla_{ b })^{  \beta }_{\mu}Q_{\dot \a }^{ \mu }\right ) \right ] \;. \nn\\
\eea

\subsubsection{The disappearance of the $\tilde{r}- \phi$ mode in
 the classical  $S^2$}\label{nofr}

We notice that the above potential depends only on the $\tilde{r}+\phi$ combination. 
The fluctuation of the parallel scalars, expressed using
 the large $N$ limit expansion of $s_i$, is
\be
r^\alpha= rG^\a+ s^\a_\b G^\b=rG^\a+\Big(K^a_iA_a+\frac{ J_i}{N}\phi\Big)\frac{(\tilde{\sigma}_i)^\a_\b}{2} G^\b
\ee
and can be rewritten, using 
\bea
\frac{J_i(\tilde{\sigma}_i)^\a_\b}{2}G^\b&=&J^\a_\b G^\b -\frac{N-1}{2}G^\a\cr
J^\a_\b G^\b=G^\a(G^\dagger_\b G^\b)&=&G^\a N(1-\bar{E}_{11}) 
\label{rela}
\eea
and $G^\a \bar{E}_{11}=0$, as
\bea
r^\alpha &=&K^a_iA_a \frac{(\tilde{\sigma}_i)^\a_\b}{2}G^\b+rG^\a+\phi\frac{G^\a}{N}\Big(\bar{J}- \frac{N-1}{2}\Big)\cr
&=&K^a_iA_a \frac{(\tilde{\sigma}_i)^\a_\b}{2}+rG^\a+\phi\frac{G^\a}{N}\Big(\frac{N+1}{2}-N \bar{E}_{11}\Big) \cr 
&\rightarrow &K^a_iA_a \frac{(\tilde{\sigma}_i)^\a_\b}{2}G^\b+\frac{\tilde r+\phi}{2}G^\a\;.
\eea
Thus at the classical level, there is no $\tilde r- \phi$ fluctuation at all, only $\tilde r+\phi$.  This fact is not an accident: The disappearance of this mode is related to the fact that expanding around the ground-state solution triggers a form of the Higgs mechanism, which renders a linear combination of the CS gauge fields dynamical, as we will see next. In that context, the combination $\tilde r- \phi$ plays the role of a Goldstone boson.

A finite $N$ version of the above calculation, keeping track of the commutators between 
 $ x_i = { J_i \over \sqrt { N^2 -1 } } $ and $ \phi $, leads  to 
\bea 
r^{\a } =   K_i^a { \cal A  }_{ a} \frac{( \ts_i )^{ \a}_{\b}}{2} G^{\b}+ { \varphi \over 2 }  G^{ \a}   \;,
\eea 
where 
\bea 
&& \varphi = \Big(  2 r  +   \sqrt { { N+1 \over N-1} } \phi \Big) = 2r + \phi \Big( 1 + { 1 \over N } +  { 1 \over 2 N^2 }   + \cdots   \Big) 
 \cr 
&& { \cal A  }_{ a} =  A_{a } - i { \d_a \phi \over { \sqrt { N^2 -1 } } } \;.
\eea 
Note that the last line is a standard gauge transformation of 
$  A_{a }$.   

This indicates that our framework for the large $N$ action  can be extended 
to subleading orders in the $\frac{1}{N} $ expansion without a drastic change 
in field content. This is 
 consistent with the philosophy of large $N$ collective field theory 
\cite{Das:1990kaa}, where classical field theories capture large $N$ dynamics
and $ { 1\over N } $ interactions are described by ordinary field 
theoretic interaction terms. In the case at hand, the subleading 
corrections will be related to the geometry of the fuzzy $S^2$. 
This is not to say that the extension to subleading orders is trivial. 
For example the fact that $ \phi $ can be expressed as 
$ \phi = s_i J_i \sim J_i s_i $ at large $N$ is no longer true,
since we have to take in to account  $[J_i,s_i ] \ne 0 $.

\subsection{The fuzzy funnel case}

Up to now, we have focused on the theory around the fuzzy sphere solution of \cite{Gomis:2008vc}. However, it is straightforward to see that the case of
the fuzzy funnel solution of the undeformed ABJM theory is exactly the
same as far as the action for fluctuations is concerned. Indeed,
the Hamiltonian for the funnel solution with $Q^{\dot\alpha}=0$ and $R^\alpha\neq 0$ is given by \cite{Hanaki:2008cu}
\be\label{hamiltonian}
 H\propto\int dx_1 ds\;
\Tr\left|\dot{R}^\alpha-\frac{2\pi}{k}R^{[\alpha}R^\dagger_\beta R^{\beta]}\right|^2+{\rm
  topological \;\; term} \;,
\ee
where $s$ is the M2 worldvolume direction extending away from the M5-brane in the M2$\perp$M5 funnel and a dot denotes differentiation with respect to that coordinate. The BPS condition is solved by 
\be
R^\alpha= f(s) G^{\alpha}\qquad\textrm{where}\qquad
f(s) = \sqrt{\frac{k}{4 \pi s}}\;.
\ee
The background field plus fluctuations will be
\be
R^\alpha=f(s)\Big(G^{\alpha}+rG^\alpha +s^\a_\b
G^\b\Big)\equiv f(s)R^{\alpha}_0\;,
\ee
with  $R^\a_0$ the fluctuating fuzzy sphere field.\footnote{We could  also have  chosen to rescale the fluctuations $r^\alpha$ by $f(\mu)$ in the mass deformation calculation done up to now.} We also have 
\be
\dot{R}^\a = -\frac{2\pi}{k} f(s)^3\Big(G^{\alpha}+rG^\alpha +s^\a_\b
G^\b\Big)\equiv -\frac{2\pi}{k}f(s)^3R^{\alpha}_0\;. 
\ee
Then the Hamiltonian picks up a common prefactor that we can ignore, since it will just normalise the overall coefficient
\be
 H\propto \frac{4\pi^2}{k^2} f(s)^6 \int dx_1 ds\;
\Tr\left|R^\alpha_0-\frac{2\pi}{k}R^{[\alpha}_0R^\dagger_{0\,\beta} R^{\beta]}_0\right|^2+{\rm
  topological \;\; term} 
\ee
and after expanding the above the matrix form of every term can be exactly related back to the sextic, quartic and quadratic potential terms of \cite{Gomis:2008vc}.

Hence, there is no difference between the fuzzy funnel and the mass-deformed cases, except for the profile function being a function of $s$ in the former, $f(s)$, while a function of the mass-deformation parameter $\mu$ in the latter, $f(\mu)$, and our calculation will go through essentially unchanged.

\section{The CS-Higgs system and the Higgs mechanism}\label{CSHiggs}

We now turn our attention to the fluctuations analysis of the CS-Higgs action in the first two lines of (\ref{abjmaction}), that is
\be\label{CSkin}
\int d^3 x\left[\frac{k}{4\pi}\epsilon^{\mu\nu\rho} \Tr[A^{(1)}_\mu\d_\nu
  A^{(1)}_\rho + \frac{2i}{3}A^{(1)}_\mu A^{(1)}_\nu A^{(1)}_\lambda - A^{(2)}_\mu\d_\nu
  A^{(2)}_\rho -   \frac{2i}{3}A^{(2)}_\mu A^{(2)}_\nu
  A^{(2)}_\lambda ]-\Tr[D_\mu C^\dagger_I D^\mu C^I]\right]
\ee
containing the gauge fields of the $\U(N)\times \U(\bar{N})$ group.
The covariant derivative acts on the bifundamental fields as 
\be
D_\mu C^I = \partial_\mu C^I + i A^{(1)}_\mu C^I - i C^I A^{(2)}_\mu\;.
\ee
Hence we have for the fluctuations
\be
-D_\mu C^{\dagger}_ID^\mu C^I=-(\partial_\mu r^\dagger_\a-ifG^\dagger_\a A^{(1)}_\mu  +ifA_\mu ^{(2)}G^\dagger_\a)
(\partial_\mu r^\alpha+ifA^{(1)}_\mu G^\a -ifG^\a A_\mu ^{(2)})\label{ccdag}\;.
\ee
In order to obtain a dynamical gauge field from the above CS expressions we will make use of a novel implementation of the Higgs mechanism, in accordance with \cite{Mukhi:2008ux}. In that reference, the Higgs mechanism was studied in the context of 3-algebra BLG theories.\footnote{See also \cite{Ezhuthachan:2009sr} for an exposition in terms of the bifundamental notation of \cite{VanRaamsdonk:2008ft}.} The generalisation  to the ABJM model is straightforward and was carried out in \cite{Pang:2008hw, Li:2008ya}. The mechanism involves a linear combination involving the difference of the gauge fields becoming massive, as in the usual Higgs mechanism, due to the coupling between gauge fields and scalars. However, in the absence of a kinetic term said combination can be integrated out rendering the other linear combination dynamical. The surviving gauge field transforms in the adjoint of the diagonal subgroup of $A^{(1)}_\mu$ and $A^{(2)}_\mu$. The bifundamental scalars also get promoted to adjoint fields.

In the approach of \cite{Mukhi:2008ux} a Higgs vev was developed by a single scalar and was proportional to the identity matrix.  In our case however, all the $R^\alpha$'s get a background value, which is proportional to $G^\alpha$.  The parallel with \cite{Mukhi:2008ux} is obtained by remembering the observation made in Section\,\ref{fuzzystr}, that in the large $N$ limit both the fuzzy spherical harmonics $ Y_{lm}(J_i)$, coming from $\U(N)$, and the fuzzy spherical harmonics $ Y_{lm}(\bar J_i)$, coming from $\U(\bar{N})$, give rise to two $\U(1)$ gauge fields. For these fields we will perform a Higgsing procedure analogous to \cite{Mukhi:2008ux}, so that only the diagonal $\U(1)$ gauge field will remain. In this limit, the Higgs vev will effectively act as a $\U(1)$ vev.

At large $N$ the operators  $J_i$ and $\bar J_i$ become the classical sphere coordinates
\bea 
\sum_i  x_i^2 =  \sum_i  \bar x_i^2 = 1 \;,
\eea 
with $ x_i = { J_i \over \sqrt { N^2 -1 } } $ and 
$ \bar x_i = { \bar J_i \over \sqrt { N^2 - 2N  } } $.  
Then to  leading order in $N$ 
\be
J_i\rightarrow N x_i,\qquad \bar{J}_i\rightarrow N\bar{x}_i\;, 
\ee
but we will actually also have to consider ${ 1\over N} $ corrections in the definitions of $ x_i  , \bar x_i $ for our derivation of the large $N$ action, as we will need to understand subleading orders in the fields for a subtle term.  For the spherical harmonics 
\bea\label{normalisations}
Y_{lm}(J_i) &\rightarrow& Y_{lm}(x_i) = (N^2-1)^{\frac{l}{2}}\sum_{i} f_{lm}^{(i_1\ldots i_l)} x_{i_1}\ldots x_{i_l}\cr
 Y_{lm}(\bar J_i)&\rightarrow&  Y_{lm}(\bar x_i) = (N^2-2N)^{\frac{l}{2}}\sum_{i} f_{lm}^{(i_1\ldots i_l)} \bar x_{i_1}\ldots \bar x_{i_l}\;.
\eea

We remind that the CS gauge fields $A_\mu^{(1)}$ are
 functions of the fuzzy sphere coordinates $J_i$, whereas
 $A_\mu^{(2)}$ are functions of $\bar{J}_i$. It will also 
be useful for us to define fields with $J_i$  and  $\bar J_i$ interchanged
\bea\label{hatmap} 
A_\mu^{(1)} = \sum_{l=0}^{N-1} a_\mu^{lm}  Y_{lm}(J_i)
&\textrm{,}& \hat A^{(1)}_\mu = \sum_{l=0}^{N-1} a_\mu^{lm}
 Y_{lm}(\bar J_i)\nn\\ 
A^{(2)}_\mu = \sum_{l=0}^{N-2} \bar a_\mu^{lm}  Y_{lm}(\bar J_i)
&\textrm{,}& \hat A_\mu^{(2)} = \sum_{l=0}^{N-2} \bar a_\mu^{lm}
 Y_{lm}(J_i)\;,
\eea
 We can view $ A^{(1)}_{\mu} \to \hat A^{(1)}_{\mu} $ as a map from $ End ( \bVp ) $ to $ End ( \bVm ) $, which preserves the $\SU(2)$ transformation properties. Likewise $ A^{(2)}_{\mu} \to \hat A^{(2)}_{\mu} $ is a map from $ End ( \bVm ) $ to $ End ( \bVp ) $. In the classical limit, the 
four fields above are determined by the degrees of freedom 
 $ a^{l m }_{\mu }  , \bar a^{lm}_{\mu}  $ of two classical fields 
on $S^2$. We will define such classical fields
 $ {\bf A}^{(i)} $   in the course of the derivation.

\subsection{Higgs mechanism}

As a first step in the analysis, we will look at the Higgs mechanism for the gauge field components in the $\mu=0,1,2$ membrane worldvolume directions. 
In the presence of the background scalar field $C^I=(fG^\a,0)$, the CS action for the two $\U(1)$ fields on 
the classical (large $N$) 2-sphere becomes a Yang-Mills action for a single $\U(1)$ dynamical gauge field.

In more detail, the pure gauge ($A_\mu^{(i)}$) terms in (\ref{ccdag}) give 
\be
 -f^2\Tr[J A^{(1)\mu} A_\mu^{(1)}]-f^2\Tr[\bar{J}A^{(2)\mu}A_\mu^{(2)}] +2f^2\Tr[ A^{(1)\mu}G^\a A^{(2)}_\mu G^\dagger _\a]\label{gaugeacti}\;.
 \ee
The first trace is over $ \bVp $ and equals 
\bea
-f^2\Tr[J A^{(1)\mu} A_\mu^{(1)}] &=&
 -f^2 (N-1) \Tr_{\bVp} [  A^{(1)\mu} A_\mu^{(1)}] \cr 
&\simeq& - f^2 N \Tr_{\bVp} [  A^{(1)\mu} A_\mu^{(1)} ] \;.
\eea 
The second trace is over $ \bVm $, but can be mapped using (\ref{hatmap}) 
and can be written as 
\bea 
 - f^2 \Tr[ \bar{J}A^{(2)\mu} A_\mu^{(2)}]  
 &=& - f^2 N \Tr_{ \bVm } [   A^{(2)\mu}  A_\mu^{(2)}] \cr 
&\simeq& - f^2 N \Tr_{\bVp} [ \hat A^{(2)\mu } \hat A_{\mu}^{(2)}  ] \;.
\eea 
Using $\Tr_{ \bVp} \to  N \int d^2 \s \sqrt { \hat h } $ and $A^{(1)}\to {\bf A}^{(1)}$ along with  
$\hat{A}^{(2)}\to {\bf A}^{(2)}$ we have,
 always in the strict large $N$ limit,\footnote{Note here that we could have instead expressed everything in terms of $\Tr_{ \bVm} $ and then used   
 $\Tr_{ \bVm} \to  N \int d^2 \s \sqrt { \hat h } $, along 
with $A^{(2)}\to {\bf A}^{(1)}$, 
$\hat{A}^{(1)}\to {\bf A}^{(2)} $,  to get the same large $N$ action,
 as was the case for the matter fluctuations. This is a reflection of 
a $\mathbb Z_2$ symmetry inherent in the derivation of the
 classical sphere action.}
\be
 -N^2f^2\int d^2 \s \sqrt { \hat h }( {\bf  A}^{(1)\mu}  {\bf A}^{(1)}_\mu
+  {\bf A}^{(2)\mu}  {\bf A}^{(2)}_\mu) \;.  
\ee
 For the third term in (\ref{gaugeacti}) we need to commute $G^\a$ past the fuzzy spherical harmonics. Using (\ref{offdiagtrans}) one can show the following exact relation, which is valid for any finite value of  $N$
\be
G^\alpha  Y_{lm}(\bar{J}_i) G^\dagger_\alpha = 
J   Y_{lm}(J_i) - l  Y_{lm}(J_i)\label{exacteq}\;.
\ee
This uses the fact that the expansion of the spherical harmonics involves symmetric and traceless combinations of the $J_i$'s in a crucial way.\footnote{We expand on the derivation of (\ref{exacteq}) in Appendix \ref{derivation}.} Being careful about applying the correct normalisation factors, the above becomes in terms of the coordinates on the sphere
\bea
G^\alpha  Y_{lm}(\bar{x}_i) G^\dagger_\alpha &=& 
\Big(\frac{N^2-1}{N^2-2N} \Big)^{\frac{l}{2}}\Big((N-1)  Y_{lm}(x_i) - l   Y_{lm}(x_i)\Big)\nonumber\\
&=&(N-1)  Y_{lm}(x_i) -\frac{l(l+1)}{2N} Y_{lm}(x_i)+\cO\Big( \frac{1}{N^2}\Big)\label{rescaledcoo}\;.
\eea
The reader might be puzzled about the fact that we have taken extra care to keep track of the subleading term in the above expression, given that we are interested in the large $N$ limit. The reason for this will become apparent 
very soon, in Eq.\,(\ref{extra}). For the moment, note that 
\be
-l (l+1) Y_{lm}(x_i)= \hat \Box Y_{lm}(x_i) \;.
\ee
 As a result, one has that at large $N$ the term in question becomes\footnote{Once again, one could have chosen to expresses everything in terms of $\bVm$ and $\bar{x}_i$ instead to obtain the same action, as explained in Appendix \ref{otherway}.}
\bea\label{inquestion}
2f^2\Tr[ A^{(1)\mu}G^\a A^{(2)}_\mu G^\dagger _\a] 
&=&  2 f^2 N \Tr_{\bVp}  [ A^{(1)} \hat A^{(2)\mu } +
 f^2   A^{(1)}_{\mu}  \hat \Box \hat A^{(2)\mu } ] \cr 
& \to &  f^2\int d^2 \sigma \sqrt{\hat h} 
\;\Big( 2N^2 {\bf A}^{(1)}_\mu  {\bf A}^{(2)\mu} + {\bf A}^{(1)}_\mu \hat \Box {\bf  A}^{(2)\mu}\Big)\;.
\eea
The combined part of the scalar kinetic term involving only gauge fields gives  
\be
 - \int d^3x \;\Tr \Big(D_\mu C^{\dagger}_ID^\mu C^I\Big) \rightarrow f^2 N^2 \int d^3x \; d^2 \s \sqrt{\hat h}\Big( \frac{1}{N^2}{\bf A}^{(1)}_\mu \hat \Box {\bf  A}^{(2)\mu} -( {\bf A}_\mu^{(1)}-  {\bf A}_\mu^{(2)})^2 \Big)\;.  \label{termul}
\ee
 We now form linear combinations of the two U(1) gauge fields on the classical $S^2$,
 \bea\label{lincomb}
 A_\mu &=&\frac{1}{2}({\bf A}^{(1)}_\mu +  {\bf A}^{(2)}_\mu) \cr 
B_ \mu&=& \frac{1}{2}( {\bf A}^{(1)}_\mu -  {\bf  A}^{(2)}_\mu)\;,
 \eea
so that (\ref{termul}) becomes
\be
f^2 \int d^3x d^2\sigma\; \sqrt { \hat h }\Big( A_\mu \hat \Box A^\mu - B_\mu \hat \Box B^\mu -4N^2  B^\mu B_\mu \Big)\;. 
\ee
Setting that aside, the Chern-Simons action for $A_\mu^{(1)}$ and $A_\mu^{(2)}$  gives in the classical limit
\bea
S_{\textrm{CS}} &=&\int d^3 x
\frac{k}{4\pi}\epsilon^{\mu\nu\rho}\Tr[A^{(1)}_\mu\d_\nu A^{(1)}_\rho  - A^{(2)}_\mu\d_\nu A^{(2)}_\rho]\nonumber\\
&\rightarrow&
N\frac{k}{2\pi}\int d^3 x\; d^2 \s \sqrt { \hat h }\Big(\epsilon^{\mu\nu\rho}B_\mu F_{\nu\rho}\Big)\;,
\eea
where  $F_{\mu\nu}$ is the usual abelian field strength, defined as
\be
F_{\mu\nu} = \d_\mu A_\nu - \d_\nu A_\mu\;.
\ee
Thus the total action for $A_\mu$ and $B_\mu$  at leading $N$ order is 
\be
\int d^3 x \; d^2 \s \sqrt { \hat h }\;\Big(N\frac{k}{2\pi} \epsilon^{\mu\nu\rho}B_\mu F_{\nu\rho}-4f^2N^2 B_\mu B^\mu + f^2  A_\mu \hat \Box A^\mu - f^2 B_\mu \hat \Box B^\mu\Big)\;.
\ee
We now notice that $B_\mu$ is an auxiliary field, which can be eliminated through its equation of motion. The $B_\mu \hat \Box B^\mu$ is subleading in $N$ and can be dropped so as to get
\be
 B^\mu = \frac{1}{8f^2 N }\frac{k}{2\pi}\epsilon^{\mu\nu\lambda}F_{\nu\lambda}\label{bsol}\;,
 \ee
 leading to the following expression, up to a total derivative
\be\label{extra}
\int d^3x d^2 \sigma \;\sqrt{\hat h}\;\Big(-f^2 \d^a A_\mu \d_a A^\mu  -\left(\frac{k}{2\pi}\right)^2\frac{1}{8 f^2} F^{\mu\nu}F_{\mu\nu}\Big)\;.
\ee
 Note importantly that the first term came from the $\frac{1}{N}$ corrections in (\ref{rescaledcoo}) and is a part of the gauge field strength on the sphere that otherwise would have been absent. The second term, coming from the manipulation of the CS part of the action, has become a Yang-Mills kinetic term. In 3d a Yang-Mills action contains one degree of freedom, whereas the Chern-Simons action contains none. The extra degree of freedom comes from the disappearance of a Goldstone boson as in \cite{Mukhi:2008ux}. As we have seen in Subsection\,\ref{nofr} this is the scalar mode $\tilde r- \phi$.

\subsection{Other contributions from the scalar kinetic term }

Apart from   (\ref{gaugeacti}), which contains only gauge fields, there are additional contributions coming from the scalar kinetic term  (\ref{ccdag})
\bea
- \Tr \Big(D_\mu C^{\dagger}_ID^\mu C^I\Big)&\to&\Tr\left[-\partial_\mu q^\dagger_{\dot\a} \partial^\mu q^{\dot\a}
-\frac{N-1}{4} \partial_\mu s_i \partial^\mu s_i 
-(\partial_\mu r)(\partial^\mu s_i)J_i +\frac{i}{4}
\epsilon_{ijk} J_k \partial_\mu s_i \partial^\mu s_j\right]\nonumber\\
&&-if \Tr[\d_\mu r^\dagger_\a(A_\mu^{(1)}G^\a-G^\a A_\mu^{(2)})+\d_\mu r^\a(A_\mu^{(2)}G^\dagger_\a-G^\dagger_\a A_\mu
^{(1)})]\;.
\eea
After some algebra, the terms on the first line give in the large $N$ classical limit
\be
N \int  d^3x d^2 \s \sqrt { \hat h }[- \frac{N}{2} \d^\mu Q^{\dot \beta}_\alpha \d_\mu  \bar Q^\alpha_{\dot \beta} 
-\frac{N}{4} \partial_\mu(\phi+\tilde{r}) \partial^\mu(\phi+\tilde{r})-\frac{N}{4}h^{ab}\partial
_\mu A_a \partial^\mu A_b]\;.
\ee
The terms on the second line reduce to 
\be
 \frac{if}{2}\Tr\left(\d_\mu s_i[J_i,A^{(1)\mu}]\right)
\ee
resulting in
\be
f \int d^3x \; \Tr(\d_\mu A_a)\d^a (A^\mu+B^\mu)\label{gaugecond}\;.
\ee
However, after substituting the leading result (\ref{bsol}), one observes that the term coming from  $B_\mu$ is also subleading in $N$ and can be readily dropped from 
(\ref{gaugecond}). This is a nice feature since, had it not done so, it would have led to a higher derivative interaction. Hence in the classical limit the second line yields
\be\label{2ndline}
f N \int d^3x d^2 \sigma\sqrt{\hat h} \; (\d_\mu A^a\d_a A^\mu)\;.
\ee

\section{Final action for fluctuations}\label{D4interpretation}

We can now collect the above as well as  all other bosonic
 terms for the action on the classical 2-sphere, to obtain the
 bosonic piece of the action for fluctuations 
\bea\label{initaction}
S^B&=&\int d^3x d^2\sigma \;\sqrt{ \hat h} 
 \Big[-\left(\frac{k}{2\pi}\right)^2\frac{ 1}{8 f^2} F_{\mu\nu}
 F^{\mu\nu} -\frac{N^2 \mu^2}{2}F_{ab}F^{ab} -\frac{N^2}{4}\partial
_\mu A^a \partial^\mu A_a  +Nf \d_\mu A_a\d^a A_\mu\cr
&&\qquad\qquad\qquad\quad -f^2 \d^a A_\mu \d_a A^\mu  + N^2 \mu^2 F_{ab}\omega^{ab}\Phi  - \frac{N^2}{4} \partial_\mu  
\Phi \partial^\mu \Phi- N^2 \mu^2 \d_a \Phi \d^a \Phi- N^2 \mu^2 \Phi^2  \cr
&& \qquad\qquad+ 2N^2 \mu^2 \left( (\nabla_{a})_{ \gamma}^{\alpha }    Q^{\dot \a }_{\alpha }   
\right) (\delta^\gamma_\beta + x_i (\tilde \sigma_i)^\gamma_\beta)
\left( ( \nabla^a)^{  \beta }_{\mu}Q_{\dot \a }^{ \mu  }\right)- \frac{N^2}{2} \d^\mu Q^{\dot \alpha}_\alpha (\delta^\alpha_\beta + x_i (\tilde \sigma_i)^\alpha_\beta) \d_\mu  
Q^\beta_{\dot\alpha}\Big]\;,\nonumber\\
\eea
where we have renamed $(\tilde r + \phi) = \Phi$. This action does not have the canonical form of a  5d theory but we will perform a set of rescalings for the fields, which will bring the right relative factors for the various
terms.

In \cite{Hanaki:2008cu} it was observed that in order to convert the $C^I$ kinetic term of the undeformed  ABJM
theory in (\ref{abjmaction}) 
\bea 
S = -\int d^3 x \;\Tr( D_\mu C^I  D^\mu C_I^{\dagger})  
\eea 
to the physical form 
\bea 
S_{phys} = -T_2  \int d^3 x ( D_\mu X^I D^\mu X_I^{\dagger} )\;,
\eea 
where $T_2 = [l_p^3 (2 \pi)^2]^{-1}$ is the membrane tension and $X^I$ are spacetime coordinates 
with dimensions of length, one needs the redefinition $ C^I = T_2^{1/2} X^I $. Note that the scalar fluctuations will also have dimensions of length in this language.

The classical solution $ C^I = (R^\alpha, Q^{\dot \alpha}) = (f G^\alpha, 0) $ implies that the physical theory has a solution $ X^I = (T_2^{-1/2} f G^\alpha,0) $, hence $f\to T_2^{-1/2} f$. What this means in practice, is that all terms originating from the potential and scalar kinetic term will pick up a factor of $T_2$ apart from (\ref{2ndline}), which will pick up a $T_2^{1/2}$. The $F_{\mu\nu}^2$ and $(\d_a A_\mu)^2$ contributions remain the same, since they came from terms which did not involve scalars. We can encode this into a `physical' form of the action\footnote{We keep the names for the various fields as before. We hope that this will not cause confusion.}
\bea\label{physaction}
S^B_{phys}&=& T_2\int d^3x d^2\sigma \;\sqrt{ \hat h}  \Big[-\frac{k^2  T_2^{-1}}{32\pi^2  f^2} F_{\mu\nu}
 F^{\mu\nu} -\frac{N^2 \mu^2}{2}F_{ab}F^{ab} -\frac{N^2}{4}\partial
_\mu A^a \partial^\mu A_a   +\frac{Nf}{ T_2^{1/2}} \d_\mu A_a\d^a A_\mu\cr
&&\qquad\qquad\qquad\qquad -\frac{f^2}{T_2} \d^a A_\mu \d_a A^\mu  + N^2 \mu^2 F_{ab}\omega^{ab}\Phi  - \frac{N^2}{4} \partial_\mu  
\Phi \partial^\mu \Phi- N^2 \mu^2 \d_a \Phi \d^a \Phi  \cr
&& \qquad\qquad\qquad\qquad\quad - N^2 \mu^2 \Phi^2 + 2N^2 \mu^2\left( (\nabla_{a})_{ \gamma}^{\alpha }  
  Q^{\dot \a }_{\alpha }   \right) (\delta^\gamma_\beta + x_i (\tilde \sigma_i)^\gamma_\beta)
\left( ( \nabla^{ a })^{  \beta }_{\mu}Q_{\dot \a }^{ \mu  }\right)\nonumber\\
&&\qquad\qquad\qquad\qquad
- \frac{N^2}{2} \d^\mu Q^{\dot \alpha}_\alpha(\delta^\alpha_\beta + x_i(\tilde\sigma_i)^\alpha_\beta) \d_\mu 
 Q^\beta_{\dot \alpha} \Big]\nn\\
\eea
and  perform the following rescalings of the fields
\be
 A_\mu \to
 A_\mu\, (4 \pi l_s)\frac{1 }{T^{-1/2}_{2}f}\;,\quad A_a \to A_a \,(4 \pi l_s)\frac{1}{N} \;,\quad \Phi\to
\Phi \,(4 \pi l_s)\frac{1}{N\mu} \;,\quad Q^{\dot \beta}_\alpha \to Q^{\dot \beta}_\alpha \,(4 \pi l_s)\frac{ 1}{N\mu}\;.
\ee
We also rescale the metric on the sphere defining 
$ h_{ab} = \mu^{-2} \hat h_{ab} $ 
so that $ \sqrt { h } =  \mu^{-2} \sqrt { \hat h } $,
 as well as the worldvolume coordinates $x^\mu\to \frac{1}{ 2} x^\mu$, leading to $\d_\mu \to  2 \d_\mu$ and the scaling of the measure  $d^3 x \to 2^{-3}d^3 x$. The action takes the form
\be
\begin{split}
S^B_{phys} = \frac{1}{g_s l_s}
\int d^3 x  d^2 \s \sqrt { h } 
& \Big[-\frac{1}{4}  F_{\mu\nu}  F^{\mu\nu}-\frac{1}{4}F_{ab}F^{ab}
 - \frac{1}{2} \d^a A_\mu \d_a A^\mu -\frac{1}{2}\partial_\mu A^a \partial^\mu A_a   +\d_\mu A_a\d^a A^\mu\\ 
&   
+\Big( (\nabla_{a})_{ \gamma}^{\alpha }  
  Q^{\dot \a }_{\alpha }   \Big) (\delta^\gamma_\beta + x_i (\tilde \sigma_i)^\gamma_\beta)
\Big( ( \nabla^{ a })^{  \beta }_{\mu}Q_{\dot \a }^{ \mu  }\Big)- \d_\mu Q^{\dot \alpha}_\alpha (\delta^\alpha_\beta + x_i (\tilde \sigma_i)^\alpha_\beta) \d^\mu   Q^\beta_{\dot\alpha}\\
&\qquad-\frac{1}{2} \partial_\mu 
\Phi \partial^\mu\Phi- \frac{1}{2}\d_a \Phi\d^a \Phi - \frac{\mu^2}{2} \Phi^2 
+ \frac{\mu}{2}\; \omega^{ab} F_{ab}\Phi  
\Big]\;.
\end{split}\label{split2}
\ee
Note that the rescalings that we have performed are precisely the ones needed to yield the correct $N$ dependence for $(\d_a A_\mu)^2$. This was the term that we got by keeping track of subleading corrections in (\ref{rescaledcoo}). In fact, had it appeared at leading $N$ order in that expression it would have dominated the whole action after the rescalings.

The above action resembles closely the canonical form of an abelian 5d Yang-Mills on a two-sphere, which is the low-energy limit of a D4-brane theory with $g^2_{YM}=g_s l_s $. Using the index $A = \{\mu,a\}$ to include both the flat and angular variables, and the metric $ g_{AB} = ( \eta_{\mu \nu } , h_{ab} ) $ the action finally becomes
\be 
\begin{split}
S^B_{phys} = \frac{1}{g_{YM}^2}\int d^3 x d^2 \s \sqrt { h }\; 
& \Big[-\frac{1}{4}  F_{AB}  F^{AB}-\frac{1}{2} \partial_A 
\Phi \partial^A\Phi - \frac{\mu^2}{2} \Phi^2 - \d_\mu Q^{\dot \alpha}_\alpha  (\delta^\alpha_\beta + x_i (\tilde \sigma_i)^\alpha_\beta) \d^\mu   Q^\beta_{\dot \alpha}\\
 & + \Big( (\nabla_{a})_{ \gamma}^{\alpha }  
  Q^{\dot \a }_{\alpha }  \Big) (\delta^\gamma_\beta + x_i (\tilde \sigma_i)^\gamma_\beta)
\Big( ( \nabla^{ a })^{  \beta }_{\mu}Q_{\dot \a }^{ \mu  }\Big)  + \frac{\mu}{2}\; \omega^{ab} F_{ab}\Phi\Big]\;.
\end{split}\label{split}
\ee
The size of the 2-sphere seen by the fluctuations is set by $ \mu^{-1} $.

Some comments are in order: The fact that the transverse scalars are naturally transformed to a representation in terms of commuting spinors is related to the issue of the exact interpretation of the final action in terms of branes and is a familiar feature of D-brane worldvolume theories on compact spaces. We elaborate on this in Section\,\ref{discussion}. We are also getting a mass term for $\Phi$ as well as an $F\Phi$ interaction term. This is a sign of the nontrivial geometry seen by the fluctuations.

\section{M5 to D4 through Hopf}\label{sechopf} 

We have shown in the previous section that the analysis of 
fluctuations in the membrane theory leads to a $\U(1)$ theory 
on $ \mathbb { R}^{2,1} \times S^2 $. This naturally suggests an interpretation  
 as a D4-brane action. In this section, we will develop the 
D4-brane interpretation further. We start  by looking at the energy 
of  funnel solutions and show that they match the D4-brane interpretation 
just as well as the M5-brane interpretation \cite{Hanaki:2008cu}. We explain the reduction from M5 to 
D4 by using the Hopf fibration of $S^3$ over $S^2$ and the $\mathbb Z_k$ quotient 
 action along the $S^1$ fibre. We then show that the structure of the 
matrices of \cite{Gomis:2008vc} and in fact the field content of the ABJM 
theory can be anticipated by considering   finite matrix 
constructions in fuzzy geometry inspired by 
 the Hopf fibration. We extend these considerations to
speculate on the problem of seeing the classical $S^3$ of M2-M5 
systems in flat space from some appropriate large $N$ membrane theory.

\subsection{Physical radius and D4-brane energy for the funnel }

In the case of the ground-state (zero energy) solutions for the massive deformation, where we have set up our calculation so far, we have seen that the form of the fluctuation action is a field theory on $ \mathbb { R}^{2,1} \times S^2 $.  A simple calculation in favour of the $S^2$ D4-brane interpretation is also available in the energy of the funnel solution.  Following the literature on brane polarisations, it is possible to define a physical radius at large $N$ as
\bea\label{physrad}  
R_{\textrm{ph}}^2 = { 2 \over  N } \Tr( X^I X^\dagger_I) = 8\pi^2 f^2 N l_p^3 =
 2\pi N l_p^3 { k \over s } \;,
\eea 
where $f^2=\frac{k}{4 \pi s}$. The energy of the funnel can then be
 easily extracted from (\ref{hamiltonian}) and is given by
\bea \label{EM5}
E = \frac{T_2^2}{2\pi} \int 
 \frac{2\pi^2}{k} R_{\textrm{ph}}^3dR_{\textrm{ph}} dx_1 \;.
\eea 
The authors of \cite{Hanaki:2008cu} interpreted the
 $R_{\textrm{ph}}^3$ as indicating a 5-brane
wrapping an $S^3/\mathbb Z_k$ in M-theory, 
with $T_{5}=\frac{T_2^2}{2\pi}$ the M5-brane tension and 
$\frac{2\pi^2}{k}$ the volume of an $S^3/\mathbb Z_k$ of unit
 radius ($V_{S_3}=2\pi^2$ is divided in $k$ units).
However, the above formula is equally 
compatible with a 4-brane interpretation in Type IIA, 
which is in fact the one 
 supported by the presence of the fuzzy $S^2$
 obtained from our fluctuation analysis. Eq.\,(\ref{EM5}) can 
be  rewritten as
\bea 
E = { 1 \over g_s^2 l_s^6 (2 \pi )^4 } 
 \int 4\pi \frac{R_{\textrm{ph}}^2}{4}{ R_{\textrm{ph}} \over k } 
dR_{\textrm{ph}} dx_1 \;.
\eea 
Now the ratio $ { R_{\textrm{ph}} \over k  } $ is 
\be
 { R_{\textrm{ph}} \over  k } =
 \sqrt{\frac{ 2 \pi N l_p^3}{k s }}  \qquad \hbox{ (funnel)}\;.
\ee
For the ABJM theory to be weakly coupled one needs $ { N \over k }\to 0  $, so that $k\to\infty$. 
Hence, with $l_p, s $ fixed, the ratio 
$ { R_{\textrm{ph}}  \over k } \rightarrow 0 $. In this limit it is natural to go to a 
IIA picture, with 
\be
 { R_{\textrm{ph}} \over k} \equiv R_{11} = g_s l_s \label{r11}\;.
\ee
We can then write 
\bea
E &=&   { 1 \over g_s l_s^5 (2 \pi)^4}  \int 4\pi  \Big(\frac{R_{\textrm{ph}}}{2}\Big)^2dR_{\textrm{ph}} dx_1\cr
&  =& T_4 \int 4\pi  \Big(\frac{R_{\textrm{ph}}}{2}\Big)^2dR_{\textrm{ph}} dx_1\;,
\eea 
as is expected for a D4-brane  in IIA involving an $S^2$ of radius $\frac{R_{\textrm{ph}}}{2}$ with the correct tension $T_4^{-1} = g_s l_s^5 (2 \pi)^4$. 

For the case of the mass-deformed theory we can similarly define the physical radius of the classical solution, which should correspond to the size of the brane in the D4 description. We have that $ f(\mu)^{2} = \frac{\mu k}{2\pi}$ and therefore
\be\label{Rphys}
R_{\textrm{ph}}^2 =  4\pi N l_p^3  k \mu  \;.
\ee
It is then also easy to  obtain
\be 
 R_{11} = { R_{\textrm{ph}} \over k }  =  \sqrt{\frac{ 4\pi N l_p^3 \mu}{k}   }  \qquad \hbox{ (massive) }
\ee 
and  with this definition of the M-theory radius one can express (\ref{Rphys}) as
\be
R^2_{\textrm{ph}} = R_{11} 4\pi  k N\mu \,l_s^2  = R_{\textrm{ph}} 2 N\mu (2\pi l_s^2)  \;.
\ee
Hence the radius of the two-sphere, evaluated according to the definition (\ref{physrad}), is
\be
\frac{R_{ph}}{2} = N\mu (2 \pi l_s^2)\;.
\ee
Note that one would naturally  expect $R_{\textrm{ph}}$ to be the radius for the sphere that appears in the action for small fluctuations. However, we have explicitly seen that what actually appears in Eq.\,(\ref{split}) is $\mu^{-1}$. This is interesting and we will return to discuss it further 
 in Section\,\ref{discussion}.

\subsection{$\mathbb Z_k$ reduction of  $S^3$}

We have just shown that there is a natural connection (\ref{r11}) between the radius of the proposed M-theory 3-sphere $R_{\textrm{ph}}$ and the M-theory radius $R_{11}$. We now explain these relations by analysing the classical geometry of $ S^3/\mathbb Z_k$. As a warm-up, we start by looking at the case of flat space solutions.

The moduli space of the $\U(N)\times \U(\bar N)$ level $k $
ABJM theory is  the same as that for $N$ M2-branes on $\mathbb C^4/\mathbb Z_k$ \cite{Aharony:2008ug}, namely 
$[(\mathbb C^4/\mathbb Z_k)^N]/S_N$. Here the action of $\mathbb Z_k$ in the target space is 
\be
Z^i\rightarrow Z^i e^{\frac{2\pi i}{k}}\qquad\textrm{with}\qquad i=1,...,4\label{refl}\;.
\ee
As argued above, in order to have a weakly coupled (perturbative) interpretation, we need to take $k\rightarrow \infty$.
Let us first understand the case of $\mathbb  C/\mathbb Z_{k\rightarrow \infty}$. For that example 
\be
Z\rightarrow Z e^{\frac{2\pi i}{k}}\simeq Z\Big(1+2\pi i \frac{1}{k}+...\Big)
\simeq Z+2\pi i \frac{Z}{k}\;.
\ee
Expanding around $Z=v+i 0$, with $\frac{v}{k}\equiv r$ we have that 
\be
Z\rightarrow Z+2\pi i r
\ee
is an invariance, or if $Z=X^1+iX^2$ that $X^2$ is compactified with radius $r$. 

We can do the same analysis for $\mathbb C^4/\mathbb Z_{k\rightarrow\infty}$, expanding around 
the background
\be
Z^1=v+i0\qquad\textrm{with}\qquad Z^2=Z^3=Z^4=0\;,
\ee
which means that if
\bea
Z^1 &=& X^1+iX^2\cr
Z^2&=&X^3+iX^4\;,
\eea
then $X^2$ is again compactified with radius $r$.

Now let us move on to the theory around the fuzzy sphere (or fuzzy funnel) solution, by imposing a 3-sphere constraint on the coordinates $Z$, for instance $|Z^1|^2+|Z^2|^2=R^2$.  One obtains an $S^3/\mathbb Z_k$ solution, as also considered in \cite{Hanaki:2008cu}, since the action of $\mathbb Z_k$ in (\ref{refl}) preserves this sphere constraint.  We can then expand around the vacuum with $Z^1=R+i0$, which means (exactly as above) that $X^2$ is compactified with radius $r=\frac{R}{k}$, and the fluctuations $X^1,X^3,X^4$ take us in the direction of an $S^2$ of radius $\frac{R}{2}$. Thus as $k\rightarrow \infty$, we have the spacetime result $S^3_R/\mathbb Z_k \rightarrow S^2_{R/2}\times S^1_{R/k}$, compatible with the worldvolume picture of the last section.

\subsection{Hopf fibration and the classical limit of the fuzzy two-sphere}\label{Hopffib}

We next try to find a link between the classical spacetime and matrix (finite $N$) descriptions. The ground-state solution of \cite{Gomis:2008vc} was  $C^{I}=(R^\a,Q^{\dot{\a}}) $ with $Q^{\dot{\a}}=0$ and
\bea 
 R^{ 1 } &=& G^{1} \cr 
 R^{2} &=& G^2 \;.
\eea 
In the classical interpretation of the fuzzy (matrix) coordinates $C^I$, we have
\bea 
 R^1 &=& X^1 + i X^2 \cr 
 R^2 &=& X^3 + i X^4 \cr
 Q^1 &=& X^5 + i X^6 \cr 
 Q^2 &=& X^7 + i X^8 \;.
 \eea
We know that that the fuzzy coordinates satisfy $ G^1 G^{\dagger}_1 + G^2 G^{\dagger}_2 = N-1 $ and $G^1 = G^{\dagger}_1 $.  The first one suggests a fuzzy 3-sphere, but the second is an extra constraint which reduces the geometry to a 2d one and we have indeed established that we have a fuzzy $S^2$ through the fluctuation analysis.

The construction  of the fuzzy $S^2$ 
was obtained by
\bea 
 J_i &=& ( \tilde \sigma_i )^{\a}_{\b} G^{\b} G^{\dagger}_{\a} \cr 
 x_i &=& { J_i \over \sqrt{ N^2-1} }\; \Rightarrow
\left\{ \begin{array}{ll} 
x_1 = { J_1 \over \sqrt{ N^2-1} } = { 1 \over \sqrt{ N^2-1} } ( G^1 G^{\dagger}_2 +
 G^2 G^{\dagger}_1) & \\ 
  x_2 = {  J_2 \over \sqrt{ N^2-1} } =  
{ i  \over \sqrt{ N^2-1} } (   G^1 G^{\dagger}_2 -  G^2 G^{\dagger}_1 ) &  \\
x_3 = {  J_3 \over \sqrt{ N^2-1} } =  {1 \over \sqrt{ N^2-1}} ( G^1 G_1^{\dagger} - G^2 G_2^{\dagger} ) &
\end{array} \right.  \;.
\eea 
Following the standard Matrix Theory logic, the $G^i$ should become 
coordinates in spacetime. On the other hand we are saying that 
bilinears in $G^i$ are also spacetime coordinates. 
How can both be true? 
The answer is given by the classical Hopf map  $S^3\stackrel{\pi}{\rightarrow} S^2$.

The description of the Hopf map in classical geometry 
 is precisely that we start with the Cartesian coordinates on the unit $S^3$, $X_1, X_2, X_3, X_4 $, with
\be
 X_1^2 + X_2^2 + X_3^2 + X_4^2 =1
\ee
 and then  go to a set of `twistor-like' variables $ Z^1 = X_1 + i X_2$, $Z^2 = X_3 + i X_4 $, which are related to the Cartesian coordinates on the unit $S^2$ by  
\bea\label{classHopf}  
x_i = ( \tilde \sigma_i )^{ \a}_{\b} Z^{ \b} Z^*_{\a} \;,
\eea 
to get 
\be
 x_i x_i =  ( \tilde \sigma_i )^{ \a}_{\b} 
( \tilde \sigma_i )^{ \mu  }_{\nu } Z^{ \b} Z^*_{\a}
  Z^{ \nu } Z^*_{\mu } = 1\;.  
\ee
The $Z$'s obey $Z^\alpha Z^*_\alpha = 1$. 

The various pieces of our geometric analysis are now falling into place: For an M5 wrapping the $S^3$ the relations between the radii, which we uncovered in a simple way from the spacetime picture, are in fact the ones prescribed by the full realisation of the three-sphere as the Hopf fibration $S^1\hookrightarrow S^3\stackrel{\pi}{\rightarrow} S^2$, with $R_{S^2} = \frac{R_{S^3}}{2}$ and the M-theory direction being identified with the $S^1$ fibre of radius $R_{S^1} = R_{S^3}$.\footnote{For a concise   summary of the facts pertaining to the $S^3$ Hopf fibration see   \cite{Nakahara:1990th,Ishii:2008tm}.} The latter is a natural choice since this is the smallest circle that appears in the problem. For large $R_{S^3}$ the orbifold $Z^i\sim e^{\frac{2\pi     i}{k}}Z^i$ acts by shrinking the size of the fibre and hence the M-theory radius. In the limit of $k\to \infty$, $R_{11}$ is very small and one is left with the $S^2$ base which supports a D4-brane in the IIA theory.  This is consistent with the fact that our large $N$ fluctuation action is an action for fields on $S^2$.

Going in the other direction, one should start seeing the structure of the M-theory circle at finite $k$ with the D4-brane becoming an M5-brane wrapping an $S^1/\mathbb Z_k\hookrightarrow S^3/\mathbb Z_k\stackrel{\pi}{\rightarrow} S^2$.  However, as we have explicitly shown and we will discuss again in the next subsection, this is not something that can be captured by the large $k$ semiclassical fluctuation analysis around the solution of \cite{Gomis:2008vc}.  Nevertheless, the calculation of physical quantities will be sensitive to the size of the $S^1$, hence to $k$.  Due to the high degree of supersymmetry, some quantities such as the energy (\ref{EM5}) that we calculated on the classical solution at large $k$, give the correct finite-$k$ form. Hence, they admit both the expected large-$k$ Type IIA as well as a flat space ($k=1$) M-theory interpretation. For general physical quantities one needs new finite-$k$ methods to calculate properties of the the flat space M5-brane from the ABJM theory. In order to see the geometry of the extra circle more explicitly one would have to consider nonperturbative effects which we will come back to in Section\,\ref{discussion}.

\subsection{Hindsight is 20/20 : ABJM structure from fuzzy Hopf  } 
In retrospect all the key properties of $G^{ \a} , G^{\dagger}_{\a} $ as summarised in Section\,\ref{fuzzystr} are determined by their interpretation as a fuzzy (matrix) realisation of the $S^2$ base of the Hopf fibration of $S^3$. Indeed, suppose we want to lift the equations (\ref{classHopf}) from classical geometry to finite matrices. We need matrices for $Z^{\a}$ which we call $ G^{\alpha }$. Now the coordinates $x_i$ transform in the spin-$1$ representation of $\SU(2)$. If we want to build them from bilinears of the form $G^{ \dagger} G $ we need $G , \Gd $ to transform in the spin-$\frac{1}{2}$ representation. In the usual fuzzy 2-sphere, the $x_i$ are operators mapping an irreducible $N$-dimensional $\SU(2)$ representation $ V_N $ to itself. It is possible to do this in an $\SU(2)$-covariant fashion because the tensor product of spin-$1$ with $V_N$ contains $V_N$. Since $G^{\alpha } $ are spin-$\frac{1}{2}$, and $ { 1\over 2 } \otimes V_{N} = V_{N+1} \oplus V_{N-1} $ does not contain $V_N$, we need to work with reducible representations.  The simplest thing to do would be to consider $ V_N \oplus V_{N-1} $.  The next simplest thing is to work with $ V_{N} \oplus ( V_{N-1} \oplus V_1 ) $. The latter possibility is chosen by the construction of \cite{Gomis:2008vc} and allows a gauge group $\U(N) \times \U(N)$ which has a $\mathbb Z_2$ symmetry of exchange needed to preserve parity. To realise the $G^{\alpha} , \Gd_{\alpha } $ as solutions we need to set them equal to matter fields $R^{\alpha } , R^{\dagger}_{\alpha} $ which are in the bifundamental of the $\U(N)\times \U(N)$ gauge group.

So the rather curious property of the $G$ matrices, notably the difference between the last of (\ref{GGDrels}) and (\ref{GDGrels}) which led directly to the fact that $ \bVp = V_N $ and $ \bVm = V_{N-1} \oplus V_1 $ follows from requiring a matrix realisation of the fuzzy $S^2$ base of the Hopf fibration. These in turn lead to the $\SU(2)$ decompositions of $ End ( \bVp ), End ( \bVm ) , Hom ( \bVp , \bVm ), Hom ( \bVm , \bVp ) $, for the fluctuations and the Yang-Mills action. This novel realisation of fuzzy $S^2$ is thus intimately tied to the Hopf fibration. The usual fuzzy $S^2$ has also been discussed in terms of the Hopf fibration, where the realisation of the $\SU(2)$ generators in terms of bilinears in Heisenberg algebra oscillators yields an infinite dimensional space which admits various projections to finite $N$ constructions \cite{Balachandran:2005ew}.  In that case the $x_i $ are not bilinears in finite matrices.

Further light on the geometry can be shed by considering the expectation values of the operators $x_i , G , \Gd $.  The situation for the $x_i$ is familiar \cite{Madore:1991bw,   Kabat:1997im}.  If we choose the state of maximum spin, we see $ \langle x_3 \rangle = 1 $.  If we choose states with lower spins we find $ \langle x_1^2 + x_2^2 \rangle = 1 - \langle x_3^2 \rangle $. This gives a description of the fuzzy $S^2$ as a sequence of fuzzy circles fibering a discretised axis from N-pole to S-pole. The same manipulation can be done with the $x_i$ in our construction to reveal the picture of a fuzzy $S^2$ at finite $N$. 

Such an interpretation is also desirable for the $G, \Gd $, given the usual role of D-brane transverse coordinates as matrices.  Here we need to consider $ \langle- | G |+\rangle , \langle + | \Gd | - \rangle $ to extract, at finite $N$, numbers which can be compared for example with $Z^1 = Z_1^* $. The $x_i , G, \Gd $ are operators in $ \bVp \oplus \bVm$ which is isomorphic, as a vector space, to ${ \bf V}_N \otimes V_2 $. The endomorphisms of ${ \bf V }_N$ correspond to the fuzzy sphere. The $N $ states of ${ \bf V }_N $ generalise the notion of points on $S^2$ to noncommutative geometry. The 2-dimensional space $V_2$ is invariant under the $\SU(2)$. It is acted on by $G, \Gd $ which have charge $ +1, -1$ under the $\U(1)$ (corresponding to $(J , \bar J ) $) acting on the fibre of the Hopf fibration. So in a sense the $S^1$ fibre has been reduced to a noncommutative geometry consisting of two points. It is interesting to note that this emerges naturally from our large-$N$ fluctuation analysis: The field content of the theory we derived on $S^2$ can be organised using an extension of the algebra of the Cartesian coordinates $x_i$ by $G , \Gd $ subject to the relation $ x_i = { 1 \over N } ( \ts_i )^{ \b}_{ \a } G^{\a} \Gd_{\b} $. This was seen in Subsection\,\ref{transscal} where the action for the transverse scalars in terms of bosonic spinors $Q^{\dot \a}_{ \a}$ on the sphere was packaged elegantly using $ q^{ \dot \a } = Q^{\dot \a}_{\a} G^{ \a} $ into (\ref{fullquadratic}).

We emphasise that the $S^2$ we constructed in matrix geometry from the ground-state solution really describes the noncommutative version of the base of the Hopf fibration in a finite $N$ setting. In the large $N$ limit it approaches the $S^2$ base of the standard Hopf fibration of $S^3$. The fluctuation analysis does not have enough modes to describe the full space of functions on $S^3$, even if we drop the requirement of $\SO(4)$ covariance and allow for the possibility of an $\SU(2) \times \U(1)$ description.  We elaborate on this below.
 
For $k\to\infty$, where the fluctuation action around the classical solutions of \cite{Gomis:2008vc} is valid, the $S^1/\mathbb Z_k$ fibre (parametrised by a coordinate $y$) becomes a very small circle. As we explained above,  the only remnant of the circle in the matrix construction is the multiplicity associated with having states $|+\rangle , |-\rangle$ in $ \bVp $ and $ \bVm $. A classical description of the $S^3$ metric as a Hopf fibration contains a coordinate $y$ transverse to the $S^2$. However, the matrix fluctuations of the solution are mapped to functions on $S^2$ and lead to a field theory on $S^2$.  This means we do not see the coordinate $y$ as a worldvolume coordinate for the emergent D4-brane. We do not see it as a transverse coordinate either, which is as expected since in M-theory this is an M5-brane wrapping the $y$ direction. Indeed, we have shown in Section\,\ref{D4interpretation} that only the mode  $\phi+\tilde{r}$ (and not $\phi-\tilde{r}\equiv y$) appears in the action for fluctuations.

In sum, the action we have derived does not contain $y$ neither as a transverse scalar nor as a worldvolume coordinate and it can be consistently interpreted according to Subsection\,\ref{Hopffib} as the double-dimensional reduction of the M5-brane action along the $\mathbb Z_k$ quotient of the $S^3$ Hopf fibre (with length ${R \over k }$). This is a IIA reduction of M-theory on the same small circle.

\subsection{Multi-membrane actions and $S^3$ at large $N $ } 
One might ask what kind of multi-membrane action would contain 
a fuzzy $S^3$ solution and  fluctuations 
which are appropriate for an $ \SU(2) \times \U(1)$ covariant 
description of the $S^3$. 
A dimensional  reduction  on the  $S^1$ fibre is compatible 
with the symmetry. In such a description, 
there are fields on $S^2$ with  extra index $n$, as expected 
for a simple $S^1$ reduction 
\be
\phi(x,y)=\sum_n \phi_n(x) e^{in 2\pi y/R}\;.
\ee
The blow-up of a fuzzy $S^2$ into a fuzzy $S^3$ was analysed for instance in \cite{Ishii:2008ib}. In that analysis, the fibre direction $y$ is Fourier transformed by replacing the $\d_y$ with a semi-integer $-i q$.  The vacuum solution in that case has fuzzy sphere solutions $L_i^{(j_s)}$ of various sizes $j_s$ that are therefore labelled by an extra semi-integer index $j_s$, identified with $q$ above.  Then all the multiplicities $N_s$ are taken to be equal, $N_s=N,\forall s$, with a $\mathbb Z_N$ identification of the $N$ blocks, where $N\rightarrow \infty$.  In that fashion one obtains the fuzzy 3-sphere spherical harmonics from the fuzzy 2-sphere spherical harmonics with an extra index $q$. In our case, the absence of the extra index $q$ signifies that the theory is dimensionally reduced to the $q=0$ sector, or that we have a very small circle (as we argued, in the $k\rightarrow \infty$ limit). The extra fluctuations $ \hat E_{k1} , \hat F_{1k}$ from Subsection\,\ref{harmdec} are uncharged under the $\U(1)$ action of $ ( J , \bar J ) $, so they do not provide the tower of charges corresponding to the $S^1$.

As we have argued, the structure of the ABJM theory, with fields acting on $V_N \otimes V_2 $ can be given a noncommutative geometry interpretation in terms of a discrete bundle over a fuzzy $S^2$. We may speculate that a description capable of seeing more of the geometry of the Hopf fibration would involve a vector space $ V_N \otimes V_K $, where $V_K$ is a $K$-dimensional vector space. This would give a $K$-point approximation to the fibre and would suggest a gauge group $\U(N)^{\otimes K } $, with $ K \rightarrow \infty $ giving the classical geometry of the $S^3$. This could presumably draw on  an embedding of 
 \cite{Ishii:2008ib,Ishii:2008tm} in an ABJM type membrane action. 
 Another way to construct a theory that contains a classical $S^3$, described in an $\SU(2) \times \U(1)$ covariant manner, might be to realise the standard Holstein-Primakoff construction with Heisenberg oscillators \cite{Balachandran:2005ew}.  If we are to set fields in the theory to $ a , a^{\dagger} $ in a solution, we need a $\U(\infty ) $ theory. This could be viewed as a $K \rightarrow \infty $ version of a construction with $ V_N \otimes V_K $.  The question we would like to pose is whether an appropriate field theory with $\U ( \infty ) $ gauge group and containing a realisation of $S^3$ as a Hopf fibration correctly describes the large $N$ limit of membranes. A related question is to understand the relation of such a construction to the large $N$ limit of ABJM.  These questions may well lead to a new perspective on the solutions to the BLG fundamental identity in terms of infinite dimensional algebras \cite{Chu:2008qv,Bandos:2008jv}.

\section{Discussion and future  avenues }\label{discussion}

There are a number of issues that emerge from our analysis and would need to be better understood: 
\begin{itemize}
\item {\bf The number of spherical branes}

  Since we have derived a $\U(1)$ theory in $ {\mathbb{R}}^{2,1} \times S^2$ it is clear   that the solutions of \cite{Gomis:2008vc} describe a single D4-brane. We may obtain an   extension to multiple D4-branes by considering reducible representations of the   algebra satisfied by the $G^\alpha$ matrices. Such reducible representations are easy   to construct (see appendix \ref{multi5}). Some formal similarities between the current   fuzzy $S^2$ and the $\SO(4)$-covariant fuzzy $S^3$ constructions become apparent in   the proof. If one had knowledge of how to go from spherical ($S^2$) D4-branes to   spherical ($S^3$) M5-branes by blowing up the Hopf fibre over $S^2$, this would   provide a starting point for studying {\it multiple} fivebranes in M-theory.

\item {\bf Regimes and parameters}

  As we discussed, we need $k$ large so that a semiclassical approach around classical   solutions gives a reliable account of the physics, while large $N$ is needed in order   to see a classical geometry. Moreover, for the validity of the 2+1 dimensional   Chern-Simons perturbation theory, we need its 't Hooft coupling $\lambda =   \frac{N}{k}$ to be fixed and small.  Finally, we also need $l_p \rightarrow 0 $ for   the validity of the low energy ABJM action (decoupling).

We have only analysed quadratic fluctuations, but the parameter $ f^{-2}={ 2\pi \over   \mu k } $ controls the terms in the action of higher order in the fields (\ie cubic, quartic, {\it etc.}).  Since $k $ is large with $ \mu $ fixed, the parameter is small. On the other hand, fuzzy and higher derivative corrections are of order $\frac{1}{\mu   N}$, which is also small, as $\frac{N}{k}$ is fixed.

  Note then that the limits considered here are further constrained with respect to the   limits discussed in \cite{Aharony:2008ug}. The issue of what limits can be relaxed or   modified while keeping some of the features our analysis is one that deserves further   study.
 
\item {\bf Dual brane descriptions : D2, D4, M5}

  We have given strong evidence in Section\,\ref{sechopf} that the final action   Eq.\,(\ref{split}), derived from the M2-brane worldvolume, is that of a D4-brane on $   \mathbb{ R }^{2,1} \times S^2$.  A beautiful theme in the study of D-brane   intersections is the reconstruction of the same physics from the lower dimensional   branes as from the higher dimensional ones \eg   \cite{Constable:2001ag,Andrews:2006aw,Papageorgakis:2005xr}.  It will be an elegant   consistency check of the ABJM proposal to recover the same result directly from a   4-brane probe in an appropriate spacetime background. Given the subtleties related to   the counting of vacua in the mass-deformed ABJM theory \cite{Gomis:2008vc}, solving   this problem would also shed some light on important issues of the spacetime   interpretation.  Some hints about the nature of the spacetime geometry probed by the   dual D4-brane can be inferred from the fact that preserving some supersymmetry in the   2+1 noncompact directions requires performing a Maldacena-N\'u\~nez-type twisted   compactification of the higher codimension worldvolume theory \cite{Maldacena:2000yy}.   In that case, as well as in the case of 2+1 dimensional Chern-Simons system studied in   \cite{Maldacena:2001pb}, the radial evolution of the topologically nontrivial compact   sphere was crucial to the consistency of the twisting in the spacetime background   solution. However, the example at hand seems to point towards the D4-brane wrapping a   topologically trivial but dynamically stable 2-cycle due to external flux, in the   spirit of \cite{Andrews:2005cv,Andrews:2006aw}. This expectation is supported by   considering the spacetime description of the M2-M5 system, for which the fully   backreacted geometry sourced by membranes polarised into fivebranes in the presence of   flux was found in \cite{Pope:2003jp,Bena:2004jw,Lin:2004nb}.  Using the M-theory interpretation   of the ABJM model, the geometry describing the D4-brane configuration should be   obtainable by performing the $\mathbb Z_k$ quotient of the former and taking   $k\to\infty$. Our analysis suggests that the action for fluctuations around these   D4-brane solutions should see a KK-spectrum with a scale set by the background flux   ($\mu^{-1}$) and independent of the size of the brane ($R_{\textrm{ph}}$). A similar   feature is also encountered in the study of fluctuation actions around giant graviton   solutions in $\textrm{AdS}_5\times S^5$ \cite{Das:2000st}.

  Whatever the geometric description, which we will not try to tackle in this paper, the   implementation of the twisting requires the flat space 5d Lorentz invariance to first get broken   to $\SO(1,4)\to \SO(1,2) \times \U(1)_{34}$. After replacing the 3-4 plane with a sphere there   is only a local $\U(1)_{34}$ invariance left. The theory is then twisted with a $\U(1)_R$ Cartan   subgroup of the R-symmetry group and the $Q^{\dot\alpha}_\alpha$'s of Eq.\,(\ref{split}) should   be spinors under the twisted local rotation group, $\U(1)_T$   \cite{Andrews:2005cv,Andrews:2006aw}. The full details of the twisting are closely connected to   the description of fermions, which we did not address here, so we will also leave a full   investigation as an open question for future work.  A related intriguing question, is whether   the same action we have derived from M2, which has structures expected in a Type IIA picture,   can be obtained directly from a multiple D2-action.

  Alternatively one may consider working from the beginning in a dual picture of   $5$-branes in M-theory.  The starting point would be the M2-M5 system, as described by   the M5-worldvolume. At any given fixed radius, there is an $S^3$ in the M5-worldvolume   and an $\mathbb R^4$ transverse to both the M2's and the M5. The transverse $\mathbb R^4$   contains an $S^3$ surrounding the M2-M5 system.  Consider both the worldvolume $S^3$   and the transverse $S^3$ as Hopf fibrations.  When considering small fluctuations of   worldvolume fields around a configuration where the transverse $S^3$ is of vanishing   size and the longitudinal $S^3 $ is large, the quotient acts effectively on the   longitudinal $S^3$ to give a reduction to IIA.  By considering fluctuations of the   M5-action in this quotient spacetime for $k\to\infty$, it should be possible to   recover our large $N$ action.

  Another interesting question is to recover the $\frac{1}{N}$ corrections from these   different pictures. We have sketched how these could emerge from the Matrix   description in Subsection\,\ref{nofr}. From the D4-brane point of view they are expected to   arise as corrections in a noncommutative U(1) theory, where the noncommutativity   parameter goes like $\frac{1}{N}$. Seeing $\frac{1}{N}$ corrections from the M5   perspective will likely require an improvement of our current knowledge of 5-brane   actions.

\item {\bf Fermions, Supersymmetry }

  We have focused on the bosonic part of the fluctuation action, while leaving a   detailed study of the fermionic terms for the future. As we have already mentioned,   doing so would help in elucidating the twisting and the amount of preserved   supersymmetry, hence the nature of the D4 compactification. Some of the fermions would   be expected to behave as scalars or vectors under the twisted local rotation group,   $\U(1)_T$, in order for the fields to appropriately furnish supersymmetry multiplets   in 3d.

\item {\bf Nonperturbative effects and the M-theory circle}

  One way to see the structure of the M-theory direction would be to look at momentum   modes along the circle. These are D0-branes in Type IIA.  In the standard connection   between M5-branes and D4-branes, they translate into 4d-instanton solutions embedded   in the D4-brane worldvolume theory.  Hence an avenue to seek the physics of the hidden   direction would be to study instanton sectors in the $\U(1)$ theory we have obtained.   This could be done following \cite{Hosomichi:2008ip} who investigated instanton   effects in the undeformed ABJM theory, while also showing that such effects will   start contributing at four-derivative order in an $l_p$ expansion.  Similar   conclusions in the context of scattering were reached in \cite{Hirayama:2008cu}.

\end{itemize}

\section{Summary and conclusions}\label{conclusions}

In this paper we analysed the equations, symmetries and fluctuations for the classical `M2-M5' fuzzy sphere solutions of the $\U(N)\times\U(N)$ mass-deformed ABJM theory of \cite{Gomis:2008vc}, and equivalently the fuzzy funnel solution of undeformed ABJM. We found them to correspond to a fuzzy $S^2$, as opposed to the fuzzy $S^3$ previously conjectured.

As a warm-up, we showed that a newly found set of defining equations for the Guralnik-Ramgoolam $\SO(4)$-covariant fuzzy $S^3$ are not compatible with the BPS equations in ABJM, except for the case of $\SU(2)\times\SU(2)$ gauge group, that is the BLG $\cA_4$-theory.  We then explicitly displayed the symmetries of the solution to find that the physical fluctuations are only invariant under a single $\SU(2)$, indicating a fuzzy $S^2$, as opposed to an $\SU(2)\times \SU(2)\simeq\SO(4)$ expected for a fuzzy $S^3$. The fluctuations were also found to admit an expansion in terms of fuzzy $S^2$ spherical harmonics (plus some zero modes that decoupled in the large-$N$ limit).

A detailed calculation gave the action for bosonic fluctuations in the large-$N$ limit, Eq.\,(\ref{split}), in which the fuzzy $S^2$ matrix algebra approaches the algebra of functions on the commutative (`classical') $S^2$, and the result was expressed as a U(1) Yang-Mills theory on $\mathbb R^{2,1}\times S^2$. This crucially involved a novel version of the Higgs mechanism for the three dimensional Chern-Simons-matter action. The answer is compatible with an interpretation in terms of a D4-brane wrapping the $S^2$ in Type IIA. 

We then explained how the above is in agreement with a spacetime picture in which an M5-brane wraps an $S^3/\mathbb Z_k$, with the three-sphere  given by the Hopf fibration of an $S^1/\mathbb Z_k$ over $S^2$, and identified $S^1/\mathbb Z_k$ with the M-theory circle. In the $k\to \infty$ limit one has a (double) dimensional reduction and a D4-brane wrapping the $S^2$ in Type IIA string theory, as opposed to an M5-brane wrapping the $S^3/\mathbb Z_k$ in M-theory. We also discussed how the Hopf structure persists in the finite-$N$ matrix description, in a new realisation of the fuzzy $S^2$ base manifold.

In conclusion, the problem of finding a perturbative formulation of multiple membrane theory that would obtain the full classical geometry of $S^3$ at large $N$ remains open. As seen from the above, one needs to avoid the large-$k$ $\mathbb Z_k$ projection, and that seems impossible in ABJM theory. Having a perturbative multiple membrane action at large $N$ in flat space (as opposed to $\mathbb C^4/\mathbb Z_k$) would seem to be needed for that goal.

\section*{Acknowledgements}
\noindent 
We would like to thank James Bedford, David Berenstein, David Berman, Robert de Mello Koch, Aki Hashimoto, Koji Hashimoto, Katsushi Ito, Shiraz Minwalla, Sunil Mukhi and Dan Thompson for discussions and comments.  HN would like to thank the Queen Mary Physics Department for the opportunity to visit which resulted in the beginning of this work while CP for hospitality during the course of the project.  SR is supported by an STFC grant ST/G000565/1.  HN's research has been done with partial support from MEXT's program ``Promotion of Environmental Improvement for Independence of Young Researchers'' under the Special Coordination Funds for Promoting Science and Technology, and also with partial support from MEXT KAKENHI grant nr. 20740128.

\begin{appendix}

\section{Some useful formulae for the parallel scalars }\label{scalarids}

There are several useful identities that can be used to obtain the
action for the matrix fluctuations in Section\,\ref{matrixflucts}, involving the change of basis from
traceless symmetric part of $ V \otimes \bar V $ of $ \SU(2)$ to spin-$1$, where $V$ is the fundamental
\bea 
&& s_i =  s_{\a}^{\b} (\tilde \s_i)^{\a}_{\b} \cr 
&& s^{\a}_{\b} =   { 1\over 2 }  s_i (\tilde \s_{i} )^{\a}_{\b} \cr 
&&  J^{\a}_{\b}  = { ( N-1) \over 2 } \delta^{\a}_{\b} 
+   { 1\over 2 } J_i ( \tilde \s_{i} )^{\a}_{\b} \cr 
&& J_i =  ( \tilde \sigma_i )^{\a}_{\b } J^{\b}_{\a} \;.
\eea 

\subsection{Identities for the $s_i s_i$ and $r$-$s$ fluctuations}
In order to avoid the proliferation of indices in the calculation,  one can also define shorthand notation for several combinations of fields that appear frequently. We have collected a number of such definitions and identities in the following.

\subsubsection{$s_is_i$ fluctuations}
Inner products
\bea 
&& s^{\a}_{\b} s^{\b}_{\a}  = {1 \over 2} s^2 \cr 
&& J^{\a}_{\b} J^{\b}_{\a} = N ( N-1) \cr 
&& J_i J_i = ( N^2 -1 ) \cr 
&& s^{\a }_{\b} J^{\b}_{\a} = { 1 \over 2 } s_i J_i  = { 1 \over 2 }
(s. J) \cr  
&&  J^{\a }_{\b} s^{\b}_{\a} = { 1 \over 2 } J_i s_i  = { 1 \over 2 }
(J. s) \cr
&& \Tr(s\wedge J \wedge s) = i \epsilon_{ijk}\Tr(s_i J_j s_k)\;.
\eea 
General quadratic products
\bea 
&& J^{\a}_{\mu} s^{\mu}_{\b } = { ( N-1) \over 2 } s^{\a}_{\b} +
 { (J. s) \over 4 } \delta^{\a}_{\b} -
 {i \epsilon_{ijk} \over 4 } J_i s_j ( \tilde \s_k )^{\a}_{\b} \cr
&&  s^{\a}_{\mu } J^{\mu}_{\beta } =
  { ( N-1) \over 2 } s^{\a}_{\b}
 +  {  (s .  J) \over 4 } \delta^{\a}_{\b} 
- {i \epsilon_{ijk} \over 4 } s_i J_j ( \tilde \s_k )^{\a}_{\b} \;.
 \eea
Cubic products
\bea\label{cubic}
&& J^{\a}_{\mu} s^{\mu}_{\b} J^{\b}_{\a } 
= { 1 \over 4 } ( N-1) ( J. s + s. J )   - {1 \over 4 }
J\wedge s \wedge J \cr 
&& s^{\a}_{\mu} J^{\mu}_{\beta} s^{\beta}_{\a} 
= { 1 \over 4 }  ( N-1) s^2 - { 1\over 4 } s \wedge J\wedge s \cr
&&  J^{\a}_{\mu} s^{\mu}_{\beta} s^{\beta}_{\a} 
= { 1 \over 4 } ( N-1) s^2 - { 1\over 4 } J\wedge  s\wedge  s \cr
&&s^\a_\b J^\b _\g J^\g_\a= \frac{1}{2}N (s.J) \nonumber\\
&&J^\a_\b J^\b _\g s^\g_\a= \frac{1}{2} N(J.s)\;.
\eea
Quartic traced products
\bea\label{quartictraced} 
 \Tr( J^{\b}_{\mu} s^{\mu}_{\a} J^{\a}_{\nu} s^{\nu}_{\beta } ) 
& =& -{ 1 \over 4 }(  N-1 ) \Tr ( s^2 )+ { 1 \over 8 } \Tr (( J. s)^2 +
(s. J)^2 ) \cr 
&&   -  {1  \over 4 } ( N-1)  \Tr(  s \wedge  J\wedge  s ) 
  - { 1 \over 16 } \Tr \left([ J_i ,s_j ] [ J_i , s_j ]\right) \nn\\
\Tr(s^\a_\mu s^\mu_\b J^\b_\nu J^\nu_\a )
&=&\frac{1}{4}N(N-1)\Tr(s^2) -\frac{1}{4}N \Tr(s\wedge J\wedge s)\;.
\eea 
Other useful formulae
\bea 
&& ( \tilde\s_{i} )^{\a}_{\b} ( \tilde\s_i )^{\mu}_{\nu } =  2 \delta^{\alpha
}_{\nu } \delta^{\mu}_{\b }    - \delta^{\a}_{\b} 
\delta^{\mu}_{\nu } \cr
&& G^{\a} \bar J_i \bar J_j \Gd_{\a}  = ( N-3) J_i J_j + (N-1)
\delta_{ij}  + i \epsilon_{ijk} J_k\;. 
\eea 
We further define
\bea 
&& \Box s_k \equiv  [ J_p , [ J_p , s_k ]]\;.  
\eea 
Note that
\bea 
\Tr ( s_i \Box s_i ) &=&  - \Tr [ J_k , s_i ] [ J_k ,s_i ] \cr 
J_i s_j J_i & = &  (N^2-1) s_j - { 1\over 2 } \Box s_j 
\eea 
and
\bea\label{s2identities}
\frac{1}{2} \Tr (s\wedge J\wedge \Box s) &=& 2 (N^2-1)\Tr(s\wedge J\wedge s) -\Tr
(J\wedge s \wedge J)(J.s+s.J)\cr
&& -2\Tr(J.s)^2 + 2(N^2-1)s^2- \Tr (s.\Box s)\cr
\frac{1}{2}  \Tr (J\wedge s\wedge\Box s) &=& 2 (N^2-1)\Tr(s\wedge J\wedge s) -\Tr
(J\wedge s \wedge J)(J.s+s.J)\cr
&&  -2\Tr(s.J)^2 + 2(N^2-1)s^2- \Tr (s.\Box s)\cr
\Tr (s\wedge J\wedge\Box s) - \Tr (J\wedge s\wedge\Box s) &=& 4 \Tr ((s.J)^2 - (J.s)^2)\;.
\eea

\subsubsection{$r$-$s$ fluctuations}
We also give the following useful definitions for the
$r$-$s$ fluctuations
\bea\label{rsidentities}
J.r.J  &\equiv&  J_i r J_i \cr
J\wedge s \wedge J &\equiv & i \epsilon_{ijk} J_i s_j J_k\cr
 J_i J_j J_i  &=&   (N^2 -5) J_j \cr 
\frac{1}{2}\Box (s.J) &=& \frac{1}{2} \Box s.J + 2 (J\wedge s\wedge J
)\cr
\frac{1}{2}\Box (J.s) &=& \frac{1}{2} J. \Box s + 2 (J\wedge s\wedge J
)\cr
\frac{1}{2}(\Box (s.J) + \Box (J.s)) &=& \frac{1}{2} ( \Box
s.J+J. \Box s)+ 4 (J\wedge s\wedge J) 
 \cr 
 \frac{1}{2}\Tr (r J. \Box s) &=& (N^2-1)\Tr (r(J.s)) - \Tr (J.r.J)
 (J.s) - 2 \Tr( r (J\wedge s \wedge J))    \cr  
 \frac{1}{2}\Tr (r \Box s. J) &=&  
 (N^2-1)\Tr (r(s.J))- \Tr (J.r.J) (s.J) - 2 \Tr (r (J\wedge s \wedge
 J))\cr
\frac{1}{2}\Tr (r s.\Box J + r\Box J.s) &=& 2\Tr (r(J.s + s.J))\cr
\Tr A \Box B &=&  \Tr ( \Box A)   B \;.
\eea 
In the last line we can have $ A = r , B =  J.s + s.J $
 for example. One also has
\bea 
  J^{\b}_{\g} s^{\a}_{\mu}  J^{\mu}_{\b} J^{\g}_{\a}
 & =& -{ 1 \over 16 }  ( \Box ( s.J ) +  ( \Box s ).J  ) +  { 1  \over 4 } (N-1) (
   2N+1)(s.J)  - { 1 \over 4 }( N-1)( J.s)\cr
J^{\b}_{\g} J^{\a}_{\mu} 
s^{\mu}_{\b} J^{\g}_{\a}
& =& -{ 1 \over 16 }  ( \Box ( J.s ) + J. ( \Box s )  ) +  { 1  \over 4 } (N-1) (
   2N+1)(J.s)  - { 1 \over 4 }( N-1)( s.J) \cr 
 J^{\b}_{\g} s^{\a}_{\mu}  J^{\mu}_{\b} J^{\g}_{\a} 
    +   J^{\b}_{\g} J^{\a}_{\mu} 
s^{\mu}_{\b} J^{\g}_{\a}
& =&- { 1 \over 16 } (  \Box ( s.J + J.s  )+   ( \Box s ).J  +  J.(
\Box s )  )  + { 1 \over 2}N(N-1)  ( J.s + s.J )\;.\nn\\
\eea 

\subsection{Expressions and identities for $s_i$ decomposition}\label{decompid}

In obtaining the action on the classical sphere in Section\,\ref{fluctsbos} we made use of a set of Killing vectors $K_i^a$. The explicit formulae for the latter are given by
\bea
K_1^\theta = -\sin{\phi} && K_1^\phi = -\cot{\theta}\cos{\phi}\cr
K_2^\theta = \cos \phi~~~  && K_2^\phi = -\cot{\theta}\sin{\phi}\cr
K_3^\theta = 0~ ~~~~~~~&& K_3^\phi = 1\;,
\eea
 as given in \cite{Papageorgakis:2005xr}. The relations between Cartesian and spherical
 coordinates is
\bea
x_1 &=& \sin{\theta}\cos{\phi} \cr
x_2 &=& \sin{\theta}\sin{\phi}\cr
x_3 &=& \cos{\theta}\;.
\eea
Other formulae that we have made use of include
\bea\label{fuzuseful}
 K_i^a K_i^b &=& h^{ab} \cr
 x_i K_i^a &=& 0 \cr
 K_i^a K_i^b \partial_a x_j \partial_b x_j &=& 2 \cr
 \epsilon_{ijk} x_i K_j^a K_k^b &=& { \epsilon^{ab}\over \sin \theta } =
 \omega^{ab}\cr
 K_i^a K_j^b h_{ab} &=& \delta_{ij} - x_i x_j\cr
 K_i^a  ( \d_a K_i^b )  &=& \frac{1}{\sqrt{h}}(\d^b \sqrt{h})\cr
 K_i^a  ( \d_a K_i^b
)\d_b&=&K_i(K_i)=\epsilon_{ijk}x_j\epsilon_{ilm}\d_k(x_l)\d_m=-2x_i\d_i \;.
\eea

\section{Technical aspects of Higgsing} 

\subsection{Derivation of Eq.(\ref{exacteq})}\label{derivation}
It is easy to experimentally verify that commuting a number of $\bar J_i$'s past $G^\alpha$, while then contracting with $G^\dagger_\alpha$ gives
\bea
G^\alpha \bar J_i G^\dagger_\alpha &=& J  J_i -  J_i\cr
G^\alpha \bar J_i\bar J_j  G^\dagger_\alpha &=& J  J_i J_j - 2 J_i J_j + \delta_{ij}\cr
G^\alpha \bar J_i\bar J_j \bar J_k  G^\dagger_\alpha &=& J  J_i J_j J_k - 3 J_i J_j J_k \cr
&& +J_i \delta_{jk} + J_j \delta_{ik}+ J_k \delta_{ij}\cr
&&- i \epsilon_{ijs}J_k J_s - i \epsilon_{jks}J_i J_m- i \epsilon_{kis}J_j J_s \cr
G^\alpha \bar J_i\bar J_j \bar J_k \bar J_n G^\dagger_\alpha &=& J  J_i J_j J_k J_n - 4 J_i J_j J_k J_n\cr
&& +J_i J_j \delta_{kn} +\textrm{permutations}\cr
&&- i \epsilon_{ijs}J_k J_n J_s + \textrm{permutations}\cr
&\vdots&
\eea
and so on. Higher powers of $\bar J_i$ will follow a similar pattern involving $\delta$'s  and $\epsilon$'s for subleading terms, as these are the only invariant tensors of SU(2). However, we are mainly interested in commuting the whole set of fuzzy spherical harmonics. We remind that the latter are defined as
\be
 Y_{lm}(\bar J_i) = \sum_i  f^{(i_1\ldots i_l)}_{lm}\bar J_{i_1}\ldots\bar J_{i_l}\;,
\ee
with $f^{(i_1\ldots i_l)}_{lm}$ a symmetric traceless tensor in $i$. We then have
\bea
 f^{i}_{lm} G^\alpha \bar J_i  G^\dagger_\alpha &=& f^{i}_{lm} (J  J_i- J_i)\cr
 f^{(ij)}_{lm} G^\alpha \bar J_i\bar J_j  G^\dagger_\alpha &=& f^{(ij)}_{lm} (J  J_i J_j -2 J_i J_j) \cr
 f^{(ijk)}_{lm} G^\alpha \bar J_i\bar J_j \bar J_k  G^\dagger_\alpha &=&  f^{(ijk)}_{lm}( J  J_i J_j J_k - 3 J_i J_j J_k) \cr
 f^{(ijkn)}_{lm}G^\alpha \bar J_i\bar J_j \bar J_k \bar J_n G^\dagger_\alpha &=&  f^{(ijkn)}_{lm}(J  J_i J_j J_k J_n - 4 J_i J_j J_k J_n)\cr
& \vdots&\cr
 f^{(i_1\ldots i_l)}_{lm}G^\alpha \bar J_{i_1}\ldots \bar J_{i_l}  G^\dagger_\alpha &=&  f^{(i_1\ldots i_l)}_{lm}(J  J_{i_1}\ldots J_{i_l}- l J_{i_1}\ldots J_{i_l})\;,
\eea
that is only the first two terms survive in each expression, as the $\bar f_{lm}$'s project out all the remaining contributions. Summing over the above, this leads to the exact relation 
\be\label{exactapp}
G^\alpha  Y_{lm}(\bar J_i) G^\dagger_\alpha = J  Y_{lm}(J_i) - l  Y_{lm}(J_i)\;.
\ee

\subsection{Higgsing  from trace over $ \bVm $  }\label{otherway}  
We could have chosen to commute $Y_{lm}(J_i)$ past the $G^\alpha$'s in the third term of (\ref{gaugeacti}). This would have led to the same final answer for the action for fluctuations, but involving a trace over $ \bVm $ in the intermediate steps.\footnote{More precisely this restricts to a trace over $V_{N-1}^- $ of   $ \bVm   $.} In order to see this one needs to make use of the following exact expression, analogous to (\ref{exactapp}),
\be
G^\dagger_\alpha  Y_{lm}(J_i) G^\alpha =  \bar J Y_{lm}(\bar{J}_i) + l Y_{lm}(\bar{J}_i)\;,
\ee
which can then get converted to classical sphere variables using the normalisations (\ref{normalisations})
\bea
G^\dagger_\alpha  Y_{lm}(x_i) G^\alpha &=& \Big(\frac{N^2-2N}{N^2-1}\Big)^{\frac{l}{2}}
(\bar J Y_{lm}(\bar{x}_i) + l Y_{lm}(\bar{x}_i))\cr
 &=&N Y_{lm}(\bar{x}_i) -\frac{l(l+1)}{2N} Y_{lm}(\bar{x}_i)+\cO\Big( \frac{1}{N^2}\Big)\nonumber\\
 &=&  N  Y_{lm}(\bar x_i) + \frac{1}{2N} \hat\Box  Y_{lm}(\bar x_i) 
+ \cO\Big( \frac{1}{N^2}\Big)\label{finally}\;.
\eea
After plugging the above back into (\ref{gaugeacti}) we end up with the same subleading term in (\ref{inquestion}). This is very satisfying, since  we expect to obtain a single answer for the classical action regardless of whether we use $x_i$ or $\bar{x}_i$, although it was not obvious that this would happen at the outset of the calculation. The result reflects the $\mathbb Z_2$ symmetry inherent in the derivation of the classical action. Note however that beyond the strict large $N$ approximation the difference between $x_i$ and $\bar x_i$ (and consequently also the difference between $A^{(i)}$ and $\hat{A}^{(i)}$) could become manifest. To probe that regime one ought to first get a better understanding of the finite $N$ definitions of the fields.

\section{Towards multiple M5-branes}\label{multi5} 

In the case of the usual Myers effect \cite{Myers:1999ps} (multiple D0-branes expanding into spherical D2-branes), the equations of motion for $N$ D0-branes give
\bea 
[ \Phi_i , \Phi_j ] = i f \epsilon_{ i j k } \Phi_k \;,
\eea 
where $\Phi_i$ are the transverse scalars on the D0 worldvolume with $i=1,2,3$ and the rest of them set to zero. These can be solved by $\Phi_i=X_i$ an $N\times N $ irrep of $\SU(2)$, resulting into a single spherical D2-brane, or by $\Phi_i = \Xone_i \oplus \Xtwo_i $,
where $\Xone_i$ is an irrep of size $N_1$ and $\Xtwo_i$ is an irrep of size $N_2$, with $N=N_1+N_2$, resulting into two concentric spherical D2-branes with radii that depend on $N_1,N_2$. This is the case because the sum of two irreps of $\SU(2)$ is a reducible representation 
\be
[  \Xone_i \oplus \Xtwo_i  ,  \Xone_j  \oplus \Xtwo_j ] 
 = [  \Xone_i ,  \Xone_j ] \oplus  [ \Xtwo_i , \Xtwo_j ]  
= i \epsilon_{ijk } (  \Xone_k  \oplus \Xtwo_k ) \;.
\ee
A similar property arises in our case, for the equation of motion
\be
R^{\a} = R^{ \a} \Rd_{\b} R^{ \b} -   R^{ \b} \Rd_{\b} R^{ \a}
\ee
and solutions defined on the space $ \bVp \oplus \bVm $, where $ \bVp = V_{N}^+ $ and $ \bVm = V_{N-1}^- \oplus V_1^- $.
The single 4-brane solutions we have been describing in this paper, with a fuzzy $S^2$, can be written as 
\bea 
 G^{ \a} = \cP_{ V_{N}^+ } G^{ \a}  \cP_{ V_{N -1 }^-  } &&
 G^{\dagger}_{\a} =        \cP_{ V_{N -1 }^-  } G^{\dagger}_{ \a}      \cP_{ V_{N}^+ }\cr 
  G^{ \b }  G^{\dagger}_{\b} = ( N -1 )  \cP_{ V_{N}^+ } &&
  G^{\dagger}_{\b} G^{ \b } = N \cP_{ V_{N-1 }^-  }\;,
\eea 
where $\cP$'s are projectors $ \cP^2 = \cP $ on each subspace of the total space, giving 
\be 
 G^{ \a } G^{\dagger}_{\b}  G^{\b } = N \cP_{ V_{N}^+ }  G^{ \a }
 \cP_{ V_{N-1 }^- }\;,\qquad
  G^{ \b }  G^{\dagger}_{\b} G^{ \a } = ( N-1)  \cP_{ V_{N}^+ }  G^{ \a } 
\cP_{ V_{N-1 }^- }\;.
\ee 
Then the multi-4-brane solutions are defined on a space 
\bea 
&&  \bVp = V_{N_1}^+ \oplus  V_{N_2 }^+  \cr 
&& \bVm =   ( V_{N_1 -1 }^- \oplus V_1^- ) \oplus  
( V_{N_2 -1 }^- \oplus V_1^- )
\eea 
by the ansatz
\bea
&& G^{\a}  = G_1^{\a}   \oplus G_2^{\a}  \cr 
&&  G^{\dagger}_{\a}  = G^{\dagger}_{1~ \a}  \oplus G^{\dagger}_{2~ \a } \;,
\eea 
where as before
\bea 
&& G_{i}^{ \a} = \cP_{ V_{N_i}^+ } G_i^{ \a}  \cP_{ V_{N_i -1 }^-  } \cr 
&& G^{\dagger}_{i  \a} =        \cP_{ V_{N_i  -1 }^-  } G^{\dagger}_{i \a}    
 \cP_{ V_{N_i }^+ }\;.
 \eea
The equations of motion are trivially satisfied since 
\bea 
G^{ \a}  G^{\b}   G^{\dagger}_{\b} & = &  
( N_1 -1 ) G^{ \a}_1  \cP_{ V_{N_1}^+ } \oplus  ( N_2 -1 ) 
G^{ \a}_2  \cP_{ V_{N_2}^+ } \cr 
& =& ( N_1 -1 )  \cP_{ V_{N_1-1 }^- } G^{ \a}_1  \cP_{ V_{N_1}^+ }
     \oplus ( N_2 -1 )   \cP_{ V_{N_2-1 }^- }   G^{ \a}_2  \cP_{ V_{N_2}^+ } 
\cr 
 G^{\dagger}_{\b}  G^{ \b } G^{\a}   & = &  
N_1  \cP_{ V_{N_1-1  }^- }  G_2^{\a} \cP_{ V_{N_1}^+ } 
\oplus  N_2 \cP_{ V_{N_2-1  }^- }  G_2^{\a} \cP_{ V_{N_2}^+ } \;.
\eea   
One could also think of more general bock-diagonal combinations with $N_1 + N_2 + N_3+... = N$, corresponding to concentric spherical D4-branes, the radius of which depends on the dimension of each representation for a fixed $s$, as per (\ref{physrad}) in the funnel case.  These are also all the zero energy solutions for fixed $\mu$ in the mass-deformed case and the collection of partitions of $N$ parametrise the set of classical vacua of the mass-deformed theory, as discussed in \cite{Gomis:2008vc}.\footnote{There is a puzzle related to this counting involving   possible partitions corresponding to solutions with both $R^\alpha \neq 0,Q^{\dot     \alpha}\neq 0$.}

Of particular interest are the possibilities with $m$ copies of $N_m \times N_m$ equally sized blocks, where $m N_m = N$, since in that case the branes are coincident and there is a worldvolume gauge symmetry enhancement to $\U(m)$.  Clarifying how to go to spherical ($S^3$) M5-branes by blowing up the Hopf fibre over $S^2$, perhaps along the lines of \cite{Ishii:2008tm}, in combination with the above could provide a starting point for studying multiple fivebranes in M-theory.

\end{appendix}

\bibliographystyle{utphys}
\bibliography{abfuz}

\end{document}